\documentclass[11pt,a4paper,dvips]{article}
\usepackage{a4p}
\usepackage{cite,mcite}
\usepackage{subfigure}
\usepackage{atlasphysics}
\usepackage{atlas_title,ifthen}
\usepackage{mathptmx}
\usepackage{helvet}
\usepackage{epsfig}
\usepackage{graphics}
\usepackage{graphicx}
\usepackage{epic}
\usepackage{floatflt}
\usepackage{wrapfig}
\usepackage{amsmath}
\usepackage{amssymb}
\usepackage{a4wide}
\usepackage{xspace}
\usepackage[dvips]{color}
\def\thedate{\today}
\newlength{\capindent}
\newlength{\capwidth}
\newlength{\figwidth}
\newcommand{\icaption}[2][!*!,!]{\hspace*{\capindent}%
  \begin{minipage}{\capwidth}
    \ifthenelse{\equal{#1}{!*!,!}}%
      {\caption{#2}}%
      {\caption[#1]{#2}}
      \vspace*{3mm}
  \end{minipage}}
\begin{document}

\begin{titlepage}

\vspace*{-6mm}
\includegraphics[width=3cm]{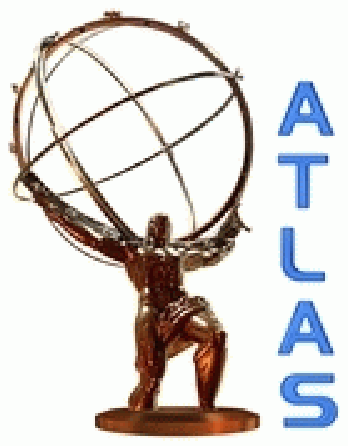} \hfill 
\begin{minipage}[b]{7cm}
\begin{center}
\end{center}
\begin{center}
\mydocversion
\end{center}
\begin{center}
\thedate
\end{center}
\end{minipage}
\hfill \includegraphics[width=3cm]{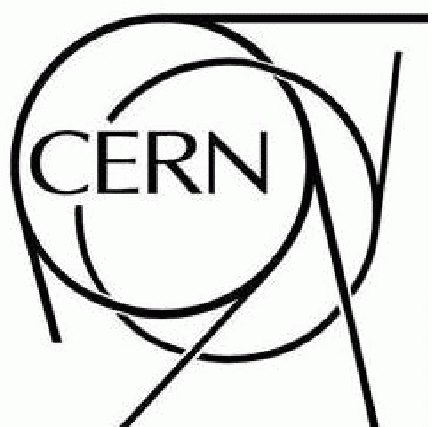}

\title{Re-evaluation of the LHC potential for the measurement of \MW}
\author{Nathalie Besson, Maarten Boonekamp (CEA-Saclay), \\
Esben Klinkby, Sascha Mehlhase, Troels Petersen (NBI)}

\begin{abstract}
  \noindent We present a study of the LHC sensitivity to 
  the \W boson mass based on simulation studies. 
  We find that both experimental and
  phenomenological sources of systematic uncertainties can be strongly
  constrained with \Z measurements: the lineshape,
  ${\mathrm d}\sigma_{\Z}/{\mathrm d}m$, is robustly predicted, and its
  analysis provides an accurate
  measurement of the detector resolution and absolute scale, while the
  differential cross-section analysis, ${\mathrm
  d}^2\sigma_{\Z}/{\mathrm d}y{\mathrm d}\pt$, absorbs 
  the strong interaction uncertainties. 
  A sensitivity $\delta\MW \sim 7 \MeV$ for each 
  decay channel ($\W \ra e\nu, \, \W \ra \mu\nu$), and for an
  integrated luminosity of 10~\ifb, appears as a reasonable
  goal. 
\end{abstract}

\end{titlepage}

\section{Introduction}
\label{sec:introduction}

The Standard Model (SM), now computed at two-loop
precision~\cite{th:mw2loop,th:sw2loop}, is a very predictive
framework. Its most precisely measured parameters $\alpha_{QED}$,
$G_\mu$, and \MZ, provide constraints on the \W boson and top
quark masses, which can be confronted with measurement.
Injecting the measured value of the \W mass and the measured \Z boson
couplings, a definite prediction is given for the top quark
mass~\cite{ex:lep1final}. This prediction, together with the discovery of
the top quark at a compatible
mass~\cite{ex:topdiscD0,ex:topdiscCDF}, has been a major
achievement in high energy physics.\\

\noindent    
The measured values of the \W boson and top quark masses are now more  
precise than their quantum predictions, and provide non-trivial
constraints on the gauge symmetry breaking sector. In the SM, this
translates into limits on the Higgs boson mass~\cite{ex:lep2prel}. Beyond
the SM, constraints are given on the contributions of other heavy
particles, like supersymmetric particles~\cite{th:susyfit}.\\

\noindent    
The \W mass has been measured at UA2~\cite{ex:ua2Mw},
LEP~\cite{ex:lep2prel}, and the Tevatron \cite{ex:tevMwRun1}.
The recent measurement by the CDF Collaboration gives  $\MW = 80.413 \pm 0.048 \gev$, 
yielding a current world average of $\MW = 80.398 \pm 0.025
\gev$~\cite{ex:cdfMwRun2}. In the SM, the resulting Higgs boson mass
uncertainty is about 50\%. Any further improvement in this measurement
will translate into more precise indirect predictions.\\

\noindent
The present paper discusses the LHC prospects for the \W mass
measurement. The expected \W cross-section at the LHC is about
20~nb~\cite{th:fewz2006}. In 10~\ifb\ of data, a benchmark for one year
of integrated luminosity during the first years of stable running,
around $4 \times 10^7$ \W events will be selected in each
exploitable decay channel ($\W \rightarrow e\nu, \mu\nu$), providing a
combined statistical sensitivity of about 1 \mev. Previous
estimates~\cite{lhc:atlasTDR2,lhc:ewlhc,lhc:cmsMw} of the systematic
uncertainties affecting this measurement 
however amount to $\delta\MW \sim 20 \MeV$ per experiment, and to a
combined uncertainty of $\delta\MW \sim 15 \MeV$. The
main sources are the imperfect determination of the experiments
absolute energy scale, and the uncertainties in the \W boson
kinematical distributions (rapidity, transverse momentum), which in
turn stem from proton structure function uncertainties and higher
orders QCD effects. \\

\noindent
The purpose of this paper is to re-investigate the possibilities to 
measure the \W mass with the greatest possible precision. As is
known from the Tevatron, the uncertainties can be significantly
reduced using \Z boson measurements; this approach will be
employed here, with modifications and improvements suggested by the
high \Z statistics expected. Although our discussion is
general, most of our arguments rely on the expected
performance of the ATLAS experiment~\cite{lhc:atlasTDR1}.\\

\noindent
The paper is structured as follows. Section~\ref{sec:methods} summarizes the
\W mass fitting procedure, lists the ingredients needed to
describe the \W distributions used in the fit, and gives a
general description of how these ingredients can be determined. The
sources of uncertainty are then discussed in turn, in
Section~\ref{sec:expunc} (experimental uncertainties),
Section~\ref{sec:theounc} (theoretical uncertainties), and
Section~\ref{sec:envunc} (backgrounds, underlying event, and effects
related to the machine operation).  Correlations between these effects
are discussed in Section~\ref{sec:correlations}, and the results are
given in  Section~\ref{sec:impact}. Section~\ref{sec:conclusions}
concludes the paper.

\section{General discussion}
\label{sec:methods}
This section discusses our technical set-up, the \W and \Z event
selection, the mass fitting procedure, and the problem of
controlling all ingredients entering in the definition of the fitted
distributions. 

\subsection{\W and \Z production. Event generation and simulation}
\label{subsec:simulation}

Throughout this paper, \W and \Z boson samples, and their distributions
and acceptances are computed using the {\tt PYTHIA} general purpose event 
generator~\cite{gen:pythia}. On top of {\tt PYTHIA}, the treatment of
photon radiation in \W and 
\Z decays is done via an interface to {\tt PHOTOS}~\cite{gen:photos}.
The size of the expected samples are computed assuming the NLO \W
and \Z cross-sections, as obtained from {\tt
RESBOS}~\cite{gen:resbos}. These choices are not unique, and the
simulation of physics processes at the LHC, in particular
non-perturbative strong interaction parameters, will obviously need to
be adjusted using the forthcoming data. In this analysis, the effects
of the corresponding uncertainties are estimated either by changing 
parameters in these programs, or by distorting the output
distributions according to our assumptions.\\ 

\noindent
When referring to ``fast simulation'', we mean a simplified simulation of
the ATLAS detector response using scale factors and Gaussian
resolution functions, applied to the generator-level information
obtained above~\cite{sim:atlfast}. When referring to ``full simulation'', we mean the
complete simulation of the ATLAS detector using {\tt GEANT4}~\cite{sim:atlasg4}. 
In our discussions below, and in the absence of real physics data, we
often treat our fully simulated event samples as data samples, and the
fast simulation samples as their Monte-Carlo simulation. The
different detector response in fast and full simulation allows to
emulate the realistic situation where the imperfect detector
simulation is adjusted during data taking. 

\subsection{Signal selection and fitting procedure}
\label{subsec:observables}

\noindent
At hadron colliders, \W and \Z events can be detected and
reconstructed in the $e\nu_e$, $\mu\nu_\mu$, $ee$, and $\mu\mu$ final
states. The hadronic modes suffer prohibitively large background from
jet production; $\tau$ modes can be detected but the $\tau$-lepton
decay produces additional undetected particles in the final state,
diluting the information that can be extracted from
these modes. In \W events, the observables most sensitive to \MW\ are:\\ 
$\bullet$ The reconstructed lepton transverse momentum, \ptl;\\ 
$\bullet$ The reconstructed \W transverse mass,
  $\mtw \equiv \sqrt{2 \ptl \ptn (1 - \cos(\phil - \phin))}$.\\

\noindent
The transverse momentum of the neutrino, \ptn, is inferred from the
transverse energy imbalance, calculated from a summation of energy in
all calorimeter cells. Electrons are measured using the inner detector 
(ID) and electromagnetic calorimeter (EMC). They are reconstructed and
identified with an efficiency of about 65\%, while rejecting
background from jets up to one part in $10^5$; in \W decays, the energy
resolution is about 1.5\%. For muons, the ID is used together with the
muon spectrometer; the reconstruction efficiency is about 95\% and the
relative momentum resolution about 2\%~\cite{lhc:atlasDetector}.\\

\noindent
The \W signal is extracted by selecting events with one reconstructed 
isolated, high-$p_T$ lepton (electron or muon), large missing
transverse energy (due to the undetected neutrino), and low hadronic activity. 
In the following, we
require \ptl$>$20 \GeV, $|\eta_\ell|<2.5$, \met$>$20~\GeV, and require the
hadronic recoil (defined as the vector sum of all calorimetric
transverse energy opposite to the reconstructed \W decay products) to
be smaller than 30 \GeV. These selections have a total efficiency
(trigger and selection) 
of about 20\%, providing a sample of about $4 \times 10^7$~events in each decay
channel. The backgrounds are at the percent level. 
Table~\ref{tab1}
summarizes these numbers. The \ptl\ and \mtw\ distributions obtained 
with fast simulation after the \W event selection are shown in Figure~\ref{fig1}.\\ 

\begin{table}
\begin{center}
\begin{tabular}{ccc}
\hline \hline
Channel & $\W \rightarrow l\nu$ & \Z~$\rightarrow ll$ \\
\hline
Cross-section (pb) & 19800 & 1870 \\
Lepton $\eta$ acceptance & 0.63 & 0.51 \\
Selection eff. & $\sim$ 0.2 & $\sim$ 0.2 \\
 (including acceptance) & & \\
Expected statistics (10~fb$^{-1}$) & $4\times 10^7$ & $3.5\times 10^6$\\
\hline \hline
\end{tabular}
\caption{\label{tab1} Cross-section, $\eta$ acceptance, total
selection efficiency (averaged for electrons and muons) and 
expected sample size for 10~\ifb, in each decay channel.}
\end{center}
\end{table}

\begin{figure}[tp]
\begin{center}
\includegraphics[width=0.48\textwidth]{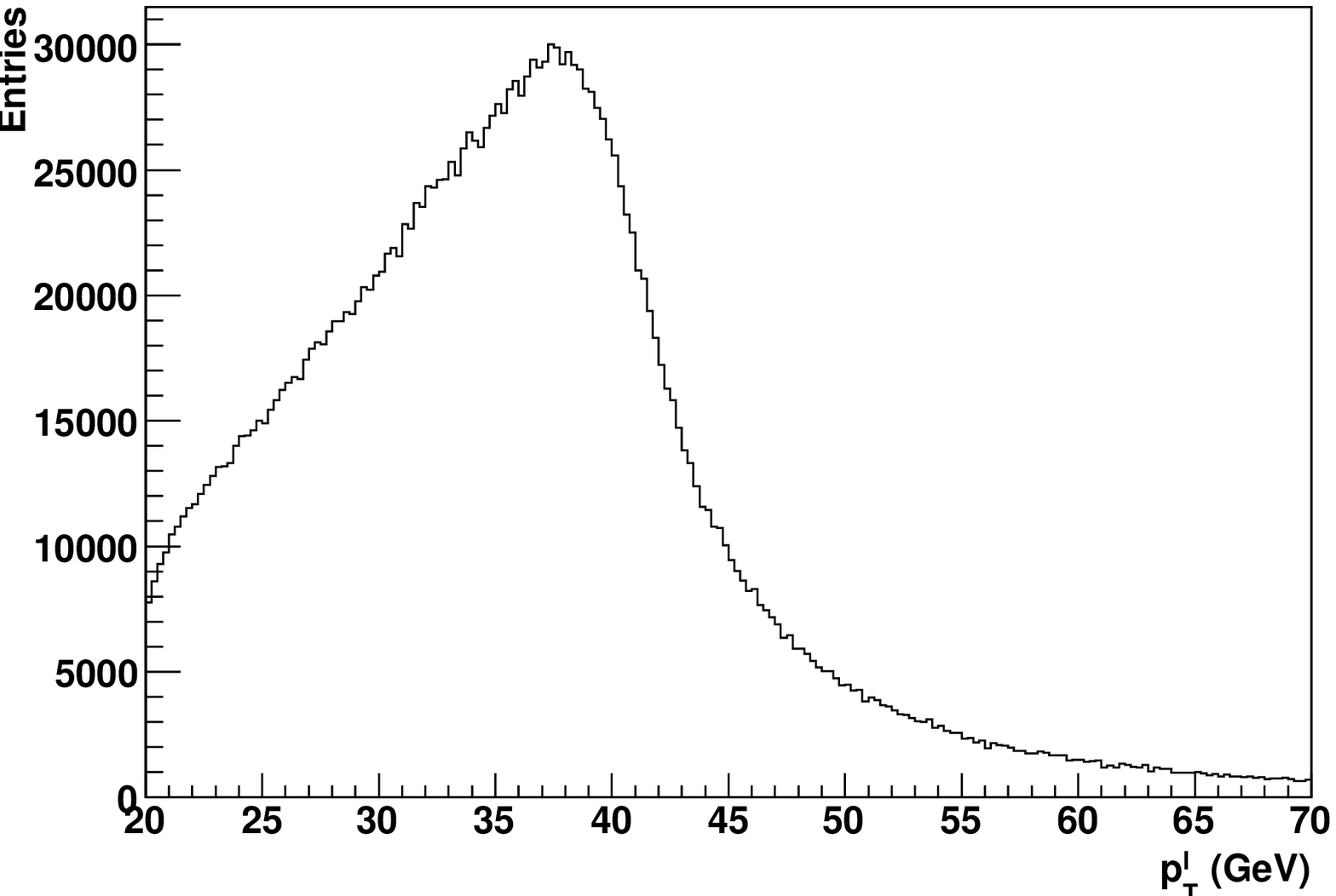}
\includegraphics[width=0.48\textwidth]{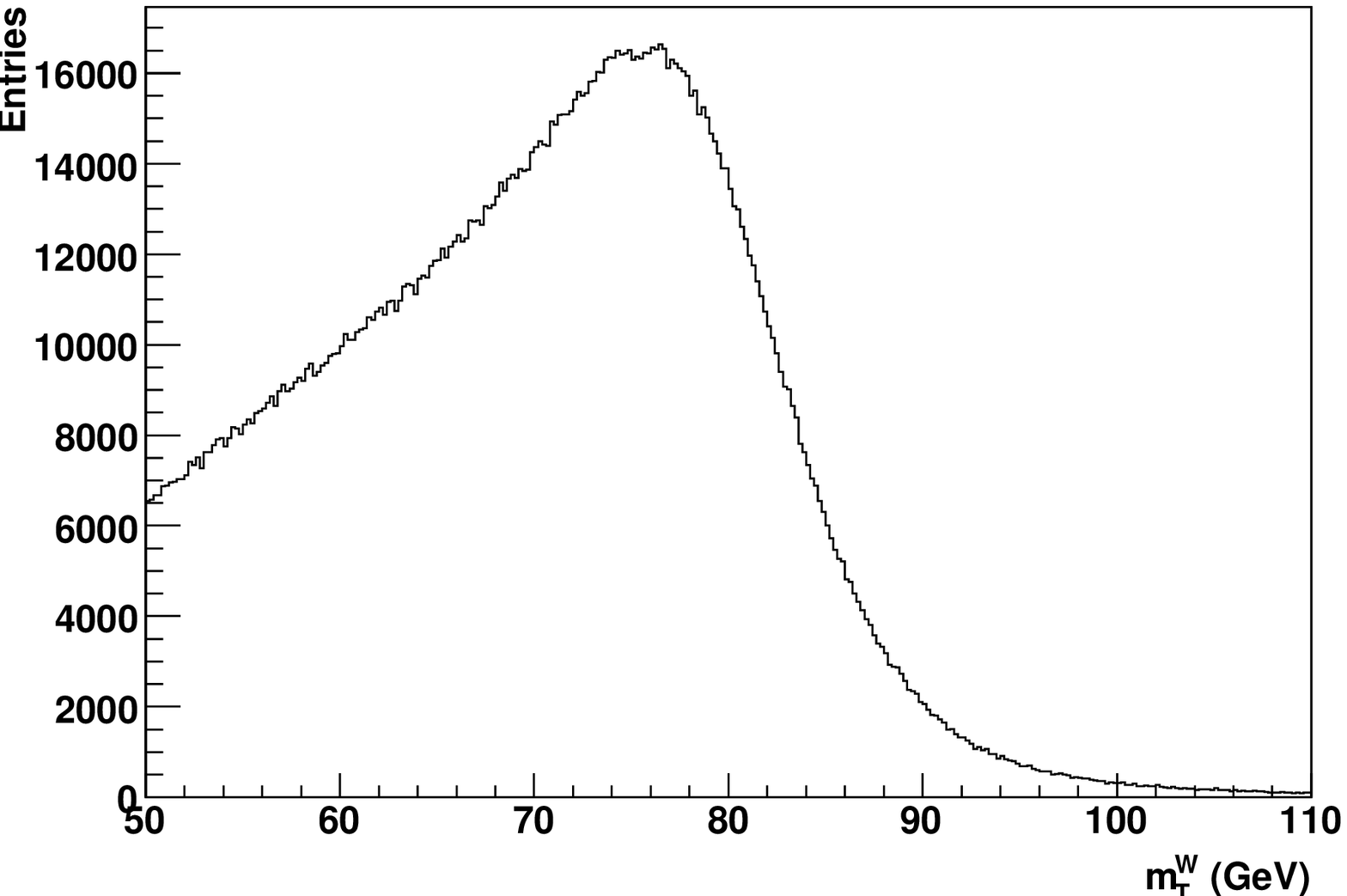}
\caption{\label{fig1} Distributions of the lepton transverse momentum,
\ptl, and of the \W transverse mass, \mtw, after typical \W event
selections (\cf\ text). The Jacobian edges in these distributions
provide sensitivity to the \W mass.}
\end{center}
\end{figure}

\noindent
Based on these distributions, \MW\ can be extracted by
comparing the data to a set of models (or template distributions)
obtained from \W event generation followed by a fast simulation of the
decay particles. The different template distributions are obtained by
varying the value of the \W boson mass parameter in the event
generation. The statistical comparison of the data to the templates
can be performed in various ways; throughout this study we will use a
simple binned $\chi^2$ test. The $\chi^2$ quantifying the
compatibility of a given template distribution with the data is
defined as follows:    

\begin{equation}
\label{eq:chi2}
\chi^2 = \sum_{i} \frac{(n^{obs}_i - n^{exp}_i)^2}{\sigma_i^2}
\end{equation}

\noindent where $n^{exp}_i$ and $n^{obs}_i$ are the number of expected and observed
events (in the template distribution and in the data, respectively) in
bin $i$ of the \ptl\ or \mtw\ spectrum, $\sigma_i$ is the expected resolution, and the 
sum extends over all bins in the fitting window. The Gaussian approximation
used above is justified for large statistics, which is the case we
consider here.\\ 

\noindent
After all $\chi^2$ evaluations, a parabola is fitted through the $\chi^2$
values as a function of \MW. The
procedure is illustrated in Figure~\ref{fig:parabola}. With the
statistics given in Table~\ref{tab1}, each channel provides a
statistical precision of about 2~\MeV\ for data corresponding to an integrated luminosity
of 10~fb$^{-1}$.\\  
 
\begin{figure}
\begin{center}
\includegraphics[width=0.6\textwidth]{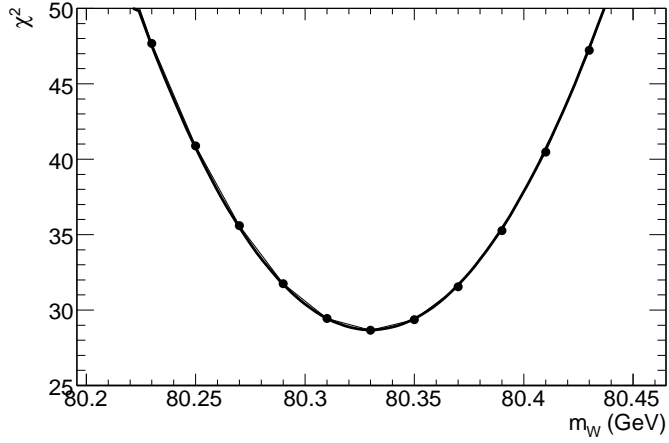}
\caption{\label{fig:parabola} $\chi^2$ value as function of the tested value
of \MW. Each dot represents a comparison between the data and the
template distribution obtained for a given \MW. The curve is the
fitted parabola.}
\end{center}
\end{figure}

\subsection{Required inputs}
\label{subsec:ingredients}

\noindent
For the above procedure to work in practice, one must predict the
\ptl\ and \mtw\ distributions as a function of the \W mass. These
distributions are however affected by many effects, which need to be
included correctly in order to avoid biases in the mass fit. The
needed inputs are listed below.\\

\noindent$\bullet$ {\bf Experimental inputs}: the energy 
scale and resolution need to be known in order to describe the
Jacobian peak correctly (position and spread). Electron and muon   
reconstruction efficiency effects also distort the spectra, if
this efficiency is \pt\ and $\eta$ dependent.\\ 

\noindent$\bullet$ {\bf Theoretical inputs}: the \W rapidity
distribution, \yw, affects the \mtw\ and \ptl\ distributions.
The transverse momentum of the \W, \ptw, directly
affects the \ptl\ spectrum; its impact is weaker on the \mtw\
spectrum. The \yw\ and \ptw\ distributions depend on the proton
structure functions and on higher-order QCD effects. The lepton   
angular distribution in the \W rest frame is of importance for both
\ptl\ and \mtw\, and changes with the \W polarization
\cite{th:tevWpol}. Finally, 
QED effects (photon radiation in the \W decay) shifts the lepton \pt\   
downwards. Since the radiated photons are mostly collinear to the
charged decay lepton, the impact on electrons and muons is different:
the measured muon momentum entirely reflects the momentum loss by
radiation, whereas the electron energy, measured essentially in the
EMC, includes most of the radiated energy.\\ 

\noindent$\bullet$ {\bf Environmental inputs}: these include, among
others, backgrounds surviving the \W selection, underlying event and
pile-up effects on reconstructed energies and momenta, random neutron
hits in the muon spectrometer (``cavern background''), and the impact
of a non-zero beam crossing angle. In all cases, imperfect modelling
of these inputs biases the event reconstruction, leading to distorted
\ptl\ and \mtw\ distributions.\\

\subsection{Propagation of systematic uncertainties}
\label{subsec:propagation}

The impact of underlying physics mechanisms affecting the \W mass determination 
is estimated by producing template distributions of \ptl\ and \mtw\
unaware of the effect under consideration, and fitting them to
pseudo-data including this effect. The resulting bias (i.e. the
difference between the injected and fitted values of \MW) gives the
corresponding systematic uncertainty.\\ 

\noindent
In the simplest case, a physics effect affecting the distributions
(for a given value of the \W mass) can be summarized by a single
parameter. In this case, the induced systematic uncertainty is simply
given by: 

\begin{equation}
\delta \MW = \frac{\partial \MW}{\partial_{rel} \alpha}(\delta_{rel}\alpha)
\label{errprop1}
\end{equation}

\noindent where $\alpha$ is the parameter controlling the parasitic 
physics effect, $\delta_{rel}\alpha$ its relative uncertainty, and
$\delta \MW$ the induced systematic uncertainty on the \W mass. When
applicable, we will quote the uncertainty $\delta_{rel}\alpha$, the
derivative ${\partial \MW}/{\partial_{rel} \alpha}$ and the estimated
$\delta \MW$. As a convention, we normalize ${\partial
\MW}/{\partial_{rel} \alpha}$ in \MeV/\%.\\

\noindent
Sometimes, however, a single parameter is not sufficient. The
uncertainty $\delta \MW$ is then the result of all parameter
uncertainties and their correlations:
\begin{equation}
{\delta \MW}^2 = \sum_{i,j}\frac{\partial \MW}{\partial_{rel}\alpha_i}
               \frac{\partial \MW}{\partial_{rel}\alpha_j}
               (\delta_{rel}\alpha_i) (\delta_{rel}\alpha_j) \rho_{ij}.
\label{errprop2}
\end{equation}

\noindent This happens when the systematic is parametrized by a 
(sometimes empirical) function. In this case, we choose to quantify
the impact by Monte-Carlo propagation: we generate random
configurations of the $\alpha_i$, within their uncertainties, and
preserving their correlations; for each configuration, we produce the
corresponding pseudo-data, and fit them to the
unaffected templates. The spread of the distribution of the fitted \MW\ values
gives the contribution to $\delta \MW$. 

\subsection{The impact of \Z boson measurements}
\label{subsec:strategy}

\noindent
The LHC will produce a large number of \Z events. Their
selection is rather straightforward, requiring two reconstructed
isolated, high-$p_t$ leptons (\ptl$>$20 \GeV, $|\eta_\ell|<2.5$), and low 
hadronic activity (hadronic recoil smaller than 30 \GeV). \\

\noindent
For each useful decay mode (\Z~$\rightarrow ee,\mu\mu$)  
and for $\sim 10$~fb$^{-1}$, around $3.5 \times 10^6$ events should survive
selections. This represents  
a factor 10 less than the expected \W statistics, but the fact that \Z
events are fully reconstructed largely compensates this deficit.
Cross-sections and statistics are summarized in Table~\ref{tab1}.\\

\noindent
The precise knowledge of the \Z mass and width will allow to determine 
the lepton energy scale and resolution precisely. Exploiting the
energy distribution from the decay leptons will also allow to
determine the scale's energy dependence (i.e, the linearity of the
detector response), and the energy dependent resolution function. Once
this is achieved, the \Z transverse momentum will also serve to scale
the measured hadronic recoil to the \Z; together with the measured
lepton transverse momentum, this defines the missing transverse 
energy. Finally, ``tag and probe'' methods~\cite{ex:d0Zxsec} will allow
to determine the lepton reconstruction efficiency.\\
 
\noindent
Although most of the QCD mechanisms affecting \W distributions carry
significant uncertainty~\cite{th:wzpt1}, they affect \W and \Z events
in a similar way. This is 
the case for non-perturbative contributions to the \W transverse momentum  
distributions, but also for parton density (PDF) effects: at the LHC,
the \W and the \Z are essentially sensitive to high-$Q^2$ sea partons, and a
variation of these parameters  will affect the \W and \Z distributions
(in particular \yw, \yz) in a correlated way. Hence, the measurement
of the \Z distributions will help to control the \W ones.\\   

\noindent
The simulation of QED radiation in \W and \Z decays was much improved
recently \cite{gen:photos,gen:horace}. Still, the
measurement of this process (through \eg\ \Z~$\rightarrow \ell\ell\gamma$)
will allow to confirm the predictions. Other sources of uncertainty
(e.g. backgrounds and underlying event) will also be controlled by
auxiliary measurements at the LHC.\\ 

\noindent
The following sections attempt to quantify the above arguments.

\section{Experimental uncertainties}
\label{sec:expunc}

This section assesses the effect of efficiency and resolution in 
the reconstruction of leptons and missing transverse energy.

\subsection{Lepton scale and resolution}
\label{subsec:lepscale}
The \Z boson resonance has been measured very precisely at the lepton
colliders during the 90's~\cite{ex:lep1final}. The \Z boson mass and
width can be exploited as an absolute reference to determine as
precisely as possible the detector energy scale, its linearity and
resolution. \\  

\noindent
The basic method is rather simple, and consists in comparing the 
position and width of the observed mass peak in reconstructed dilepton
events with the \Z boson parameters. A shift of the observed
position of the mass peak, with respect to the nominal \Z peak position,
is corrected for by scaling the detector response, hence determining
the detector absolute scale; the additional spread of the mass
distribution, as compared to the natural \Z boson width, is used to
estimate the resolution.\\

\noindent
The high statistics expected at the LHC, however, imposes a number of
refinements. First, the scale obtained as above is averaged over the
lepton kinematical spectrum, whereas an energy-dependent scale is
needed for a correct description of the Jacobian distributions in \W
events. Secondly, lepton energy resolution effects induce a small but
non-negligible shift in the di-lepton invariant mass
distribution. This shift needs to be subtracted before converting the
scale measured from the \Z invariant mass distribution into the scale
used to describe the Jacobian distributions in \W events. The resulting
method has been described in detail in~Ref.~\cite{lhc:atlasEscale}, and is
summarized below.

\subsubsection{Average detector scale}
\label{sec:averagescale}

We first illustrate the energy-independent method, providing an
average detector scale. Using the {\tt PYTHIA} event
generator~\cite{gen:pythia}, we produce a set of template histograms corresponding
to generator-level  \Z 
lineshapes. The decay leptons are smeared and decalibrated with
different energy scale factors $\alpha$ and resolution functions
$\sigma$. For definiteness, we consider calorimeter-like resolution
functions 
parametrized as $\sigma(E) = a\times\sqrt{E}$. At this stage, $\alpha$ is independent
of the lepton energy. These templates are to be
compared to the data; for our tests, we use an independently simulated
sample as pseudo-data.\\

\noindent
A $\chi^2$ test is then performed between the pseudo-data and each of
the template histograms, as in Section~\ref{subsec:observables}. This
results in a two-dimensional $\chi^2$ 
grid as a function of the smearing parameters. At the vicinity of the
minimum, a paraboloid can be fitted through the points, and the
parameters of this paraboloid give the estimates of the true values of
$\alpha$ and $a$.\\ 

\noindent
The method is tested on a fully simulated \Z~$\to ee$ sample,
corresponding to 30700 events with $85 < m_{ee} < 97$~\GeV; the mass
resolution can be treated as Gaussian over this range.
We find an average resolution parameter $a=0.142\pm 0.003$, and an average mass
scale $\alpha=1.0038\pm0.0002$. Figure~\ref{rome} illustrates the
result, where the fully simulated \Z~$\to ee$ lineshape is compared to
an example template histogram assuming $\alpha=1$ and $a=0.12$, and to 
the best fit result. Very good agreement is obtained; moreover,
the ``measured'' scale and resolution parameters coincide with the
values found when comparing the reconstructed electron energies to
their generation-level values; the electron calibration in the fully 
simulated sample underestimates the true energy by 0.4\%.\\

\begin{figure}[tp]
\begin{center}
\includegraphics[width=.7\textwidth]{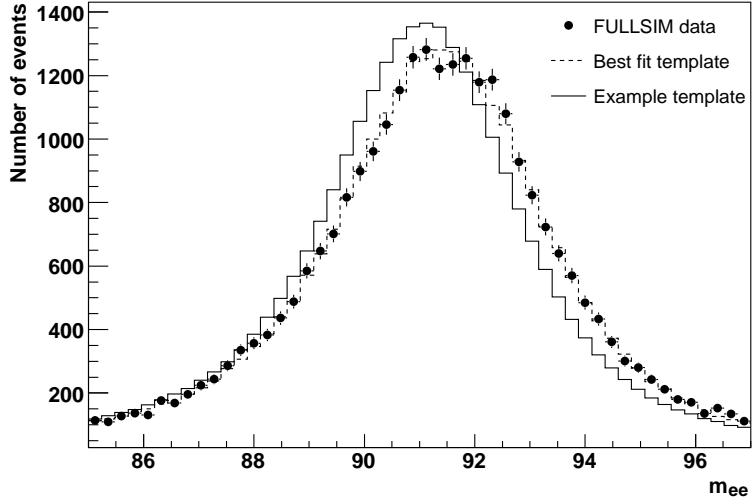}
\caption{\label{rome} Comparison of a fully simulated $\Z \to ee$ sample
(dots) to an initial template example, produced with $\alpha=1$ and
$a=0.12$, and to the best fit result.}
\end{center}
\end{figure}

\noindent
Assuming an inclusive \Z production cross-section of 2~nb per leptonic decay channel and an
integrated luminosity of 10~fb$^{-1}$, the average scale and
resolution parameters can be controlled with a relative precision of
$\delta_{rel}\alpha = 2 \times 10^{-5}$ and $\delta_{rel} a = 2 \times 10^{-4}$. 
Note that these values are not far from the actual uncertainty of the
\Z boson parameters. As far as the absolute scale is concerned, a 
correlation between the induced \W-mass systematic uncertainty and the
\Z boson mass uncertainty might finally appear.\\

\noindent
As discussed in the introduction to this section, the method
illustrated here has an important shortcoming: it only provides a
scale averaged over the \ptl\ distribution expected in
\Z events, which differs from that expected in \W
events. The averaged scale is applicable to \W events only in the
absence of any non-linearity in the detector response. In order to
correctly propagate the \Z calibration measurement to the \W sample, the
scale thus needs to be measured as a function of energy. This is discussed next. 

\subsubsection{Linearity: energy dependent scale and resolution}
\label{sec:linearity}

The above method can be extended as follows. The data and the
templates are classified as a function of the lepton energies.
This leads to templates and pseudo-data labeled $(i,j)$, corresponding
to the event categories where one lepton falls in bin $i$, and the other
in bin $j$. The scale factor $\alpha_{ij}$ and the resolution parameter
$a_{ij}$ are then fitted in every bin.\\

\noindent
In case of small non-linearities of the calorimetry response (i.e.
$\alpha_{ij}, \alpha_i, \alpha_j$ very close to 1), we can
then derive the $\alpha_{i}$ from the $\alpha_{ij}$, writing in first
order approximation that the
mass peak decalibration results from the decay lepton decalibrations:

\begin{eqnarray}
\alpha_{ij} m_{12} &=& \sqrt{ 2 \, \alpha_i E_{1} \, \alpha_j E_{2}  \,
(1-\cos\theta_{1,2}) }, \mathrm{\,\,\,\,\,or}\\
\alpha_{ij} &\sim& (\alpha_i + \alpha_j)/2
\label{eqscale}
\end{eqnarray}
\noindent Writing this for every 
$(i,j)$ gives a linear system which can be solved using least squares.\\

\noindent As for the resolution, the following linear system holds, neglecting
the small contribution from the angular terms in the expression of the
invariant mass resolution:
\begin{eqnarray}
\frac{(\delta{m_{ij}^2})^2}{{m_{ij}^4}} &=& \frac{\sigma_i^2}{E_i^2} +
\frac{\sigma_j^2}{E_j^2} 
\label{eqresol}
\end{eqnarray}
which can again be solved using least squares, yielding the
$\sigma_i$. We thus obtain the energy-dependent resolution function, 
independently of the form used to produce the templates.\\

\noindent 
Examples of results that can be achieved are shown on Figure~\ref{lin}.
With energy bins defined as intervals of 5~\GeV, and a integrated
luminosity of 10~fb$^{-1}$, the scale parameters are reconstructed
with a precision of $2 \times 10^{-4}$, as estimated from the RMS of the
$\alpha$ residuals with respect to the injected function. Similarly,
the resolution parameters are reconstructed with a precision of $2
\times 10^{-3}$.

\begin{figure}[tp]
\begin{center}
\includegraphics[width=0.48\textwidth]{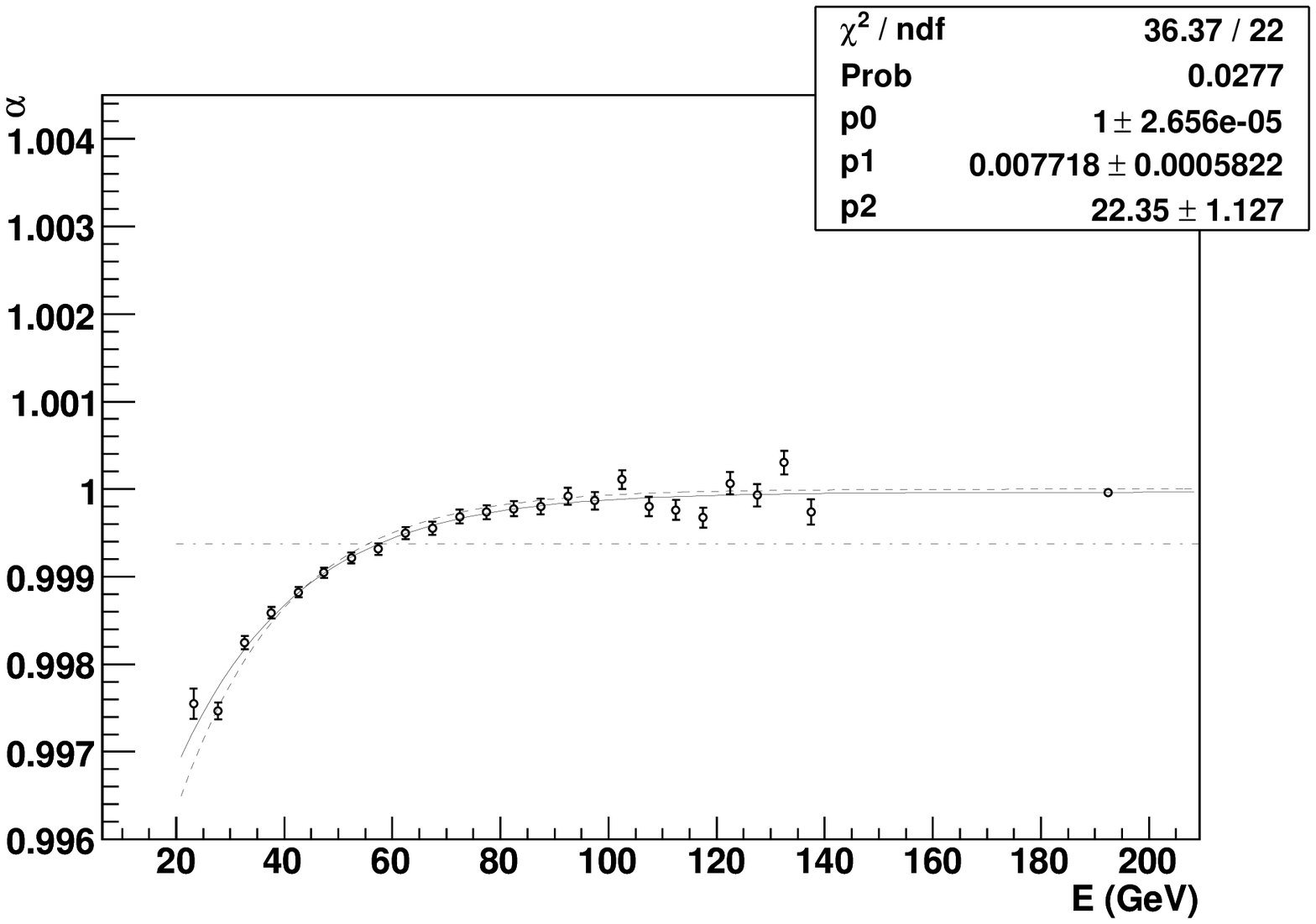}
\includegraphics[width=0.48\textwidth]{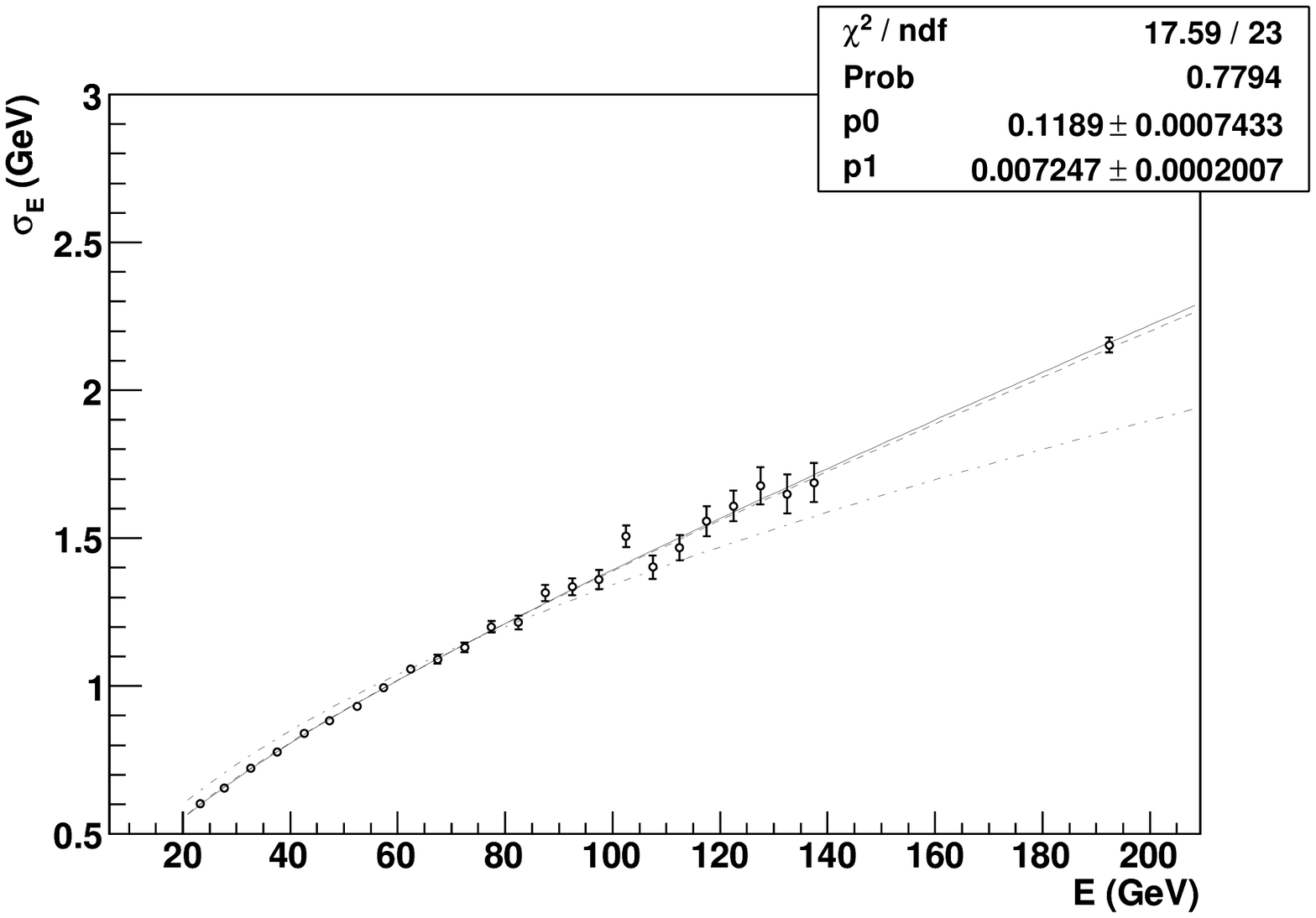}
\caption{\label{lin} Left: reconstructed absolute energy scale, as a
function of energy (points with error bars). The full line gives the
injected function, representing the effect of non simulated passive
material in front of the calorimeter. The dashed line is an empirical
function fitted through the points; the dot-dashed line shows the
result of an energy-independent analysis, missing the
non-linearities. Right: reconstructed energy resolution (points with
error bars). The full line is the true resolution function, of the
form $\sigma(E)/E = a/\sqrt{E} + b$; the dashed line is the
reconstructed function. The dot-dashed line, assuming no constant term
($b = 0$), is strongly excluded.} 
\end{center}
\end{figure}

\subsubsection{Propagation to \MW: $\delta\MW(\alpha_\ell)$,
  $\delta\MW(\sigma_\ell)$} 

Assuming that bin-to-bin variations of the scale do occur with a
spread of $2 \times 10^{-4}$, we can compute the impact of such
variations on the measurement of \MW.\\   

\noindent
As described in Section~\ref{sec:methods}, we perform a set of toy
measurements, using the electron transverse momentum as observable,
templates with varying \MW\ values but with a perfectly linear scale, and
pseudo-data with fixed \MW, but containing non-linearities.\\
 
\noindent
First of all, we can study the \MW\ bias as a function of the error on
the average absolute scale. Not surprisingly, we find a strong dependence:
$$
\frac{\partial\MW}{\partial_{rel}\alpha_\ell} \sim 800 \MeV/\%,
$$
\noindent as illustrated in Figure~\ref{bias}.\\ 

\noindent
In the case of an energy-dependent scale, the uncertainty on \MW\ is
obtained by injecting random, energy-dependent decalibrations in the
pseudo-data, with a spread corresponding to the result of the analysis
of Section~\ref{sec:linearity}. With 480 independent exercises of this type, we
obtain a distribution of \MWfit\ as shown on
Figure~\ref{mwspread}. The scale-induced \W mass uncertainty is given
by the spread of this distribution, and is $\delta\MW(\alpha_\ell) = 4 \MeV$.\\ 

\begin{figure}[tp]
\begin{center}
\includegraphics[width=0.48\textwidth]{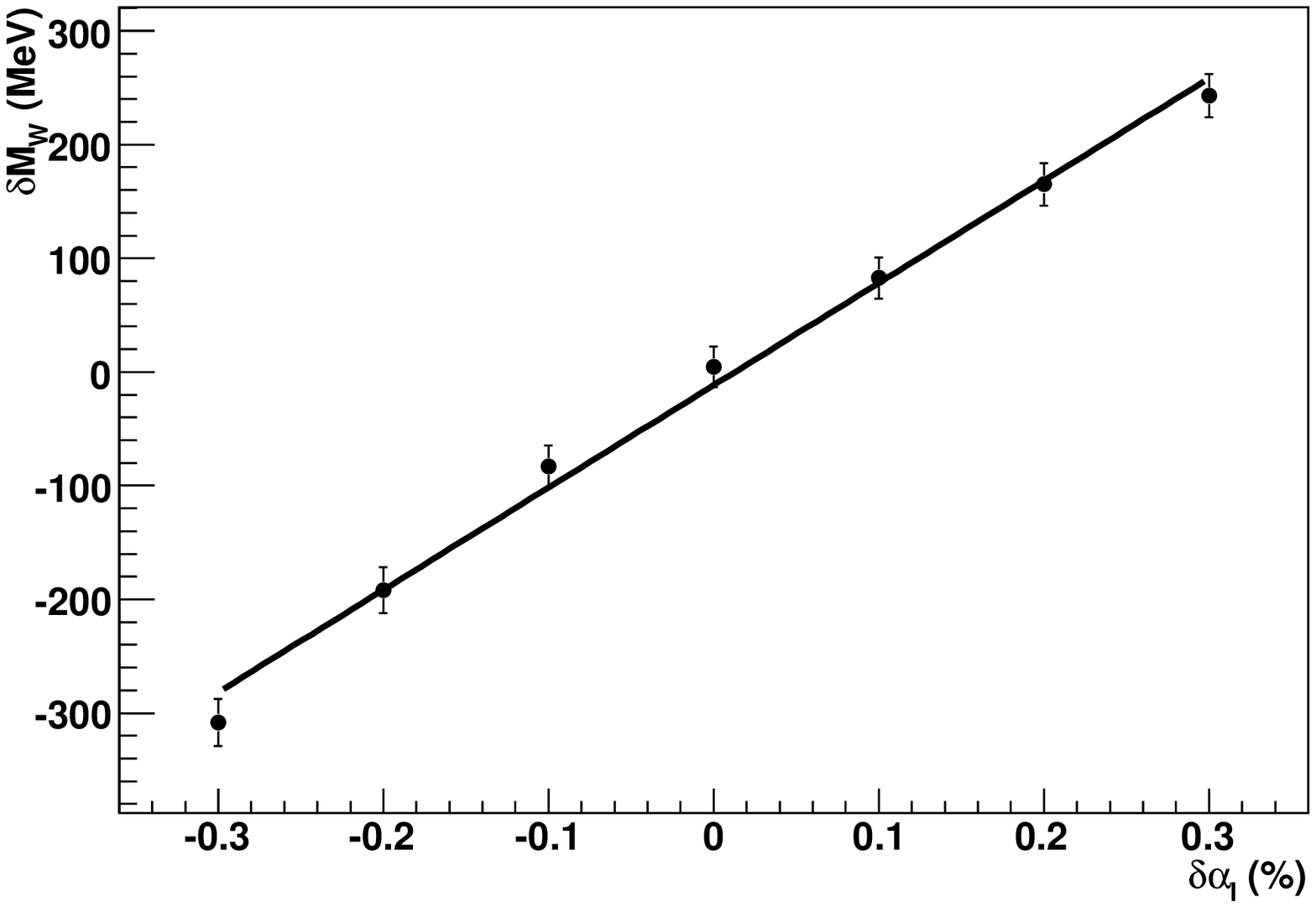}
\includegraphics[width=0.48\textwidth]{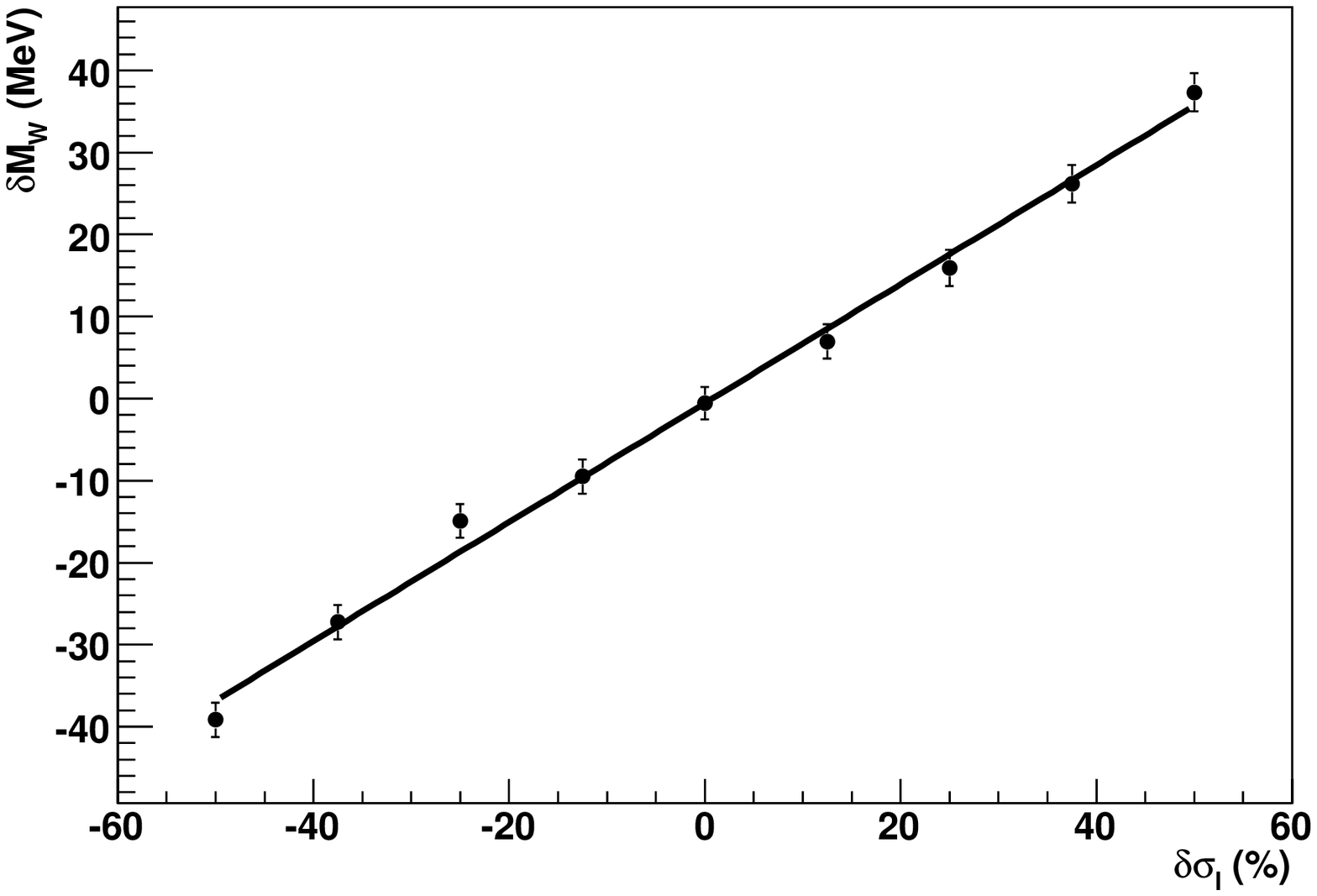}
\caption{\label{bias} Left: bias on \MW, $\MWfit - \MWtrue$, as a
function of the relative bias on $\alpha_\ell$, $\delta\alpha_\ell = (\alpha_\ell^{fit} -
\alpha_\ell^{true})/\alpha_\ell^{true}$. Right: bias on \MW\ as function of the
resolution bias, $\delta\sigma_\ell = (\sigma_\ell^{fit} -
\sigma_\ell^{true})/\sigma_\ell^{true}$. A linear dependence is observed in
each case, with $\partial\MW/\partial_{rel}\alpha_\ell = 800\MeV/\%$ and
$\partial\MW/\partial_{rel}\sigma_\ell = 0.8\MeV/\%$.}
\end{center}
\end{figure}

\begin{figure}[tp]
\begin{center}
\includegraphics[width=0.6\textwidth]{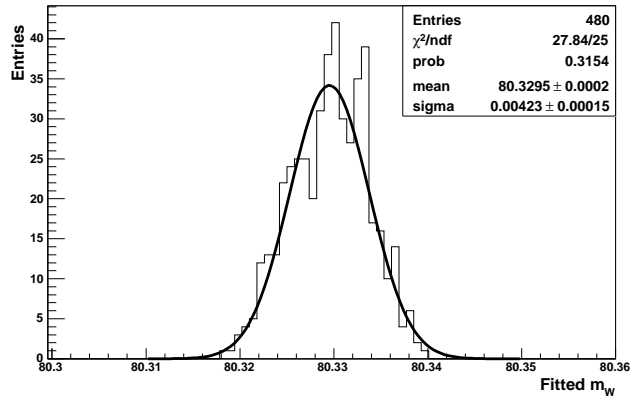}
\caption{Distribution of \MWfit, for 480 exercises with energy-dependent scale
parameters randomly drawn within their uncertainties. The true
mass is $\MW = 80.33 \GeV$; the systematic uncertainty is $\sim 4$~\MeV.}
\label{mwspread}
\end{center}
\end{figure}

\noindent
The effect of the resolution is studied by varying the resolution
parameter in the pseudo-data, fitting to templates with fixed
resolution, and collecting the corresponding values of \MWfit. This
provides the relation between the resolution bias and the resulting
bias on \MW:
$$
\frac{\partial\MW}{\partial_{rel}\sigma_\ell} = 0.8 \MeV/\%
$$
\noindent as illustrated in Figure~\ref{bias}. Injecting the
expected precision on the resolution, using the same method as above,
yields $\delta\MW(\sigma_\ell) \sim 1$~\MeV. \\

\noindent
The analysis presented here was originally done in terms of $E$
(rather than transverse energy, $E_T$) to ease comparison with the
scale and linearity measurements performed on ATLAS testbeam 
data~\cite{atlastb:linearity}. In the context of collision data, the
analysis can instead be performed in terms of $E_T$; the
energy-dependent scale is reconstructed with the same precision as
above, in the range $20 < E_T < 70$~\GeV. The propagation to
$\delta\MW(\alpha_\ell)$ and $\delta\MW(\sigma_\ell)$ is unchanged.\\

\noindent
In addition to the transverse energy dependence, the detector response
is in general also a function of the lepton pseudorapidity $\eta_\ell$,
azimuth $\phi_\ell$, and time. The physical $\phi_\ell$ distributions are however
uniform in general, and certainly for \W and \Z events. Hence it is
safe to average over $\phi_\ell$; any azimuthal dependence of the
detector scale or resolution then acts as a contribution to
the averaged detector resolution. 
Any possible time dependence of the energy response can be treated in
the same way, provided the analysed \W and \Z event samples are taken
from identical data taking periods (``runs''). It is however
beneficial to limit the impact of this time dependence on the
detector resolution by precisely monitoring its response as a function
of time.\\

\noindent
Although not strictly identical, the $\eta_\ell$ distributions
in \W and \Z events are also expected to be very similar within the
detector acceptance (the difference is below 5\% within $|\eta_\ell| < 2.5$,
cf. Figure~\ref{fig:distwz}). As a first approximation, the same
procedure can be applied; the averaging over $\eta$ then assumes that
leptons from \W and \Z are reconstructed with similar performance,
with the same averaging contribution to the global detector
resolution. 
The averaging can be improved by reweighting the $\eta_\ell$
distribution observed in \Z events, where the scale is measured, to
reproduce the distribution observed in \W events where the scale
applied. The detector response to leptons of given transverse momentum
is then identical by construction in \W and \Z decays, up to the
statistical precision of the reweighting. As will be seen in
Section~\ref{sec:correlations}, the absolute scale determination is
very stable against variations of the  underlying physics
hypotheses. In particular, it is negligibly affected by PDF
uncertainties, which are the main factor determining the physical
rapidity distribution of the \Z boson and its decay products. The
reweighting does thus not introduce hidden physics uncertainties, and
does not affect the discussion of other systematic uncertainties. \\

\begin{figure}[tp]
\begin{center}
\includegraphics[width=0.48\textwidth]{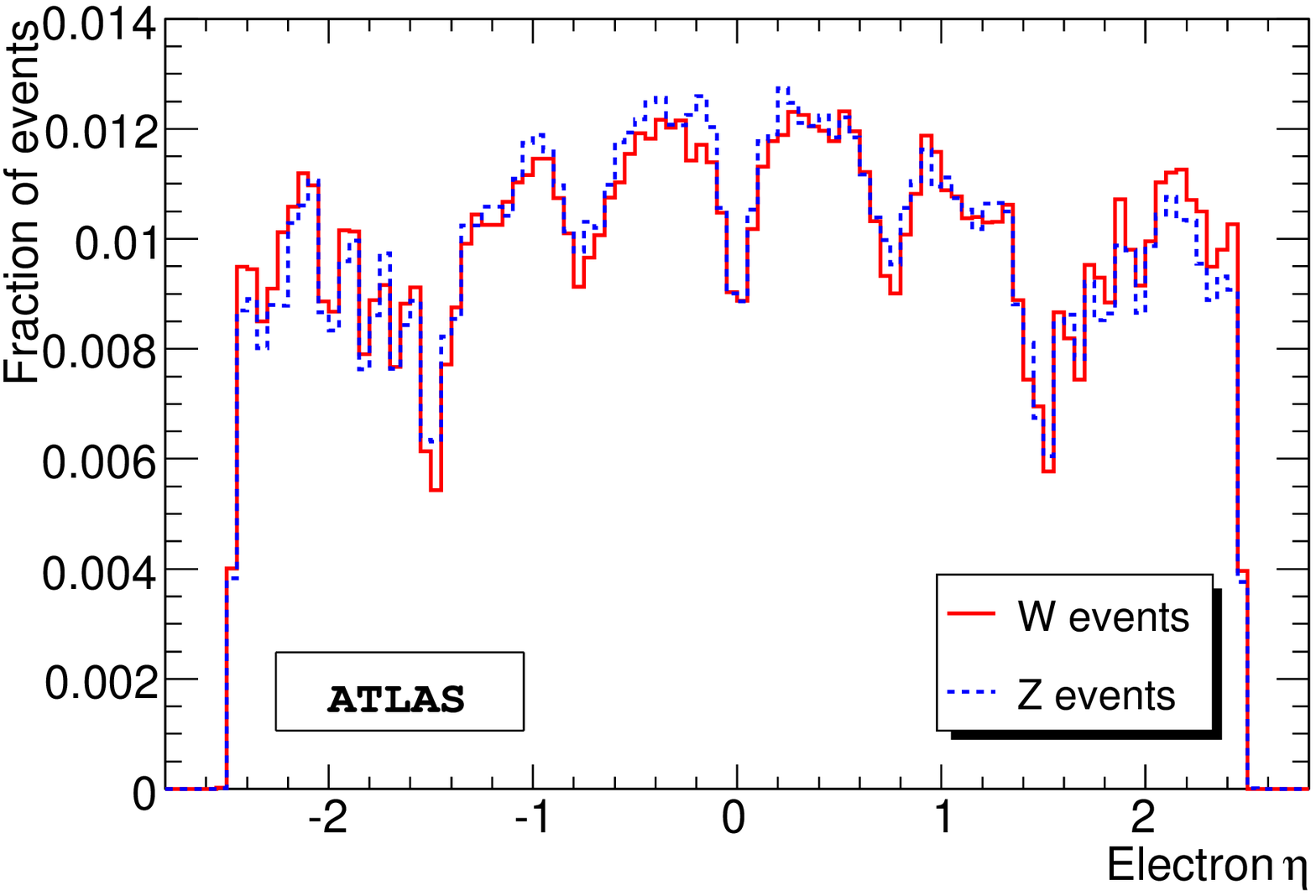}
\includegraphics[width=0.48\textwidth]{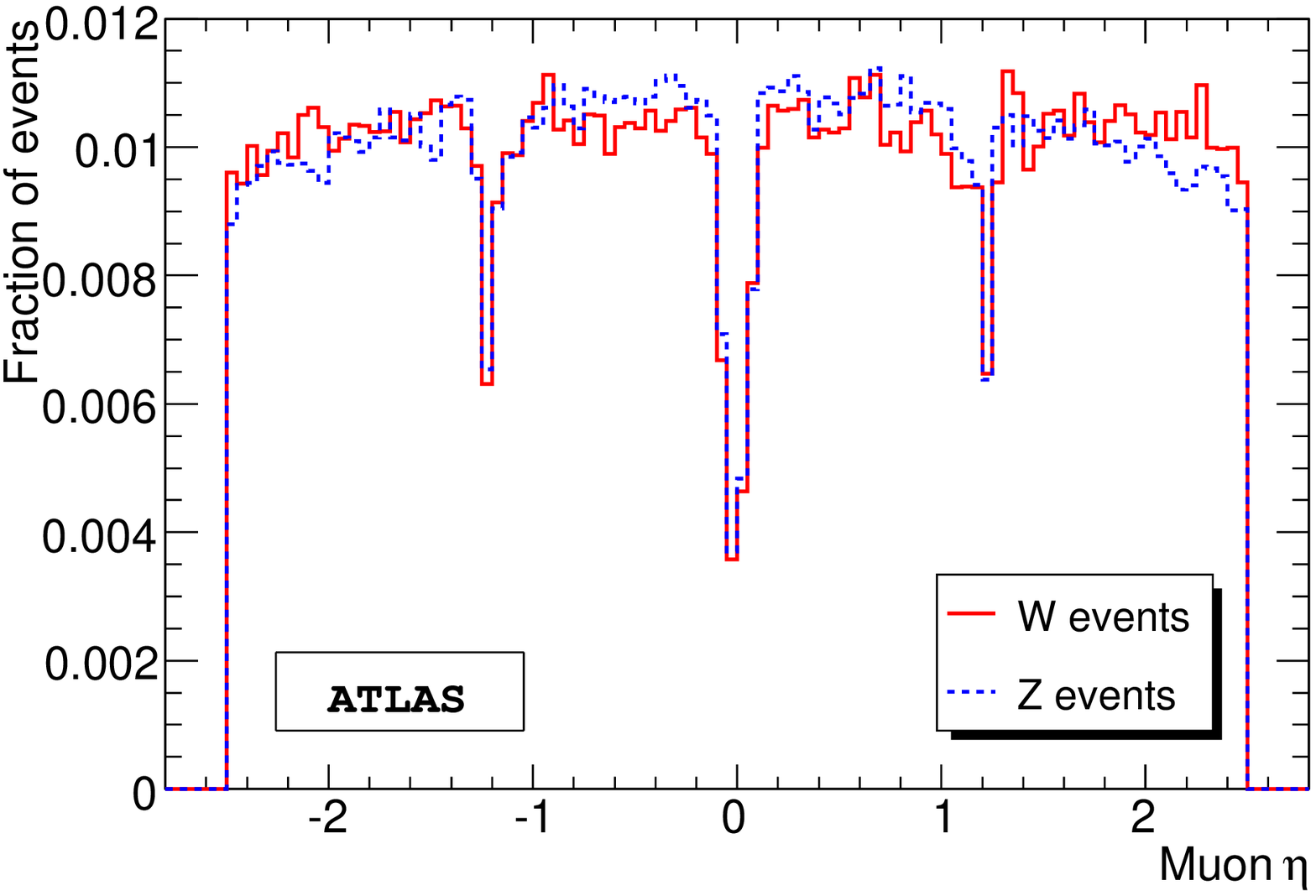}
\caption{\label{fig:distwz} Electron (left) and muon (right) $\eta$ distributions at
reconstruction level, for \W and \Z events.}
\end{center}
\end{figure}

\noindent
The above analysis is performed on the example of the electron
channels. As discussed in Section~\ref{subsec:observables}, the muon
channels provide similar statistics, and are reconstructed with
similar resolution. The present results thus equally hold in the
electron and muon channels.\\

\noindent
We end this section by noting that other well-known physics
probes of detector scale exist, such as the low-mass
vector resonances $J/\Psi$ and $\Upsilon$. An over-constrained scale
measurement can also be performed by first 
measuring the ID scale, exploiting muon final states; transporting the
ID scale to the EMC, using the $E/p$ distribution with isolated
electrons; and finally verifying that this indirect EMC scale allows
to reconstruct unbiased mass peaks for the known resonances in
electron final states. This confrontation of measurements is expected
to allow to understand the source of any observed non-linearities in
terms of magnetic field effects, imperfect alignement, excess of
passive material in the detector, etc. It will thus be possible to
confront several independent probes 
of the detector scale; compatibility between these measurements then
validates its use for the measurement of \MW. This discussion is
familiar from the Tevatron \MW\
measurements~\cite{ex:tevMwRun1,ex:cdfMwRun2}. The present analysis,
using \Z events only, quantifies the precision achievable at the  
LHC provided all measurements of the scale agree.

\subsection{Lepton reconstruction efficiency}
\label{subsec:lepeff}
The observed Jacobian distributions in \W events also reflect any 
\pt\ dependence of the lepton reconstruction efficiency. Any
difference between the simulation used to produce the templates  and
the data will induce a distortion of the spectrum and cause a bias in
the mass fit.\\

\noindent
We again take the electron channel as our main example. The
ATLAS electron identification largely exploits the shapes of their
calorimetric showers~\cite{lhc:atlasTDR1}, which have significant energy
dependence. Hence, any selection based on these will have a
\pt-dependent efficiency which has to be appropriately simulated in
the templates. Unlike the electrons, no strong \pt\ dependence affects the
muon reconstruction efficiency.

\subsubsection{Electron efficiency measurements}

Electron reconstruction efficiency can be determined from the data
with \Z events, using e.g. the so-called ``tag and probe''
method~\cite{ex:d0Zxsec}, which we briefly summarize here.\\

\noindent
Events are selected with one well-identified electron, and an
additional high-\pt, isolated track. The invariant mass of these two
objects is required to be within 10~\GeV\ from the nominal \Z boson
mass. Assuming that this selects \Z events with enough purity, the
identification efficiency is then obtained by computing the
fraction of events where the second object is indeed identified as
an electron. The efficiency of the isolation criterion is obtained
in a similar way.
Simulation studies show that the impact of backgrounds on the
estimation of the efficiency is small compared to the statistical
uncertainty.\\
 
\noindent
For the present study we use about 200000 fully simulated $\Z\to ee$
events, from which the efficiency is evaluated. The
result is shown in Figure~\ref{effiZee}, together with an empirical
function describing main features of the \pte\ dependence.  The following form:
\begin{equation}
\epsilon(\pt) = \epsilon_0 - a \exp(-b \times \pt)
\label{effparam}
\end{equation}
\noindent correctly describes the efficiency in the \pte\ range
relevant for the analysis.\\

\begin{figure}[tp]
\begin{center}
\includegraphics[width=0.6\linewidth]{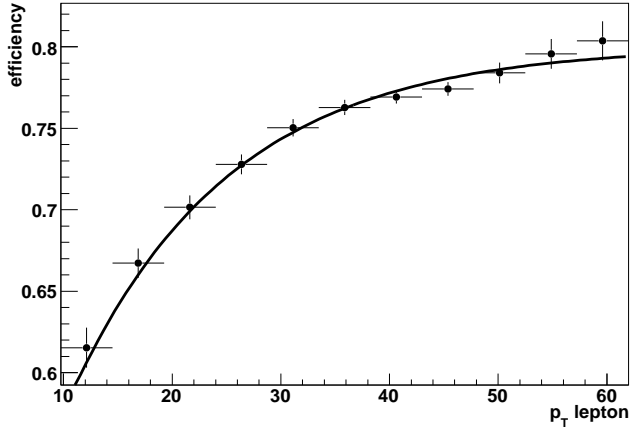}
\caption{\label{effiZee} Electron reconstruction efficiency, as
determined from fully simulated $\Z \ra ee$ events.}  
\end{center} 
\end{figure}

\subsubsection{Propagation to \MW:  $\delta\MW(\epsilon)$}

The effect of the efficiency uncertainty is estimated as in the
previous section. Template distributions are produced at generator
level, with varying values of \MW, and applying an efficiency factor
according to the best fit efficiency function obtained above.\\

\noindent
One hundred independent pseudo-data samples are generated at a fixed
mass ($\MW = 80.33 \MeV$). Efficiency functions are applied with
parameters drawn randomly within their uncertainties, as obtained in the
previous section.\\ 

\noindent
For each sample of pseudo-data, a fit is performed to the \W
mass. The fitted mass values are histogrammed, and the spread of the
histogram gives the corresponding systematic uncertainty. With the
efficiency determined using $2 \times 10^5$ \Z boson decays, the
efficiency-induced systematic \W mass uncertainty is found to be
$\delta\MW = 33 \MeV$. Other functional forms than Eq.~\ref{effparam}
yield the same result. The most sensitive parameter in Equation~\ref{effparam} is $b$, the
slope in the exponential. It is determined to be $b =
0.068 \pm 0.006$, corresponding to a precision of 9\%; in other
words, $\partial\MW / \partial_{rel} b \sim 4$~\MeV/\%.\\

\noindent
To emphasize the importance of this effect, the same pseudo-data
samples are compared to templates assuming no \pt-dependence in the
lepton reconstruction efficiency (i.e. $f(\pt) =$ constant). While the same
spread is observed, the \MWfit\ distribution indicates an average bias of
about 450 \MeV. This bias vanishes, to first order, when using the
\pt-dependent efficiency in the templates.\\

\noindent
Extrapolating to 10 \ifb, i.e. assuming $3 \times 10^6$ measured \Z boson
decays, an improvement of a factor~$\sim 4$ is expected in the
efficiency determination. Correspondingly, we obtain 
$\delta\MW(\epsilon)\sim 8 \MeV$.  
 
\subsubsection{Discussion and improvements}

As can be seen in Figure~\ref{effiZee}, the electron efficiency varies
most rapidly when $\pte\sim 20 \GeV$, and is much flatter around the
Jacobian edge. Until now, the full \pte\ spectrum, selected as
described in Section~\ref{sec:methods}, has been used in the mass fits.\\
 
\noindent
The effect of restricting the lepton \pte\ range used in the fit to
higher values is displayed in
Table~\ref{result_95sets_100tests_ptcuts}. Considering
the part of the spectrum verifying $\pte > 34 \GeV$, for example,
reduces $\delta\MW$ from 33 \MeV\ to 18 \MeV. While avoiding the region  
with strongest \pt-dependence of the efficiency, the Jacobian edge is
still fully exploited, and the statistical sensitivity is almost
unaffected. Extrapolating to 10~\ifb, we obtain a remaining uncertainty of
$\delta\MW(\epsilon) \sim 4.5 \MeV$.\\

\begin{table}[tp] 
\begin{center}
\begin{tabular}{ccccc}
\hline
\hline
\pt\ cut & $<\MWfit>$ & $<\delta\MWfit>$ & $<\MWfit>$ & $<\delta\MWfit>$ \\  
 & ($\epsilon_{ref}=1$) & ($\epsilon_{ref}=1$) &
 ($\epsilon_{ref}=f(\pt)$) & ($\epsilon_{ref}=f(\pt)$) \\  
\hline
$\pt > 20 \GeV$ & 80.78 & 0.033 & 80.34 & 0.033 \\
$\pt > 34 \GeV$ & 80.51 & 0.019 & 80.34 & 0.018 \\		      
$\pt > 37 \GeV$ & 80.44 & 0.013 & 80.33 & 0.012 \\
\hline
\hline
\end{tabular}
\end{center}
\caption{\label{result_95sets_100tests_ptcuts} Average value of \MWfit\
and its spread $\delta\MWfit$, for several lower cuts on the \pt\ range
used in the mass fit. Numbers are given as obtained from templates
assuming a flat efficiency (second and third column), and using the
efficiency measured in \Z events (fourth and fifth
column). $\MWtrue = 80.33$~\GeV.}
\end{table}

\noindent
Note that the results presented here reflect the state of the ATLAS
reconstruction software at the time of writing this paper. Significantly
improved algorithms are described in~Ref.~\cite{lhc:atlasDetector}, notably resulting in a
smaller \pt-dependence of the electron reconstruction
efficiency. The related systematic uncertainty on \MW\ should decrease
accordingly. The numbers presented here may thus be considered as
conservative.\\

\noindent
For muons with sufficient momentum to cross the whole
detector ($p > 6 \GeV$), no source of inefficiency has a strong \pt\
dependence. Hence, the corresponding induced uncertainty on \MW\ is
smaller. The above estimate is thus conservative when applied to
the muon channel.

\subsection{Recoil scale and resolution}
\label{subsec:recoilscale}
When using the \mtw\ distribution in the mass fit, \ptn\ enters the
definition of the observable. This quantity, measured experimentally
as the vector sum of the transverse energy of
all reconstructed detector signals (high-\pt\ leptons and low-\pt\
hadronic activity), needs to be precisely described by the
simulation for the same reasons as above.

\subsubsection{Sensitivity to the recoil scale with \Z events}

The \W and \Z bosons are produced through very similar
partonic processes, and thus one expects the spectator part of the
event (the underlying event) to behave similarly,
up to the small phase space difference ($\MW\neq\MZ$).\\

\noindent
Assuming that the absolute lepton scale and resolution have been measured
beforehand (\cf\ Section~\ref{subsec:lepscale}), one can measure 
the recoil scale ($\alpha_{rec}$) and resolution ($\sigma_{rec}$) in fully
reconstructed \Z events, where no significant \met\ is expected,
by comparing the measured hadronic energy $E_{rec}$, recoiling against the
\Z boson, to the reconstructed di-lepton four-momentum, $p_T^{\ell\ell}$. Specifically,
$\alpha_{rec}$ and $\sigma_{rec}$ are extracted from the peak
position and spread of the distribution of $E_{rec}/p_T^{\ell\ell}$.
The  results can then be used to correct the observed recoil, and
hence \met, in \W events.\\   

\noindent
Figure~\ref{fig:recoil1} shows the expected sensitivity to
$\alpha_{rec}$ and $\sigma_{rec}$. With 10 \ifb, these parameters can
be determined with a statistical precision of $\delta\alpha_{rec} = 5
\times 10^{-5}$ and $\delta\sigma_{rec} = 6 \times 10^{-4}$.\\ 

\begin{figure}[tp]
\begin{center}
\includegraphics[width=0.48\textwidth]{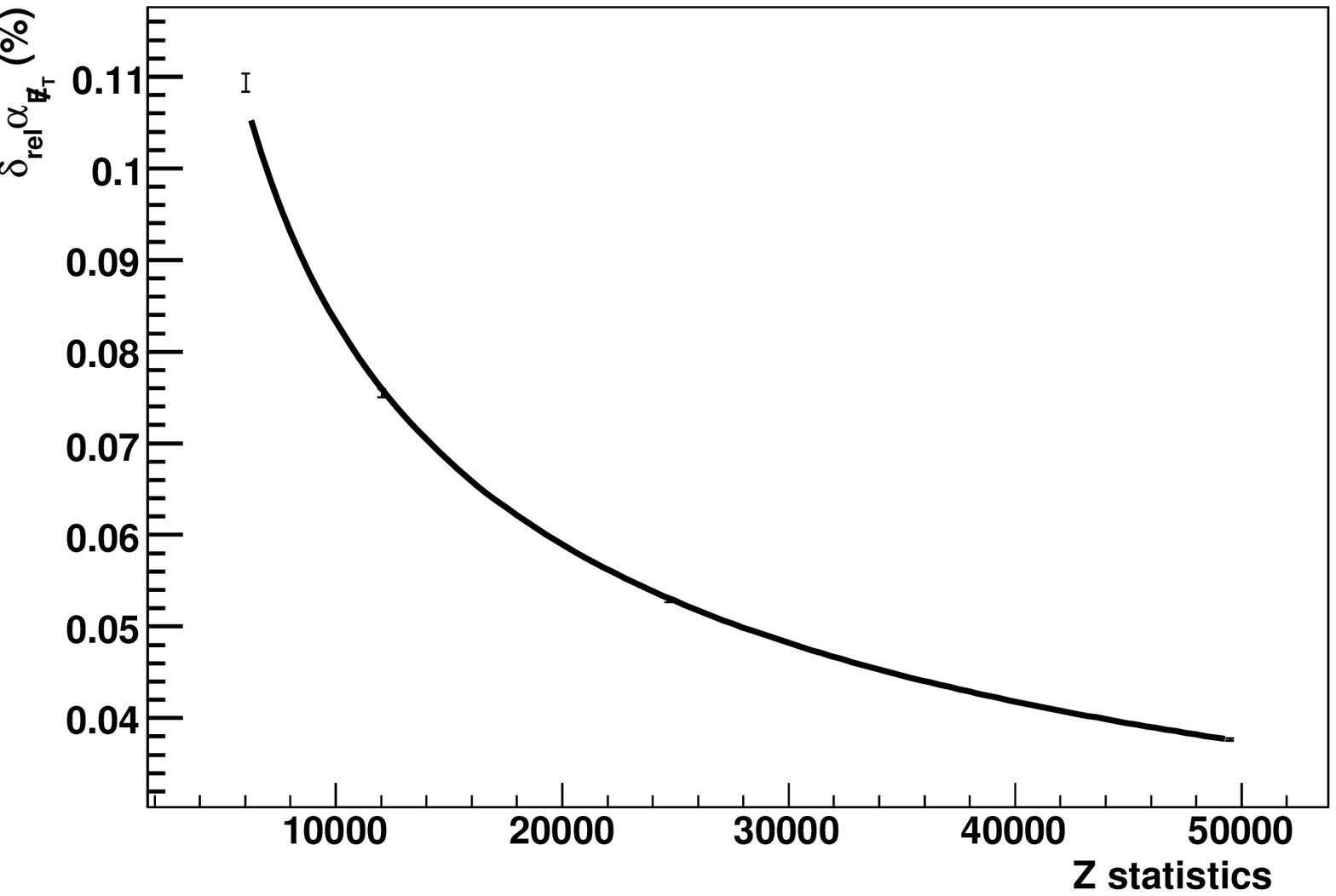}
\includegraphics[width=0.48\textwidth]{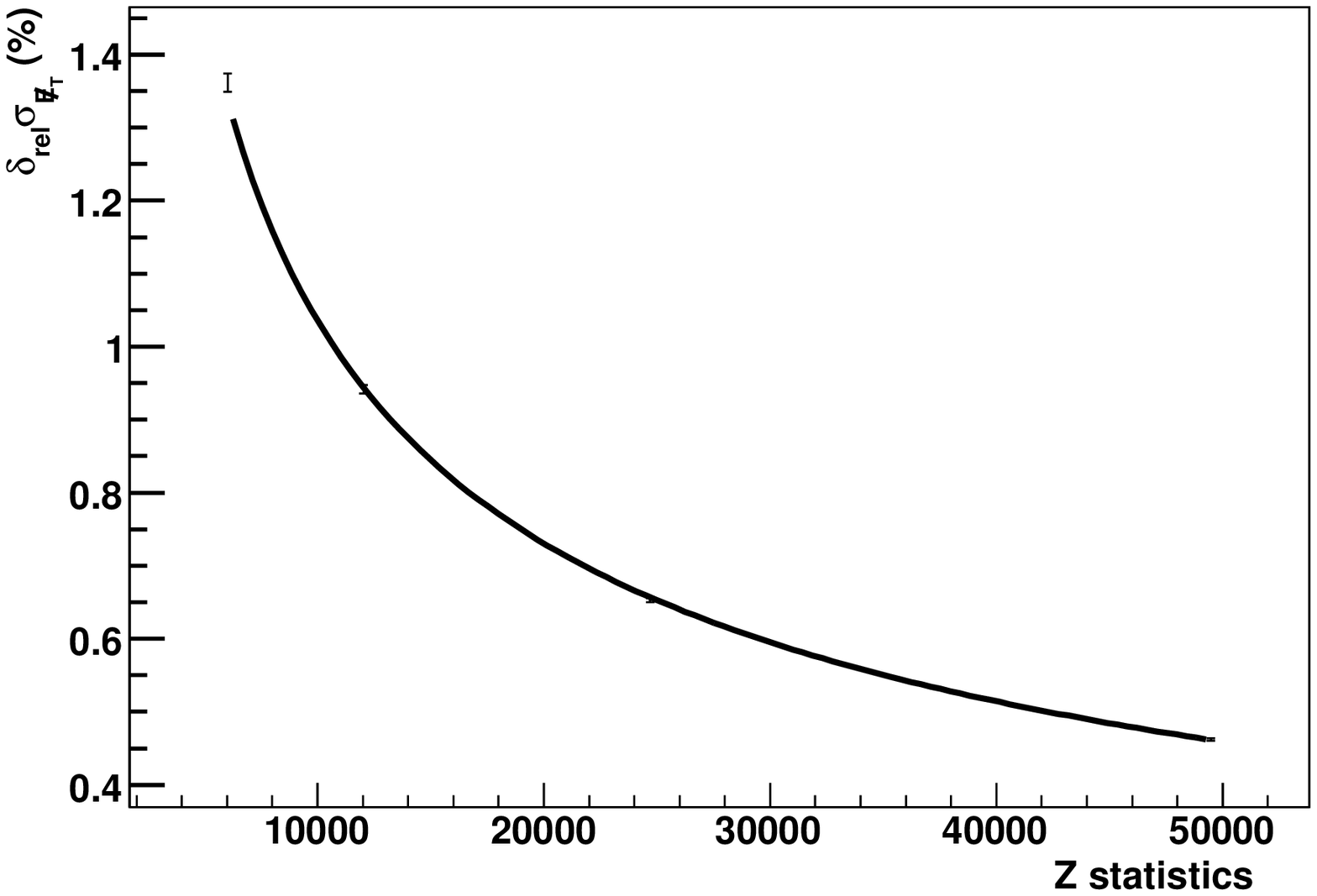}
\caption{\label{fig:recoil1} Left: statistical sensitivity to
$\alpha_{\met}$, as a function of the accumulated \Z
statistics. Right: statistical sensitivity to $\sigma_{\met}$.} 
\end{center}
\end{figure}

\subsubsection{Propagation to \MW: $\delta\MW(\alpha_{\met})$, $\delta\MW(\sigma_{\met})$}

The effect on \MW\ is evaluated by systematically varying the recoil
scale, producing corresponding pseudo-data samples as in the previous
sections, and fitting each sample to perfectly calibrated templates.
We obtain the relation between the \MW\ bias and the recoil scale and
resolution in the form of a derivative: 
$$
\frac{\partial\MW}{\partial_{rel}\alpha_{\met}} = -200 \MeV/\%
\,\,\,\,\,\,\,\,\,\,\,\,\,\,\,\,\,\,\,\,\,\,\,\,\,
\frac{\partial\MW}{\partial_{rel}\sigma_{\met}} = -25 \MeV/\%
$$
\noindent as illustrated in Figure~\ref{fig:recoil2}. Injecting
$\delta\alpha_{\met} = 5 \times 10^{-5}$, we obtain a systematic uncertainty
of $\delta\MW(\alpha_{\met}) = 1\MeV$. Similarly, we find the
contribution from the resolution to be
$\delta\MW(\sigma_{\met})=1.5\MeV$. These numbers assume that the
\Z-based calibration can be transported to the \W sample without
additional uncertainty; this is discussed further below.\\

\begin{figure}[tp]
\begin{center}
\includegraphics[width=0.48\textwidth]{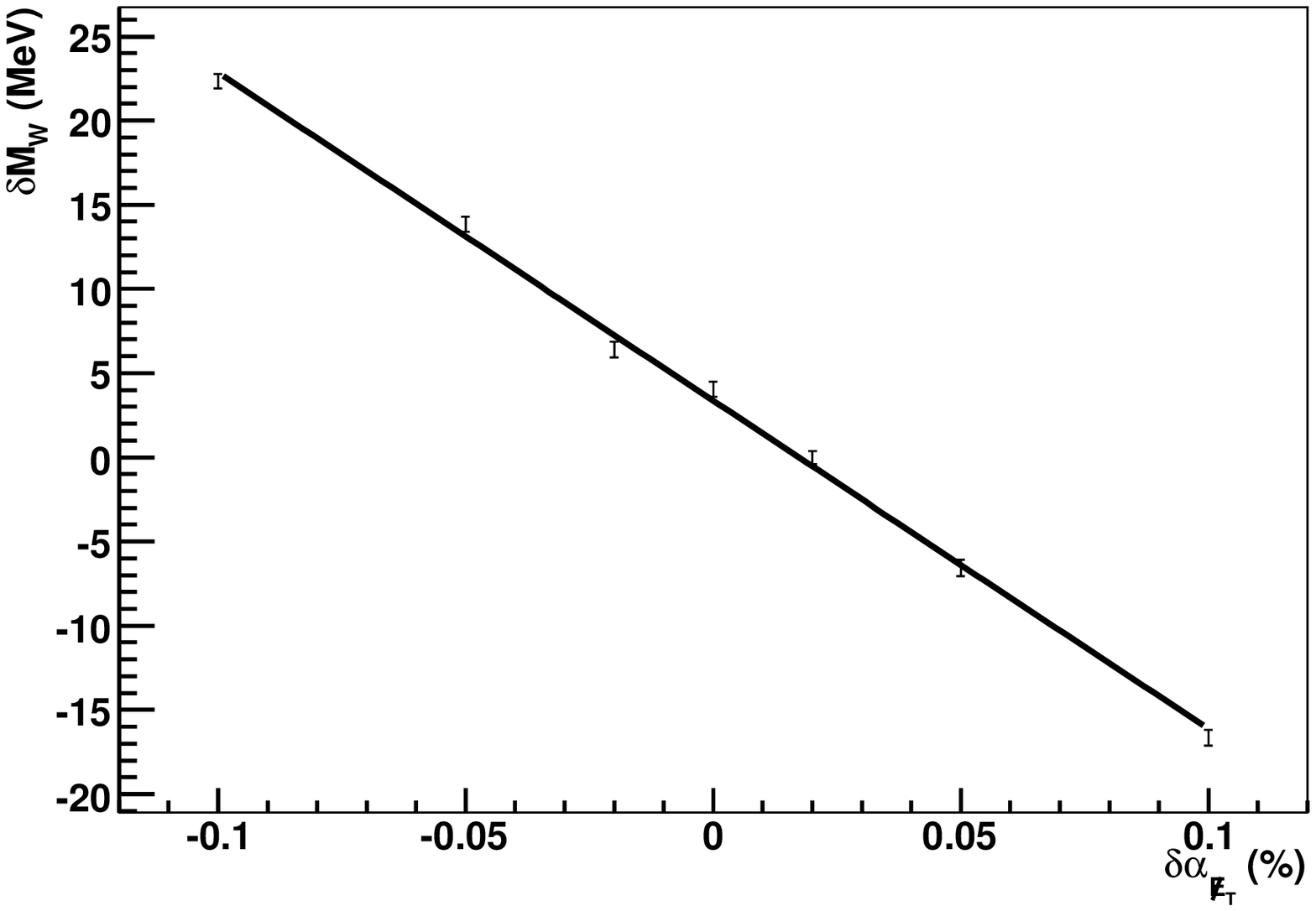}
\includegraphics[width=0.48\textwidth]{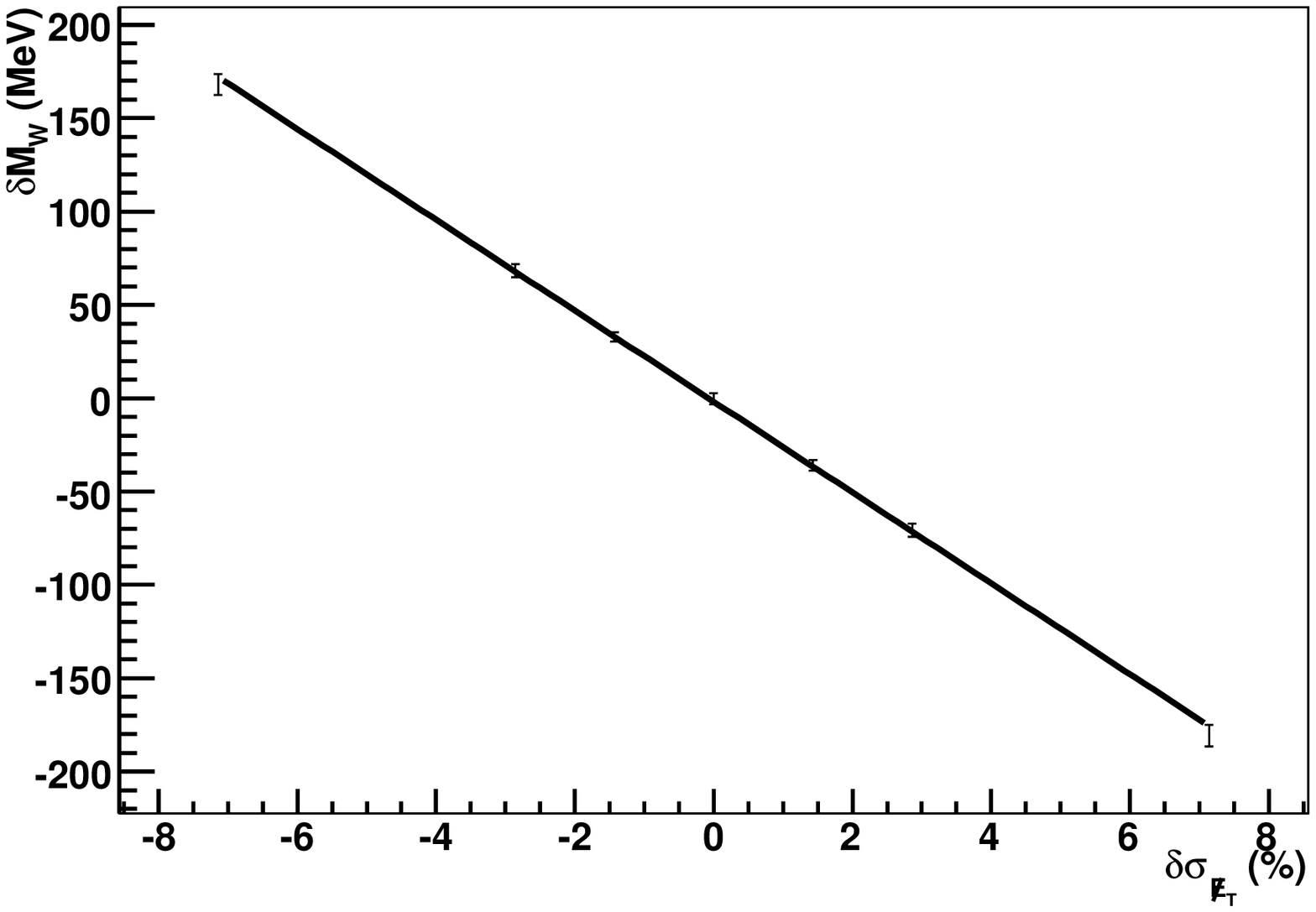}
\caption{\label{fig:recoil2} Left: Bias on \MW, $\delta\MW = \MWfit -
\MWtrue$, as a function of the bias on the recoil scale,
$\delta\alpha_{\met}$. Right: $\delta\MW$ as function of the
resolution bias, $\delta\sigma_{\met}$. A linear dependence is
observed in each case, with $\partial\MW/\partial_{rel}\alpha_{\met} = -200 \MeV/\%$ and
$\partial\MW/\partial_{rel}\sigma_{\met} = -25 \MeV/\%$.}
\end{center}
\end{figure}

\subsubsection{Further discussion}

The \met\ calibration can be studied in more detail, using real
\Z events where one reconstructed lepton is artificially
removed. In the case of electrons, the removed calorimetric energy
should be properly replaced by the expected noise. For
muons, also, the minimum-ionizing energy depositions in the
calorimeters need to be removed and replaced by the expected noise as
above. The resulting events mimic \W events and have a precisely known  
missing energy, corresponding to the energy of the removed lepton,
which can be compared to the result of the \met\ reconstruction
algorithm.\\ 

\noindent
The lepton removal requires that one can identify and remove the
electron signal from the struck calorimeter cells, while leaving a
realistic contribution from noise and hadronic background (see
Figure~\ref{fig:cluster}). Several approaches can be tried, such as
replacing the contents of the electron cluster cells by energy
measured away of any high-\pt\ object in the event (\eg\ at $90^\circ$
in azimuth), or by the average expected electronic and hadronic noise.\\

\begin{figure}
\begin{center}
\includegraphics[width=0.48\textwidth]{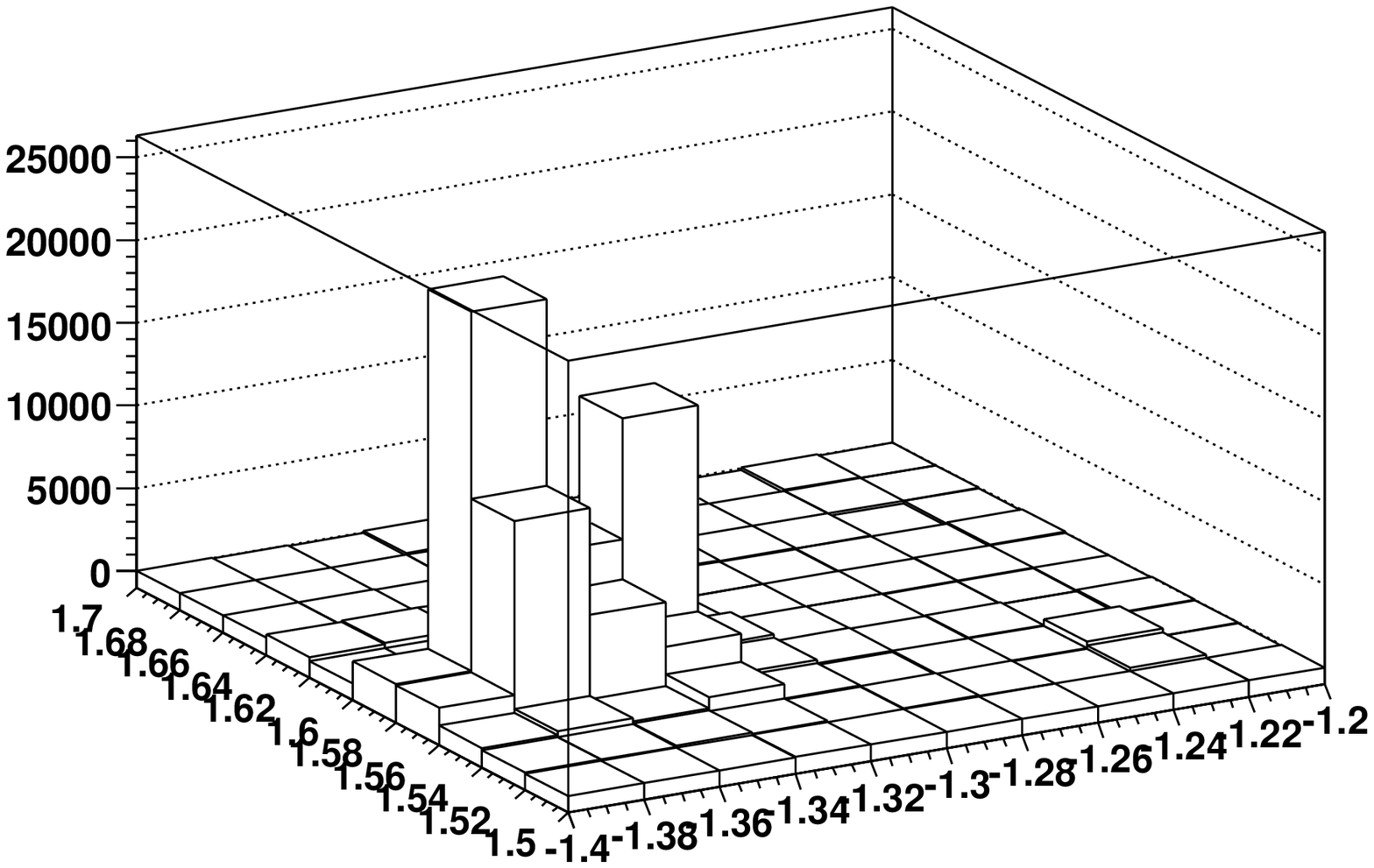}
\includegraphics[width=0.48\textwidth]{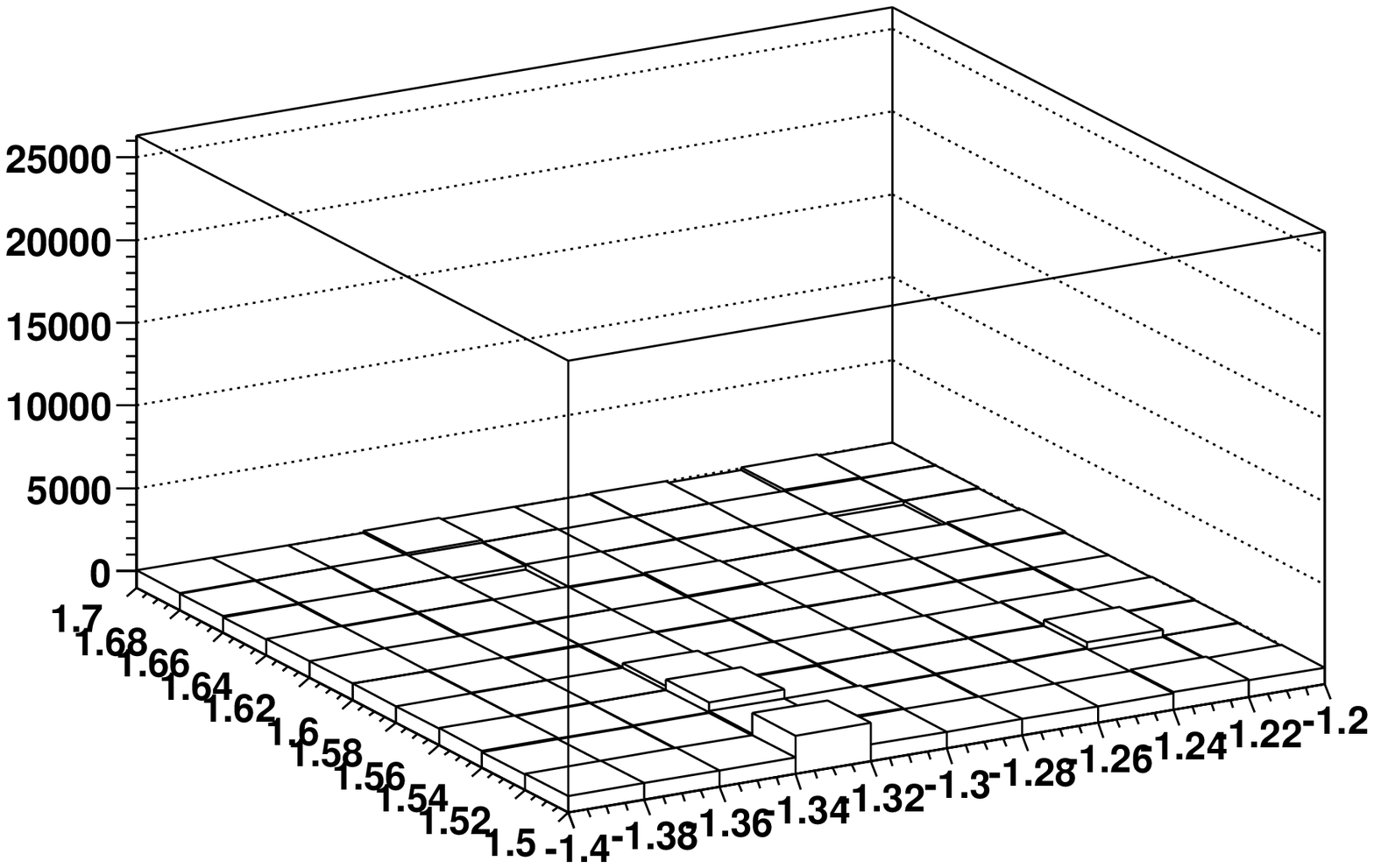}
\caption{\label{fig:cluster} Left: electron cluster in a $\Z \ra
ee$ event. Right: the same calorimeter region, after the cluster has
been removed. The energy in each cell belonging to the electron
cluster is replaced by a number drawn from a Gaussian with mean and
RMS corresponding to detector noise.}  
\end{center}
\end{figure}

\noindent
To determine the \met\ resolution and possibly correct for biases in
its measurement, we consider the reconstructed \met\ of $\Z \to \ell
\ell$ events before and after the removal of one lepton, and compare the
difference to the transverse momentum of the removed lepton. A
non-zero average value of this difference points to a bias in the
\met\ reconstruction.\\

\noindent
Rather than projecting this difference on conventional X and Y axes in
the transverse plane, it is best to consider the natural frame of the
event, with axes parallel ($\|$) and perpendicular ($\bot$) to the
\Z boson transverse momentum. Imperfect calibration
of the \met\ reconstruction will show up as biases in these
distributions, which can then subsequently be corrected for within
statistics. The axes are illustrated in Figure~\ref{fig:bosondecay}.\\

\begin{figure}
\setlength{\unitlength}{1mm}
  \begin{picture}(160,78)(0,0)
    \jput(0,0){\epsfig{file=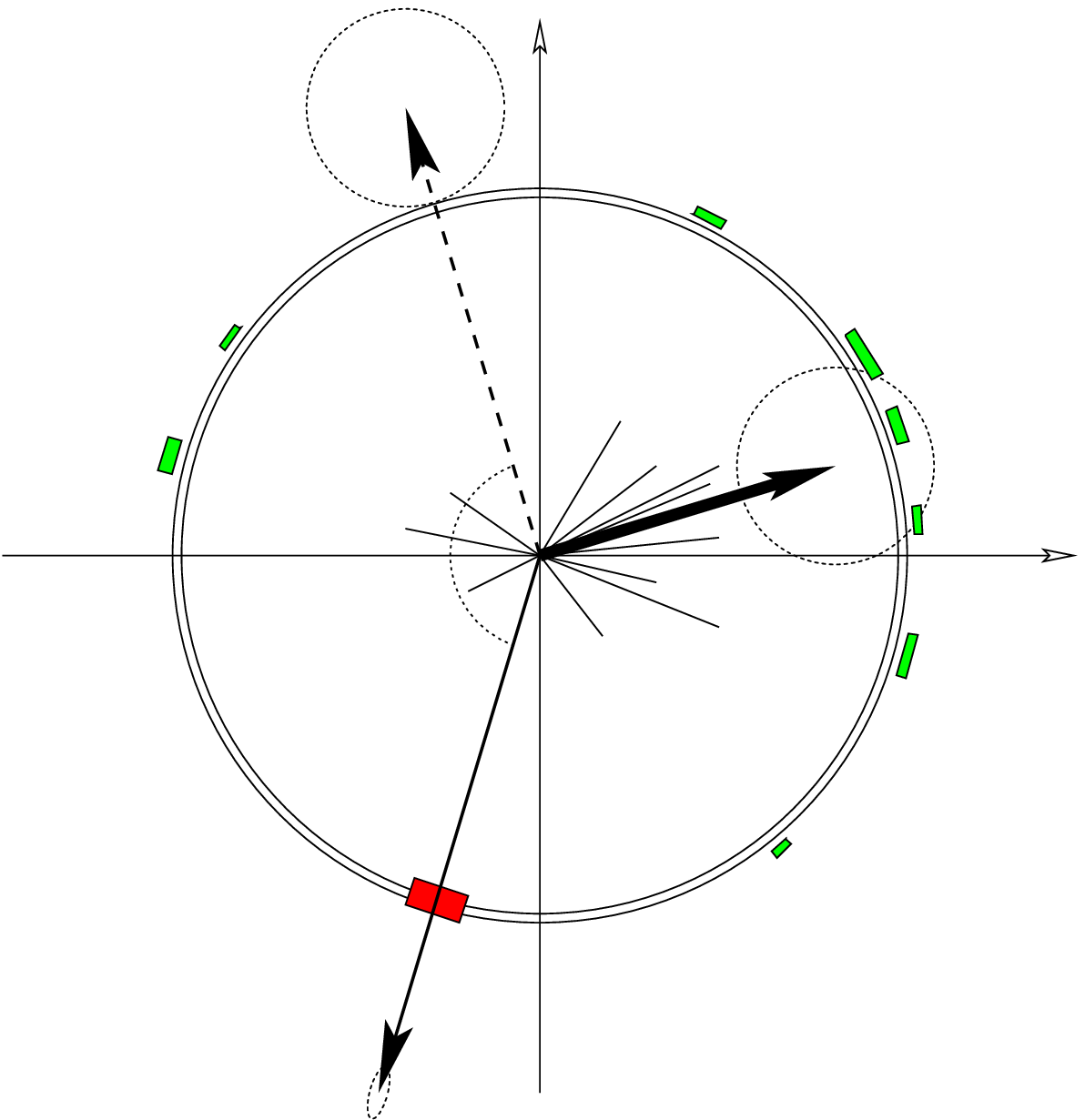,width=75mm}}
    \jput(80,0){\epsfig{file=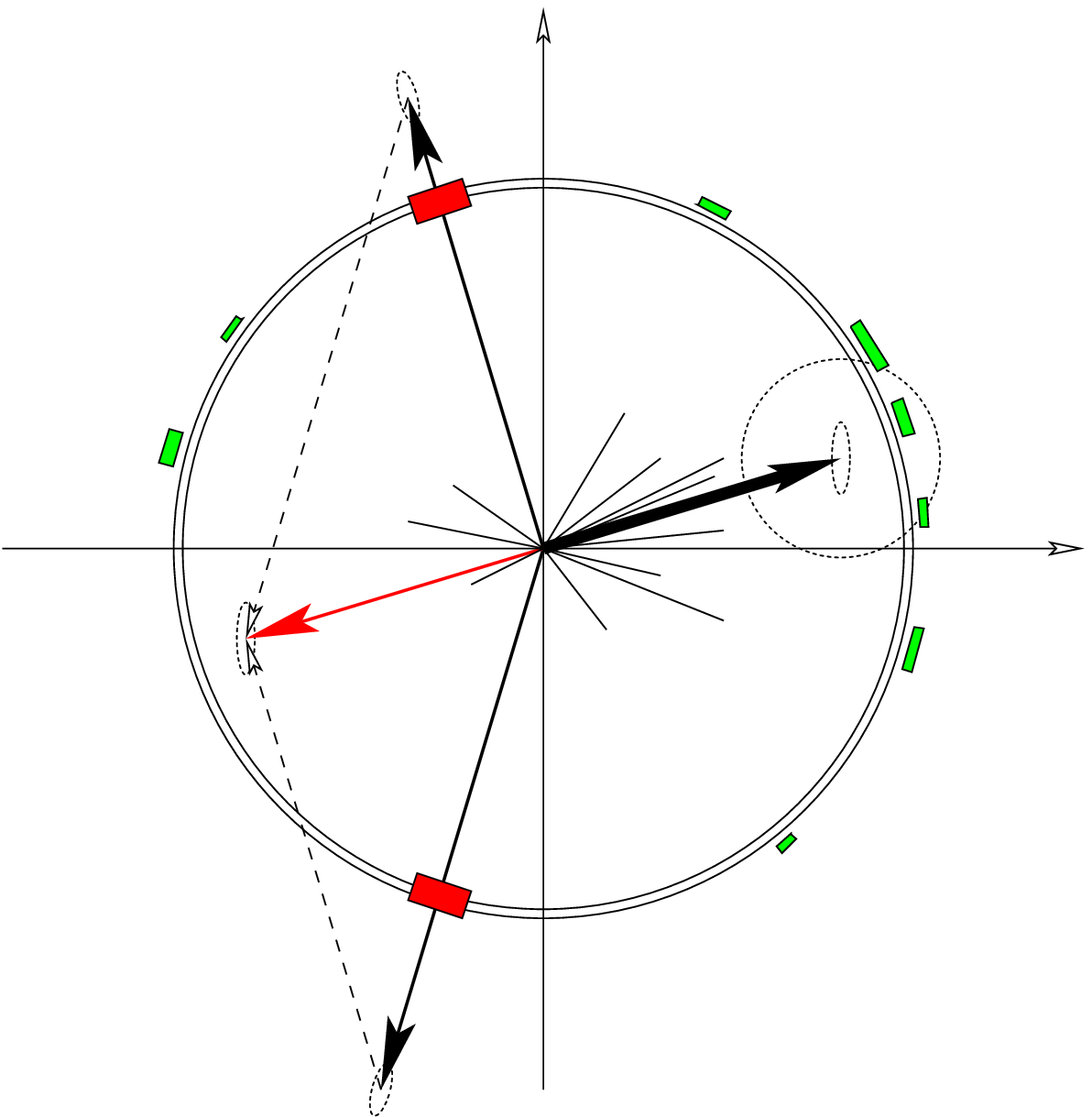,width=75mm}}
    \jput(60,72){\fbox{$W \to \ell \nu$}}
    \jput(2,2){a)}
    \jput(29,51){$\nu$}
    \jput(29,24){$\ell$}
    \jput(70,42){$p_x$}
    \jput(40,74){$p_y$}
    \jput(56,41){$u$}
    \jput(24,35){$\Delta \phi_{\ell\nu}$}
    \jput(140,72){\fbox{$Z \to \ell \ell$}}
    \jput(82,2){b)}
    \jput(109,51){$\nu$}
    \jput(109,24){$\ell$}
    \jput(150,42){$p_x$}
    \jput(120,74){$p_y$}
    \jput(136,41){$u$}
    \jput(103,32){$p_T(Z)$}
  \end{picture}
\caption{\label{fig:bosondecay}
   Transverse view of a) $\W \to \ell \nu$ and b) $\Z \to \ell
   \ell$ events. The combined transverse momentum of the recoil $u$, which
   should match that of the boson, is used to estimate the momentum of the
   undetected neutrino in the $\W \to \ell \nu$ decay. The \Z boson line
   of flight is represented, which defines the ($\|,\bot$) coordinate
   system. The size of the dotted ellipses represent the resolution on
   the reconstructed objects.}
\end{figure}

\noindent
This method is tried on a fully simulated sample of \Z $\to ee$ events,
with results illustrated in Figure~\ref{fig::etmiss}. As can be seen
in this example, a bias is observed in the \met\ reconstruction along
the \Z line of flight. No bias is observed along the other axes. In
this example, the calibration is thus correct on average, but the
\met\ reconstruction does not respond perfectly to the event-by-event topology.\\

\begin{centering}
\begin{figure}
\begin{minipage}{\linewidth}
\centering 
\includegraphics[width=0.4\linewidth]{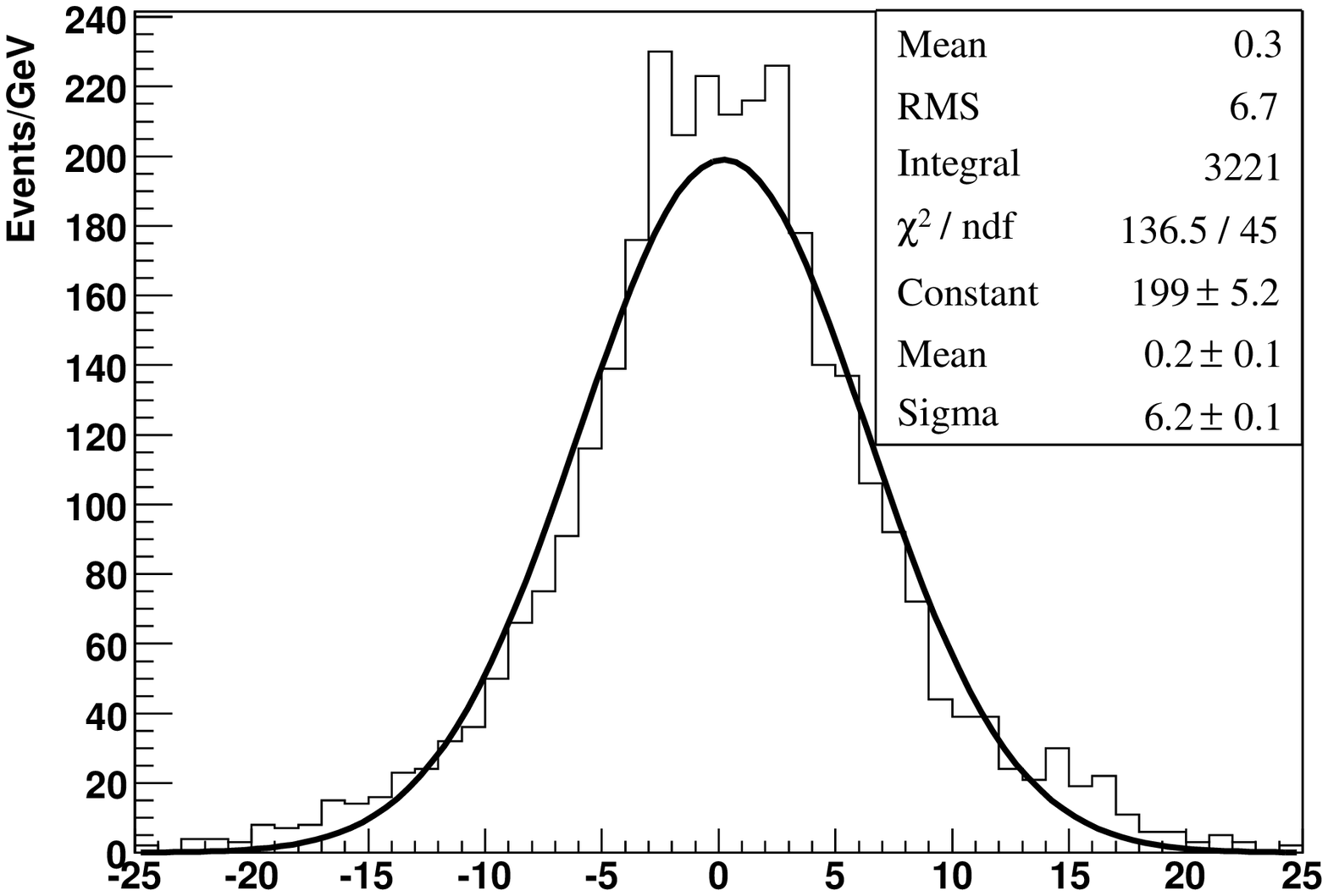}
\includegraphics[width=0.4\linewidth]{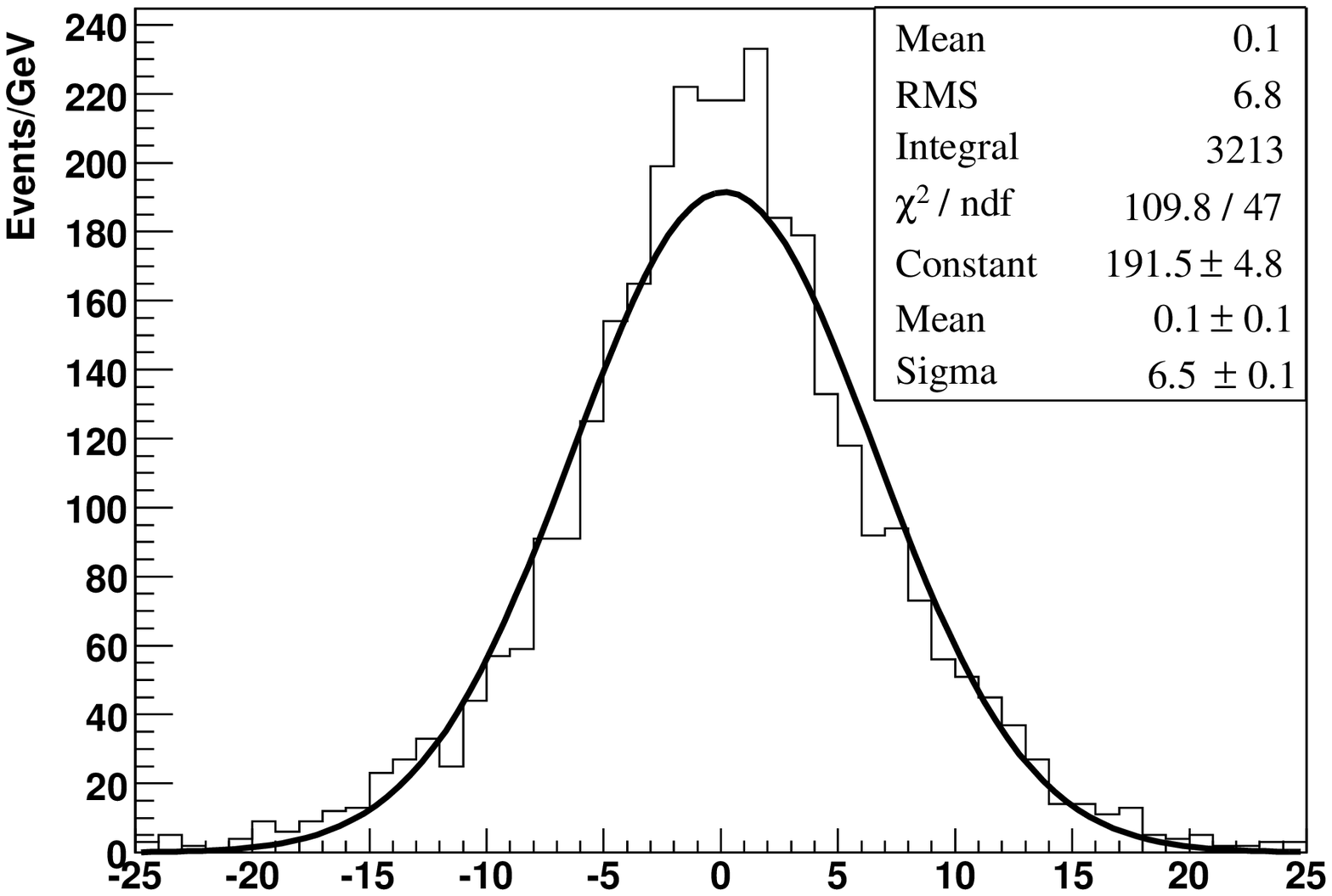}
\includegraphics[width=0.4\linewidth]{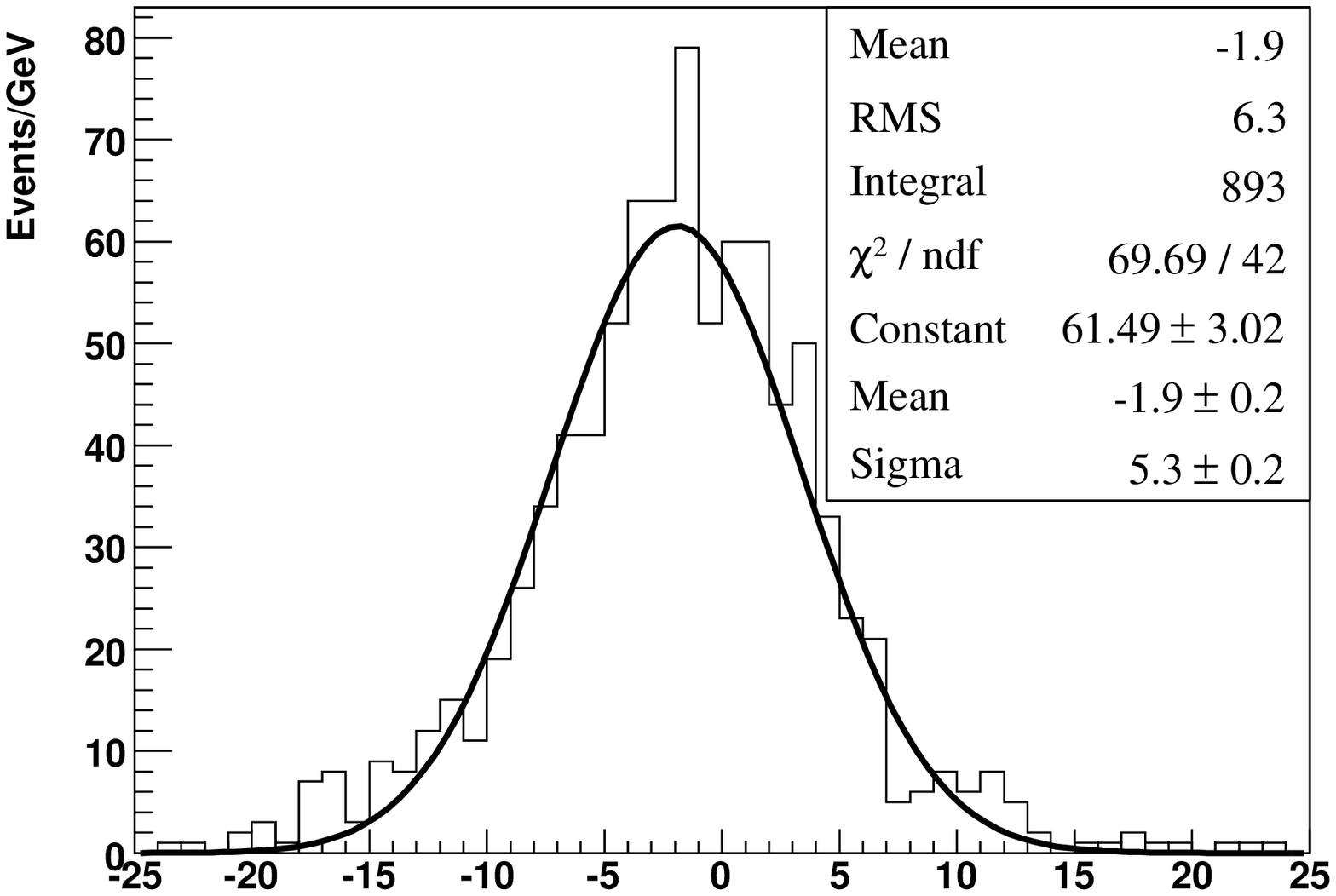}
\includegraphics[width=0.4\linewidth]{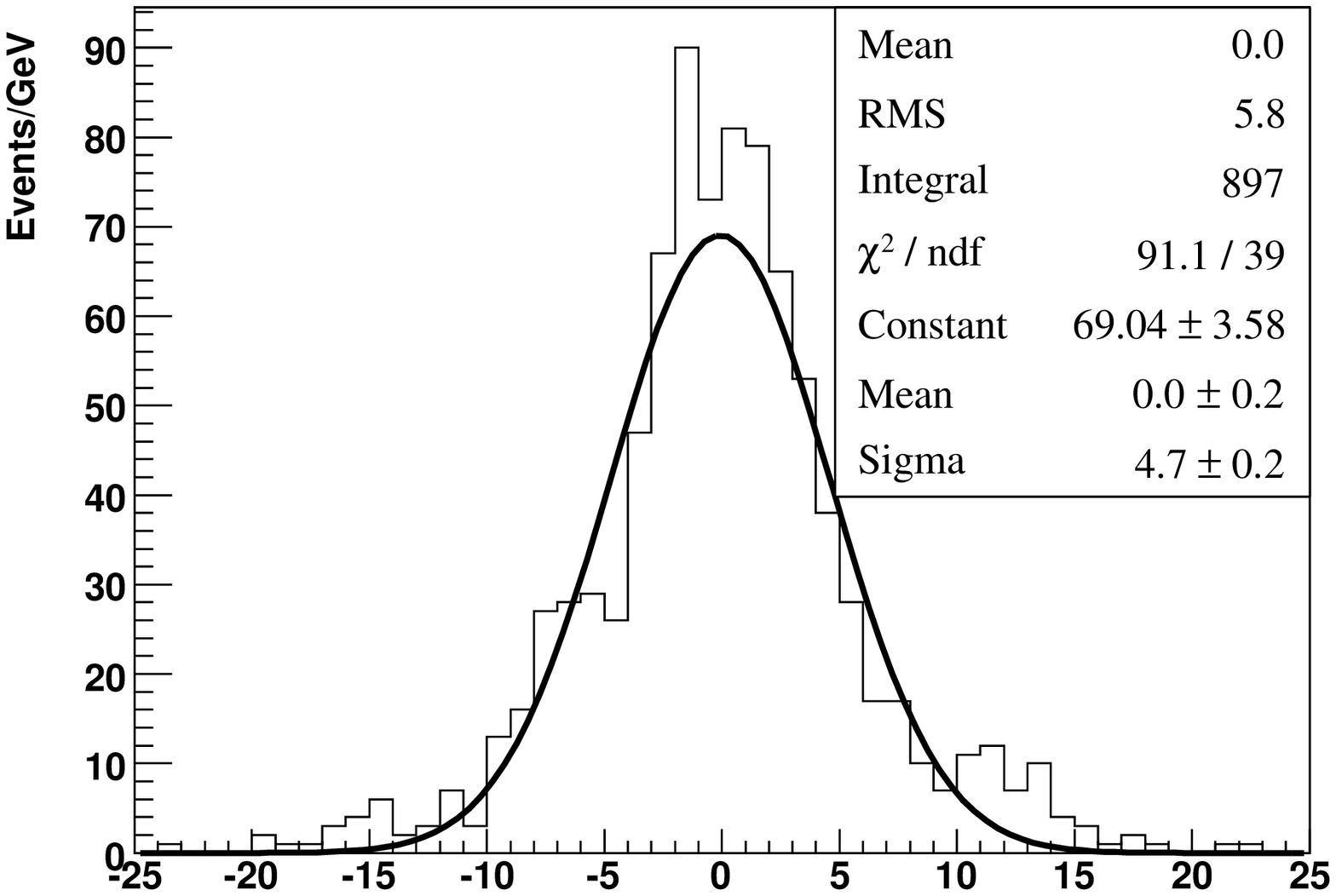}
\begin{picture}(0,0)
\put(-220,120){{\tiny \mex (GeV)}} \put(-40,120){{\tiny \mey (GeV)}}
\put(-40,-2){{\tiny \meperp (GeV)}} \put(-220,-2){{\tiny \mepara (GeV)}}
\end{picture}
\end{minipage}\hfill
\caption{Top: resolution on $\vec{\met}$, projected onto the (X,Y)
  coordinate system, for unmodified, fully simulated $\Z \ra ee$
  events. Bottom: $\vec{\met}$ resolution in the ($\|,\bot$)
  coordinate system. The absence of bias along the X and Y axes show
  that the overall calibration is correct on average, but the observed
  bias along the $\|$-axis, corresponding to the \Z line of
  flight, indicates imperfect calibration of the response to the event-by-event topology.} 
\label{fig::etmiss}
\end{figure}
\end{centering}

\noindent
As this discussion illustrates, \met\ reconstruction is a very
difficult experimental algorithm to control, especially to
the level of precision desired here. Therefore, we cannot claim at
present that the sensitivity quoted in the previous section will
indeed be reached. Instead, lacking the proof that the statistical
enhancement can be fully exploited, we assume an overall uncertainty
of $\delta\MW(\met) = 5$~\MeV. This number is a factor 3 higher than
the purely statistical sensitivity, and a factor three smaller than
the systematic uncertainty obtained in the recent CDF
measurement~\cite{ex:cdfMwRun2} based on an integrated luminosity of
200~\ipb\ and about 8000 \Z events for calibration of the hadronic
recoil.

\section{Theoretical uncertainties}
\label{sec:theounc}

We discuss below the uncertainties related to imperfect physics
modeling of \W production. The correlation of the mass measurement with
the \W width, the impact of final state radiation, and
biases in the \ptl\ and \mtw\ distributions induced by \ptw\ and \yw\
distortions are discussed in turn.

\subsection{\W boson width: $\delta\MW(\Gamma_\W)$}
\label{subsec:systgammaw}
A change in the \W width \GW\ affects the Jacobian peak, and 
can cause a bias in the \W mass measurement. To assert the size of
this effect, samples with the same \W mass but \W widths varying in
the range $1.7-2.5$ \GeV were produced and subsequently fitted.
The relation between \GW\ and \MW\ in the fit is linear, with
a slope depending on the distribution used in the mass fit. If the
\W transverse mass is used, we find:
$$
\frac{\partial\MW}{\partial_{rel}\Gamma_W} = 3.2 \MeV/\%
$$
\noindent If the lepton transverse momentum is used, we find:
$$
\frac{\partial\MW}{\partial_{rel}\Gamma_W} = 1.2 \MeV/\%
$$

\noindent
The intrinsic width of the \W resonance \GW\ has been measured
to be $2.141 \pm 0.041 \gev$, while the SM prediction is $2.0910 \pm
0.0015 \gev$ \cite{pdg2006}. It should be taken into account that the
LHC data can be expected to improve the precision on the \W width 
as well as on \MW. Earlier measurements of
\GW~\cite{ex:TevGwRun1,ex:CDFGwRun2} are affected by the same    
systematic uncertainties as those discussed in this paper. Hence,
anticipating on our results, we assume that an improvement by a 
factor five should be achievable, respectively leaving $\delta\MW(\GW) = $1.3 and 
0.5 \MeV\ for the \mtw\ and \ptl\ fits.

\subsection{QED final state radiation: $\delta\MW(QED)$}
\label{subsec:systqed}
Final state radiation causes significant distortions of the naive,
lowest order \pt\ spectrum of the \W decay leptons. We estimate
the stability of the theoretical calculation below, using the {\tt
PHOTOS} program~\cite{gen:photos} as a benchmark.\\

\noindent
The numerical importance of final state radiation is illustrated in
Figure~\ref{fsr:fig1}, which displays the distribution of the measured
lepton energy fraction (relative to their energy in the absence of
FSR). For electrons, measured $via$ calorimetric energy clusters, most
of the (collinearly radiated) photon energy is collected in the
cluster. The momentum of muons tracks, on the contrary, is measured
independently of any photon radiation. The average values of the
distributions lie at about 99\% of the original value, meaning that ignoring the effect
entirely would cause a bias on the \W mass of about 800 \MeV. The
theoretical stability of the calculation is thus of critical
importance.\\ 

\begin{figure}
\begin{center}
\includegraphics[width=0.8\textwidth]{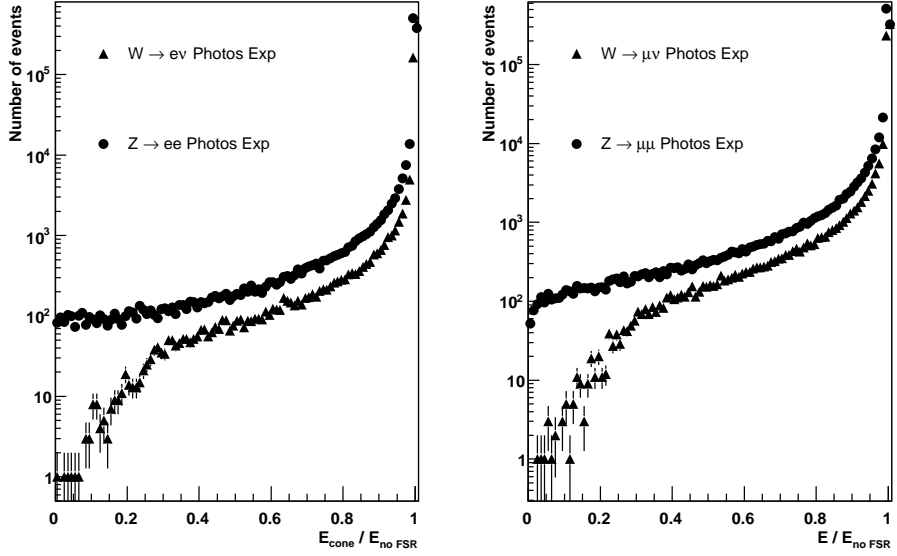}
\caption{\label{fsr:fig1} Distribution of the measured lepton energy
  fraction (\ie\ relative to their energy in absence of FSR). {\tt
  PHOTOS} is run in exponentiated mode. The energy of electrons is
  measured dressed with all photon energy radiated within a cone of
  radius 0.1, corresponding to the size of reconstructed EM clusters. 
  Muon momentum is measured bare, after FSR.}  
\end{center}
\end{figure}

\noindent
In recent versions of {\tt PHOTOS}, it is possible to switch between
several theoretical assumptions. In particular, \W and \Z
boson decays can be simulated with photon emission up to {\cal
O($\alpha$)}, {\cal O($\alpha^2$)}, {\cal O($\alpha^4$)}, or with
photon emission exponentiation~\cite{th:yfs}. To study the model
differences, we have generated about $10^6$ events for each setting,
and for each production and decay channel ($\W \ra \ell\nu$,
$\Z \ra \ell\ell$, for $\ell = e, \mu$).\\

\noindent
The average values of the energy fractions discussed above are shown
in Figure~\ref{fsr:fig2}, for successive theoretical refinements.
The different average values for electrons and muons
reflect the different ways their energy or momentum is measured.
The calculation appears stable to about 1-2$\times 10^{-4}$, the
residual differences being compatible with coming from the finite sample
statistics only. It is unfortunately not practical to further increase
the samples sizes and quantify the stability to better precision.\\

\begin{figure}[tp]
\begin{center}
\includegraphics[width=0.8\textwidth]{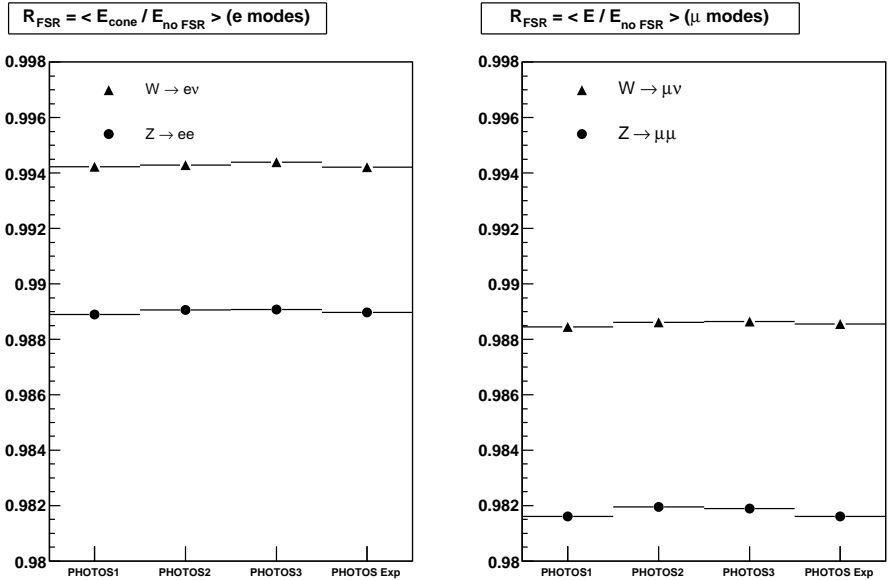}
\caption{\label{fsr:fig2} Averages of the distributions of
  Figure~\ref{fsr:fig1}, for various {\tt PHOTOS} settings (see text).} 
\end{center}
\end{figure}

\noindent
To improve on the above argument, consider the \Z boson
mass measurement at LEP1~\cite{ex:lep1final}. Similarly to our case, QED
corrections, in the form of initial state radiation off the electron
beams, have a large impact on the \Z lineshape, inducing a
decrease of the cross-section of about 
30\%, and a shift of the peak position of about
100~\MeV. Nevertheless, the theoretical uncertainty on these effects
are estimated to 0.3~\MeV, compared to a total measurement uncertainty
of 2.1~\MeV. The theory of QED radiation thus carries negligible
uncertainty.\\

\noindent
For the QED induced \MW\ uncertainty to be as small, the event
generators used to produce our templates thus need to have similar
theoretical accuracy, with the additional complication that the
present analysis requires an exclusive description of the final state
(i.e, a complete description of the photon distributions), whereas
the \Z lineshape analysis only relies on the effective energy of the
beams after radiation. In~Ref.~\cite{gen:photosnlo}, the accuracy of the
{\tt  PHOTOS} algorithm is upgraded to NLO accuracy. Similarly, the
{\tt HORACE} event generator~\cite{gen:horace} contains QED and weak
corrections to NLO accuracy. Both programs implement photon emission 
exponentiation.\\

\noindent
We thus assume that ultimately $\delta {\MW}(QED)\leq 1$~\MeV\ can be
reached. This assumption is conditioned by the availability of the
necessary tools in time for the measurement.\\

\noindent
Let us finally note that \W and \Z events behave differently
under QED radiation. The
average energy fraction in \Z events is 5-7$\times 10^{-3}$
smaller than in \W events, depending on the final state. The
energy scale measurement (\cf\ Section~\ref{subsec:lepscale}) and the
\W mass measurement should properly account for the difference in
the respective QED radiation patterns. We will come back to this point
in Section~\ref{sec:correlations}.

\subsection{\W distributions}
\label{subsec:wdist}

The \W rapidity and transverse momentum distributions result
from the interplay of the proton structure functions, and strong
interaction effects at the \W production vertex. To simplify the
discussion, we will consider the longitudinal and transverse
distributions independently, as respective results of parton distributions and QCD higher
orders.

\subsubsection{Rapidity distribution: $\delta\MW(\yw)$}
\label{subsec:systpdf}
The \W rapidity distribution is essentially driven by the proton parton
density functions (PDFs). 
Our study is based on the CTEQ6.1 structure functions sets~\cite{th:cteq6},
which provide, in addition to the global best fit, PDFs corresponding
to the variation of each diagonal parameter (i.e, the linear
combination of input parameters that diagonalize the covariance
matrix) within its estimated uncertainty. The PDF-induced
uncertainty for an observable is obtained by computing its value
with all sets, taking the central value as given by the best fit, and
quadratically summing the biases (w.r.t the best fit value) obtained
from the uncertainty sets.\\

\noindent
As illustrated in Figure~\ref{fig:pdfsyst} (see also~Ref.~\cite{lhc:cmsMw}), the
current PDF uncertainties induce an uncertainty in the \W rapidity
distributions which, through acceptance effects, propagates a
systematic uncertainty on the \W mass determination of $\sim$25~\MeV. We
present  below an attempt to estimate how this will improve with the LHC data.\\ 

\begin{figure}
\begin{center}
\includegraphics[width=0.6\textwidth]{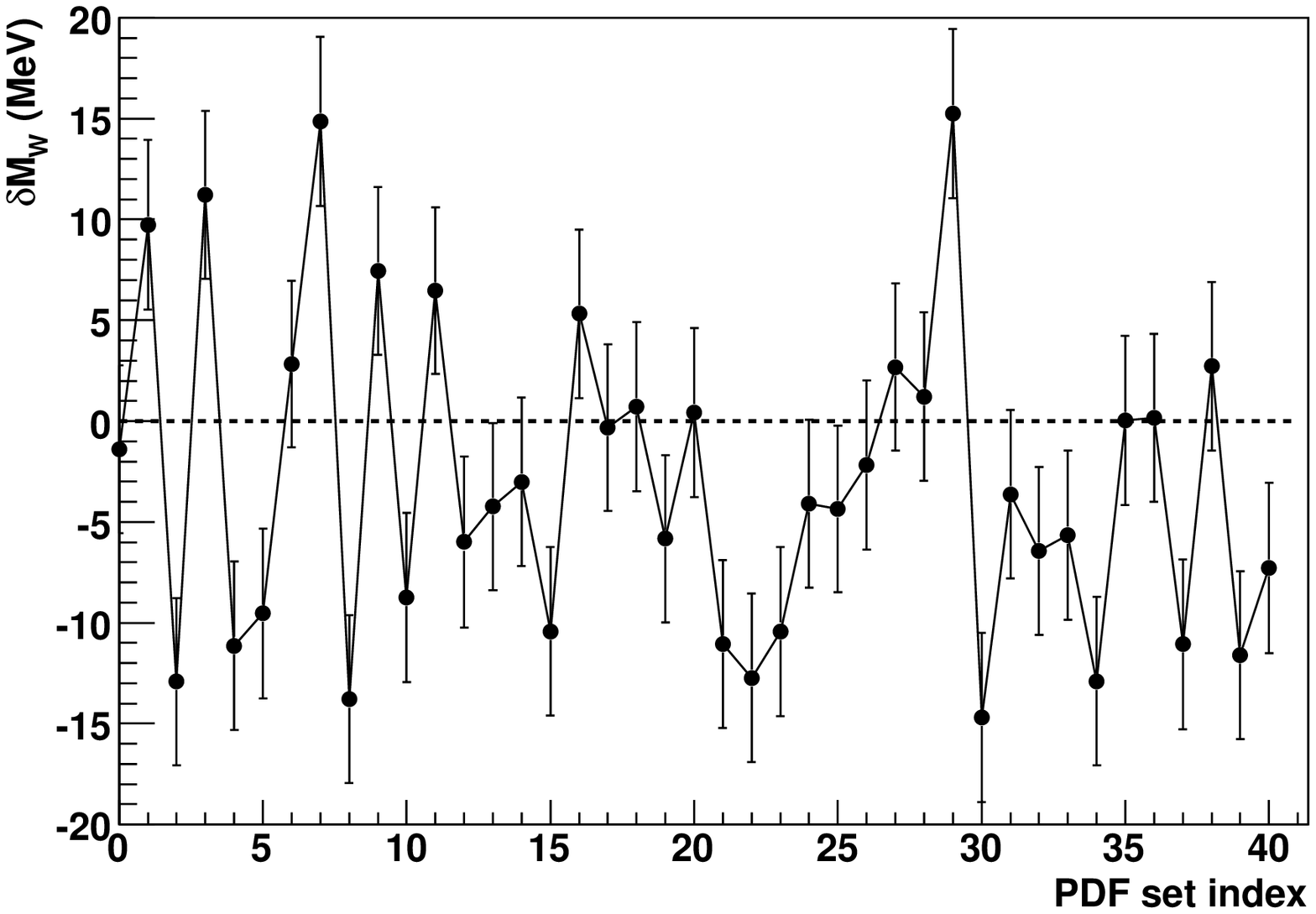}
\caption{\label{fig:pdfsyst} Bias on \MW\ obtained when varying the
proton PDFs within their uncertainties. Each point on the abscissa
correponds to a given PDF set: set 0 is the best fit, and gives 0
bias up to the statistical uncertainty of the fit; sets 1-40 are the
uncertainty sets, each inducing a given bias on \MW. The total
uncertainty on \MW\ is given by the quadratic sum of the biases,
giving $\delta\MW \sim 25 \MeV$.} 
\end{center}
\end{figure}

\noindent
At the LHC, \W and \Z particles are essentially produced through sea
quark interactions; the influence of valence quarks is small. Low-$x$,
high-$Q^2$ sea quarks mainly evolve from higher $x$, lower $Q^2$
gluons, and a consequence from perturbative QCD flavour symmetry is
that up to initial asymmetries and heavy-quark mass effects, the
different quark flavours should be represented democratically. This
then implies that the impact of sea quark PDF uncertainties on \W and \Z
production should be very similar. In other words, when varying PDFs
within their uncertainties, one expects a strong correlation
between the induced variations of the \W and \Z distributions.\\ 

\noindent
This is confirmed by Figure~\ref{pdf1}~\footnote{This plot is
reminiscent of Figure~2 in~\cite{th:Nadolsky2004}, displaying
similar correlations in the production rates. Note that for our
purpose, normalizations are irrelevant and we are interested only
in the distributions.}.  
On the left, the correlation between the widths of the \W and \Z boson
rapidity distributions is displayed. We choose to use the
distributions RMS, denoted $r_y^\W$ and $r_y^\Z$, to quantify
their width. The
current CTEQ6.1 prediction, $r_y^\Z = 2.16 \pm 0.03$, will be
refined to a precision of $\delta r_y^\Z = 0.001$. Exploiting 
Figure~\ref{pdf1} (right), which quantifies the correlation between
$r_y^\W$ and $r_y^\Z$, this can be translated into a prediction of
the \W boson rapidity distribution, $\delta r_y^\W = 0.0013$, to
be compared to the current prediction $r_y^\W = 2.24 \pm 0.03$.\\

\begin{figure}
\begin{center}
\includegraphics[width=0.48\textwidth]{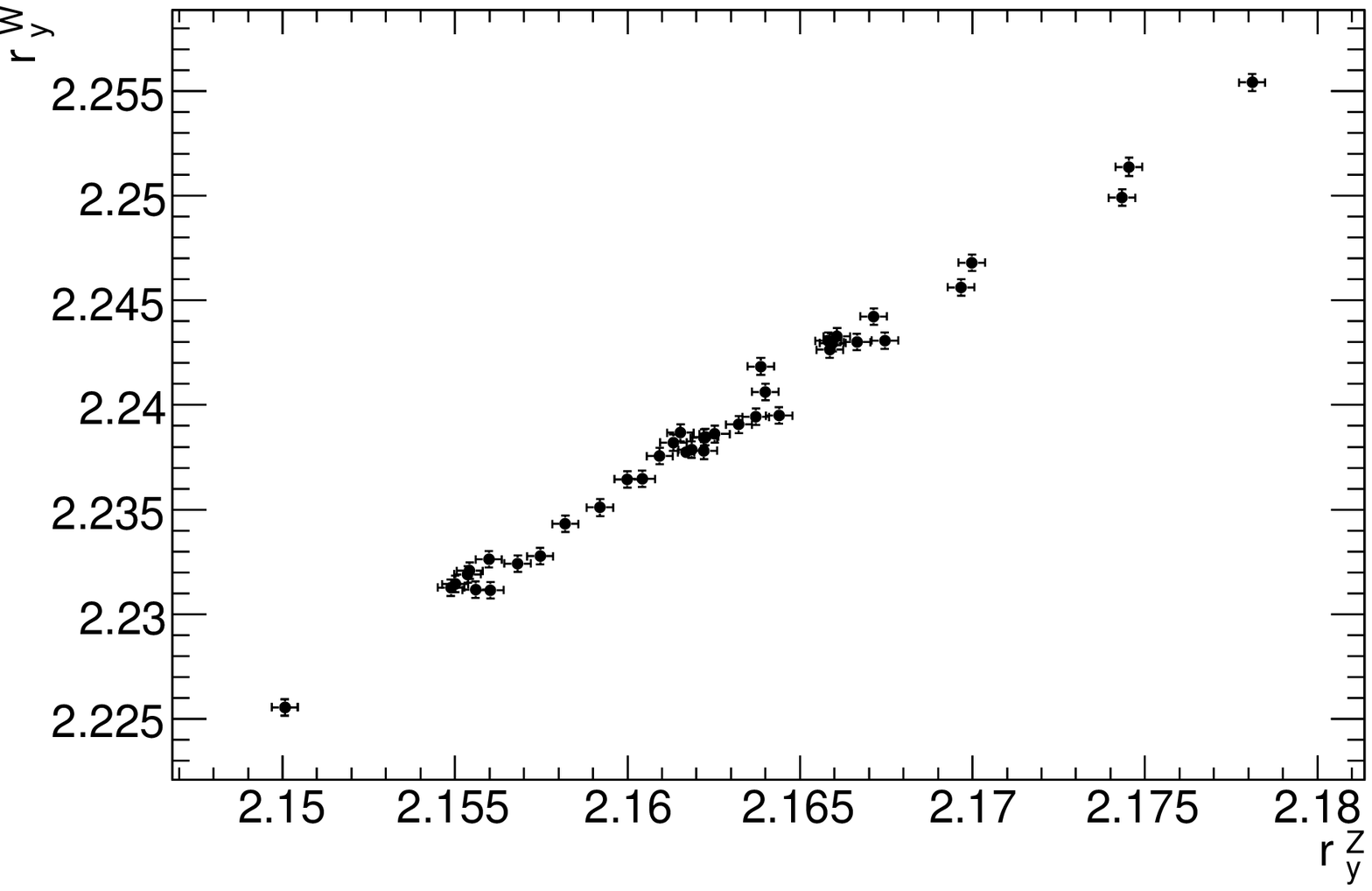}
\includegraphics[width=0.48\textwidth]{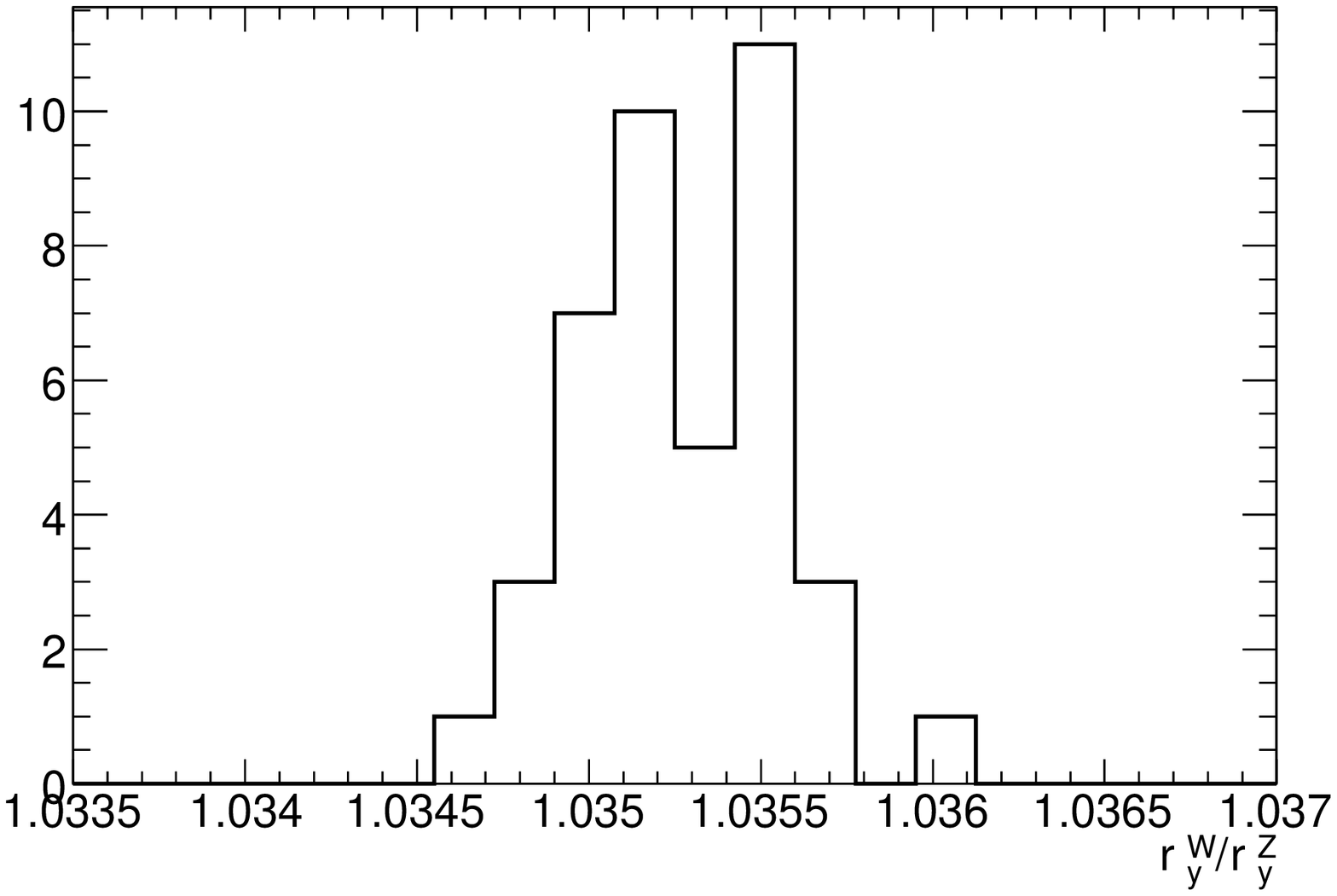}
\caption{\label{pdf1} Left : correlation between the spreads (RMS)
$r_y^\W$ and $r_y^\Z$ of the \W and \Z rapidity 
  distributions, when varying the CTEQ6.1 PDFs within their estimated 
  uncertainties. The fitted pseudo-data are scaled to an integrated
  luminosity of~10~fb$^{-1}$. Right : distribution of the ratio
  $r_y^\W/r_y^\Z$, again varying the PDFs within their
  uncertainties. The spread of the ratio distribution is $4 \times
  10^{-4}$.}  
\end{center}
\end{figure}

\begin{figure}
\begin{center}
\includegraphics[width=0.6\textwidth]{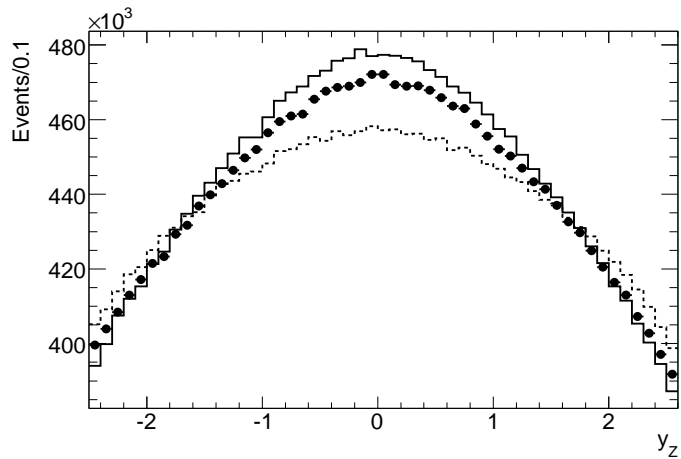}
\caption{\label{pdf2} The line histograms represent two extreme
  predictions for the \Z rapidity distribution, as given by the CTEQ6.1
  PDF sets. The points are pseudo-data, obtained with the central
  set, and scaled to an integrated luminosity of~10~fb$^{-1}$.} 
\end{center}
\end{figure}

\noindent
One thus expects an improvement on the \Z rapidity distribution by a
factor $\sim$30. This is also illustrated in Figure~\ref{pdf2}, where
two extreme predictions (with current knowledge) of the \Z rapidity
distribution are compared with an example distribution representing
the same measurement. Given the residual decorrelation between the \W
and \Z distributions, this translates into an improvement on the \W
rapidity distribution by a factor $\sim 23$.\\

\noindent
Starting with $\delta \MW(y_\W) \sim$25~\MeV, putting in a precise
measurement of the \Z rapidity distribution at the LHC, and exploiting
the strong correlation between the \W and \Z production mechanisms, we 
thus anticipate a final uncertainty from the description of the \W
rapidity distribution of $\delta \MW(y_\W) \sim$1~\MeV.\\

\noindent
In practice, the analysis will of course proceed $via$ a formal QCD
analysis to the LHC data: the measured \Z differential
cross-section ${\mathrm d}\sigma/{\mathrm d}y$, together with
other measurements (see below), will be fed to parton distribution fits, and
the systematic $\delta\MW(\yw)$ from the improved PDF sets will be
evaluated as above. The present discussion however allows to estimate
the expected improvement while avoiding these complications.\\

\noindent
Let us also note that \Z rapidity distribution can be analyzed over a
domain that fully includes the range relevant for \W  
production. In ATLAS (as in CMS), the usual \Z acceptance, given by
$|\eta_\ell| < 2.5$ for both decay leptons, can be extended in the electron
channel by allowing one of the electrons to be detected within $|\eta_e|
< \sim 4.9$. In addition, high-rapidity \Z events will be produced and
detected at LHCb (for example, the geometric acceptance of the muon
detector is approximately $2.1 < |\eta_\mu| < 4.8$). Accounting for
this, and as illustrated in Figure~\ref{pdf3}, the \W rapidity
range selected for the \MW\ measurement is entirely included in the \Z
one. This remains true in terms of the parton momentum fractions.\\ 

\begin{figure}
\begin{center}
\includegraphics[width=0.6\textwidth]{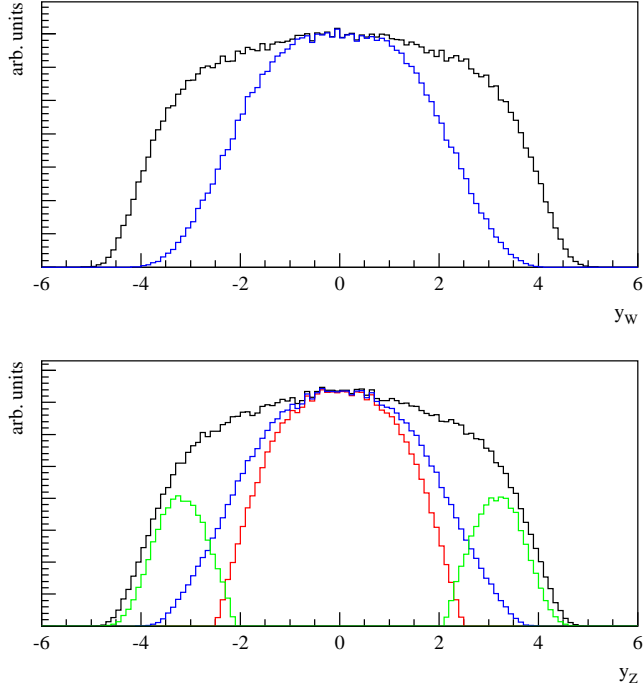}
\caption{\label{pdf3} Upper plot: the outer histogram represents the complete
  rapidity distribution for \W production at the LHC; the inner
  histogram represents the range selected by the condition $|\eta_\ell| <
  2.5$. Lower plot: the outer histogram represents the complete
  rapidity distribution for \Z events. The innermost histogram is
  obtained requiring two decay leptons within $|\eta_\ell| < 2.5$; the
  intermediate histogram is obtained when allowing one electron within
  $|\eta_\ell| < 4.9$. The two symmetric histograms at high rapidity
  correspond to the LHCb muon acceptance.} 
\end{center}
\end{figure}

\noindent
We conclude this section with some caveats. The above results partly
are a consequence of the assumed flavour and charge symmetry in the
low-$x$ proton; notably, the parton parametrisations used in the fits
used above assume that $d(x) = \bar{d}(x) = u(x) = \bar{u}(x)$ at
low-$x$, and $s = \bar{s}$ at all $x$. This implies the strong
correlation discussed above, since the \Z production rate is
proportional to $u\bar{u} + d\bar{d} + \ldots$, and the \W rate
is proportional to $u\bar{d} + d\bar{u} + \ldots$. It is thus
important to quantify the dependence of our result on these hypotheses.\\

\noindent
The anti-quark flavour asymmetry $\bar{u}-\bar{d}$ was measured to be non-0
in the region $0.015 < x < 0.35$, and $Q^2 \sim 50
\GeV^2$~\cite{ex:udbarNA51,ex:udbarE866}, in contradiction with the
flavour symmetry assumption. The relative asymmetry,
$(\bar{u}-\bar{d})/((\bar{u}+\bar{d})$, is however of the order $\sim
10^{-2}$, decreasing towards higher $Q^2$. Starting from
$\bar{u}=\bar{d}$ and full correlation between \W and \Z
production (\ie\ \W and \Z distributions have the same
rate of change under PDF variations), $\bar{u}\ne\bar{d}$ induces a
decorrelation of order $(\bar{u}-\bar{d})/(\bar{u}+\bar{d})\times
(u-d)/(u+d)$, where both factors are of order $10^{-2}$ (see for
example Figure~1 in~\cite{th:cteq6}). Hence, even in the presence of
non-vanishing $\bar{u}-\bar{d}$, the freedom of the \W
distributions is very limited once \Z ones have been precisely
measured. We thus assume that our estimates remain correct;
nevertheless, measurements of the \W charge asymmetry, sensitive
to $\bar{u}-\bar{d}$, will allow to verify this hypothesis.
Additional information will be provided by measuring \MW\ in $\W^+$ and $\W^-$
events separately.\\

\noindent
The proton strangeness asymmetry, $s(x) - \bar{s}(x)$, is constrained
by neutrino scattering 
data~\cite{ex:ssbarNUTEV,ex:ssbarBPZ,ex:ssbarCTEQ}. The relative asymmetry 
is rather small, even at low $Q^2$: $(s-\bar{s})/(s+\bar{s}) \sim
10^{-2}$ at $Q^2 = 10 \GeV^{2}$. It will only become smaller at $Q^2
\sim \MW^{2}$, where most of the strange sea is generated
radiatively. We consider, as above, that the contribution of the
asymmetry is small in terms of the overall \W production and its
uncertainty. However, the impact on the \MW\ measurement would need to
be studied specifically. At the LHC, the analysis of $\W^{-/+} +
c/\bar{c}$ production should provide additional insight.\\ 

\noindent
Finally, one may argue that the influence of heavy quark PDFs on \W and
\Z production is different, thus a source of decorrelation between the
two processes. The charm quark contribution to \W
production is significant ($\sim (V_{cs}c\bar{s} + V_{cd}c\bar{d} + c.c.)$), but
smaller for \Z production ($\sim c\bar{c}$). On the other hand, the $b$-quark
content contributes to \Z production ($\sim b\bar{b}$), but
negligibly to \W production ($\sim (V_{cb}c\bar{b} + c.c.)$), due to the smallness of
the off-diagonal third generation CKM matrix elements. These
differences are however accounted for by the present analysis, since
the heavy quark PDFs are included the CTEQ6.1 PDF sets; heavy flavours
are actually understood to cause in part the small decorrelation
between the \W and \Z boson distributions. Our conclusions thus remain
unchanged. \\

\noindent
The present study has been repeated using the MRST2001 PDF
sets~\cite{th:mrst2001}. The same correlation is observed between $r_y^\W$
and $r_y^\Z$, and the same result is obtained. Non-global parton 
density fits, such as those performed by the H1 and Zeus experiments,
are based on similar hypotheses and claim slightly smaller
uncertainties~\cite{heralhcA}, again preserving our result.
Finally, during the course of this work, CTEQ6.5 PDF sets became
available~\cite{th:cteq65}, which improves on the treatment of heavy
quark masses in the QCD evolution. The flavour symmetry assumptions
are however unchanged, so that the present discussion is not affected.

\subsubsection{Transverse momentum distribution: $\delta\MW(\ptw)$}
\label{subsec:systqcd}
The prediction of vector boson \pt\ distributions at hadron colliders
has long been an active subject~\cite{th:wzpt3,th:wzpt2,th:wzpt1}. 
It is also a crucial input
for the \W mass analysis, especially when using the \ptl\
observable. We discuss below the impact of \ptw\ uncertainties on
the \W mass  determination in this hypothesis.\\

\noindent
The measurable \ptw\ and \ptz\ distributions are the result of several
effects, most notably the repeated, partly non-perturbative parton
radiation occurring in the transition from the low-$Q^2$ proton
towards the hard process (commonly referred to as parton showers, or
soft gluon resummation). Another source is the  transverse momentum intrinsic
to the partons in the proton. We choose not to discuss these effects
separately. Rather, reckoning that although \W and \Z production differ
in several respects (the coupling to initial partons is different in
both phase space and flavour), the non-perturbative mechanisms are
universal, we evaluate how precisely their combined effect can be
measured in neutral current events, and how this  improves the \W
predictions. Notice that heavy flavour PDF have caused only a small 
decorrelation between \W and \Z events in the previous section; this is
assumed to remain true in this discussion.\\

\noindent
First, the relation between the bias in the modeling of \ptw\ and the
measurement of \MW\ is investigated by applying scaling factors to
the \ptw\ distributions in our pseudo-data, deducing the
corresponding \ptl\ distributions, and fitting \MW\ against
un-distorted templates. The bias in \MW\ appears to be a linear
function of the \ptw\ mis-modelling, with a slope of order 0.3, meaning 
a 3~\MeV\ bias on \ptw\ results in a 1~\MeV\ bias on \MW, when exploiting the
\ptl\ distribution. When \mtw\ is used, the effect is negligible.\\

\noindent
Neutral current dilepton events allow to measure the \ptll\
distribution, as a function of mass, over a large 
mass range. Assuming usual selections, this distribution will be
measured precisely for $30 < M_{\ell\ell} < \sim 200$~\GeV. This large lever
arm, in addition to the very precise determination of the \ptll\
distribution on the \Z peak, provides a precise control of $d\sigma/d\ptll$ when
$M_{\ell\ell} \sim \MW$. This is illustrated in Figure~\ref{ptw}, which
displays the dilepton mass dependence of its average transverse
momentum, $<\ptll>$, as predicted by {\tt PYTHIA}.\\ 

\begin{figure}
\begin{center}
\includegraphics[width=0.6\textwidth]{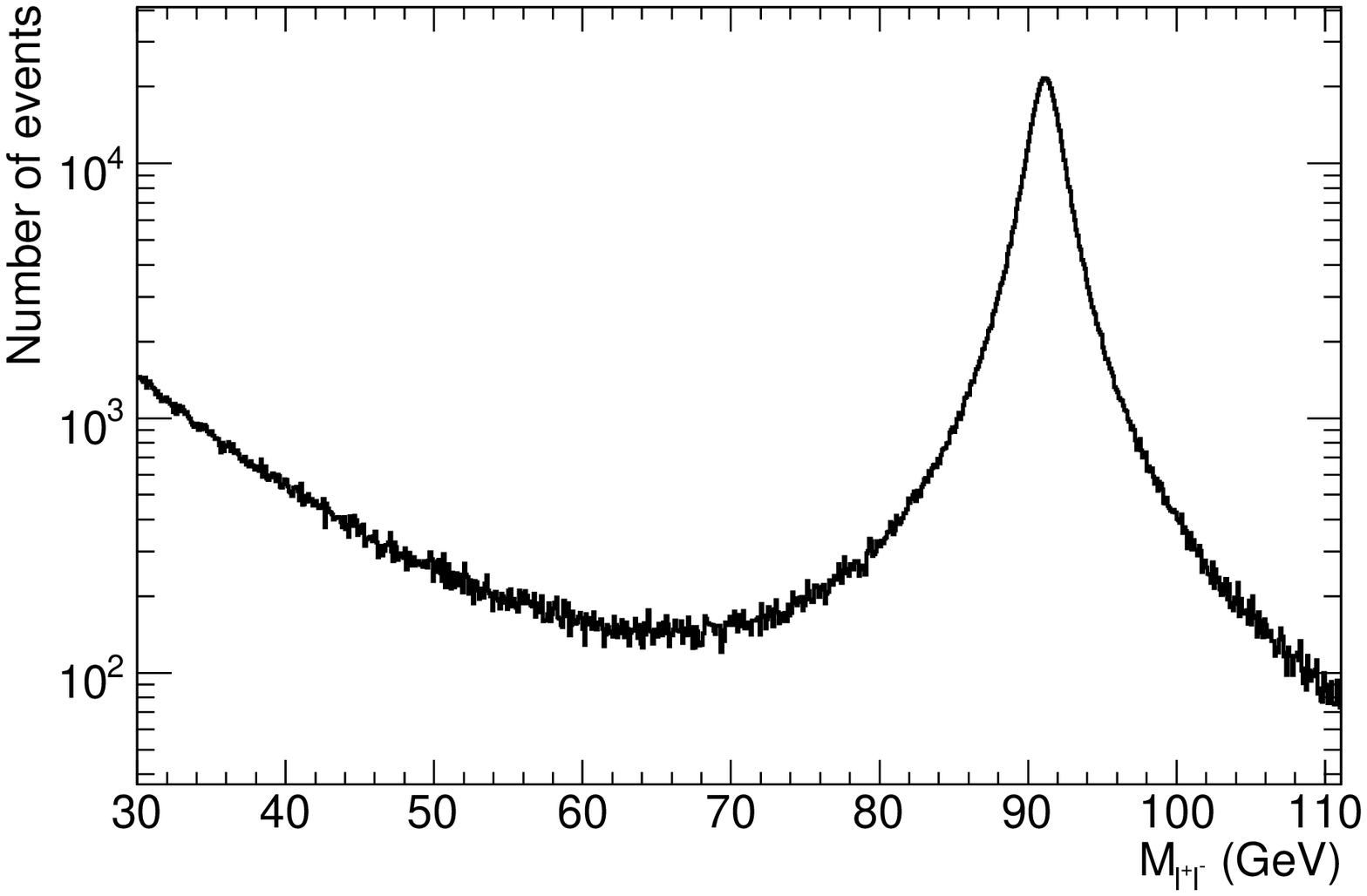}
\includegraphics[width=0.6\textwidth]{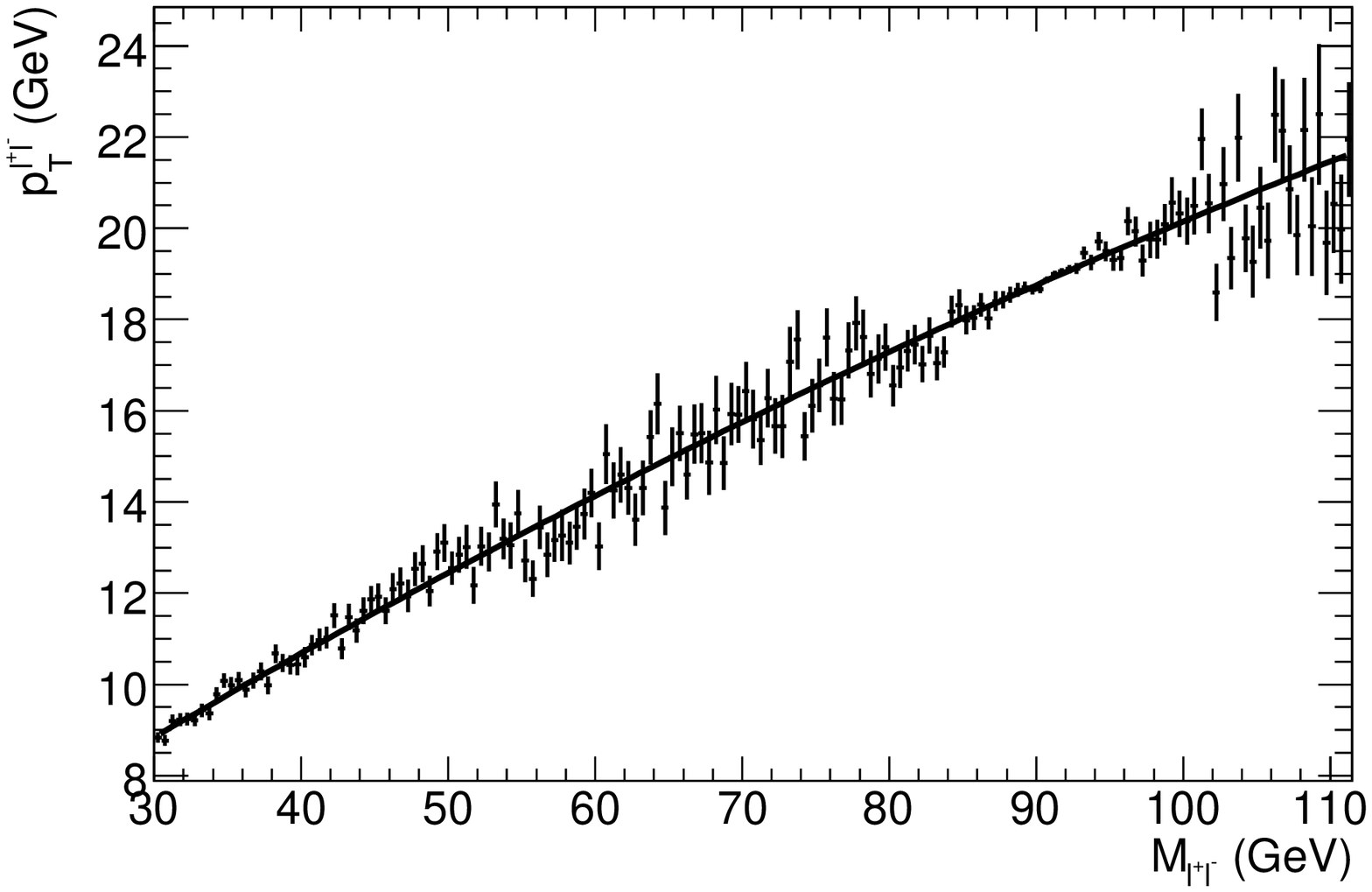}
\caption{\label{ptw} Top : Dilepton invariant mass spectrum,
from inclusive neutral current events ($\gamma$ and \Z exchange
are included). Bottom : dilepton average \pt\ as a function of the
dilepton invariant mass. The \W-mass region is strongly
constrained by the lever arm provided by the \Z peak and the
Drell-Yan rise at low mass (note the improved precision in these
regions). The points correspond to a measurement with 10~\ifb.} 
\end{center}
\end{figure}

\noindent
On the \Z peak, \ptll\ will be known to about 7 \MeV\ with an
integrated luminosity of 10 fb$^{-1}$. Thanks to the Drell-Yan
continuum, the accuracy in the region of \MW\ is still
$\sim$8~\MeV. This precision can be used to constrain the
non-perturbative parameters governing the parton shower or resummation
computations, and to predict the \ptw\ distribution with similar accuracy.
This leads to an uncertainty on \MW\ of about 3 \MeV.\\

\noindent
Arguably, the \ptw\ distribution cannot be summarized by its mean
value. However, in the low \ptw\ region (selected by the recoil cut,
cf. Section~\ref{sec:methods}), it can be empirically described by a
two-parameter function. As an exercise, the mass-dependence of the
parameters were determined on Drell-Yan events, their values and
uncertainties in the \MW\ region were used to produce \ptl\
pseudo-data as above, and  corresponding fits to \MW\ were
performed. The spread in \MW\ resulting from the uncertainty in the
empirical parameters was found compatible with the above estimate.

\section{Environmental uncertainties}
\label{sec:envunc}

\subsection{Backgrounds}
\label{subsec:backgrounds}
The leptonic \W final states benefit from low backgrounds, mostly
coming from vector boson decays; notably $\W \to \tau (\to \ell \nu \nu)
\nu$ (irreducible), $\Z \to \ell\ell$ (where one lepton is not
reconstructed), and $\Z \to \tau (\to \ell \nu \nu) \tau$. QCD dijet
events will, despite their large cross section, not be dominant. The
backgrounds from $t\bar{t}$ and $\W^+ \W^-$ events are negligible.
The systematic error on \MW\ arises from uncertainties on the
background shape and normalization in the fitting range of the \ptl\
and \mtw\ spectra.\\ 

\noindent
Uncertainties on the \W and \Z background size, relative to the signal
size, depend on cross-sections, branching fractions and
acceptances. These are obtained from the PDG 
\cite{pdg2006} and take into account the studies described in
Section~\ref{subsec:lepeff} and~\ref{subsec:wdist}. Note that in
contrast to the studies presented until now, the background
uncertainty does not scale with statistics.\\

\noindent
The background shapes are determined from simulation. They are
essentially unaffected by variations in the production, decay, and
resolution model, and play only a minor role in the overall systematic
errors. For QCD background, as a separate study, both normalization
and shape will have to be measured directly from the data.
The \ptl\ distributions, including signal and backgrounds, are
illustrated in Figure~\ref{fig:background_shapes}.\\ 

\begin{figure}
\begin{center}
\includegraphics[width=0.48\textwidth]{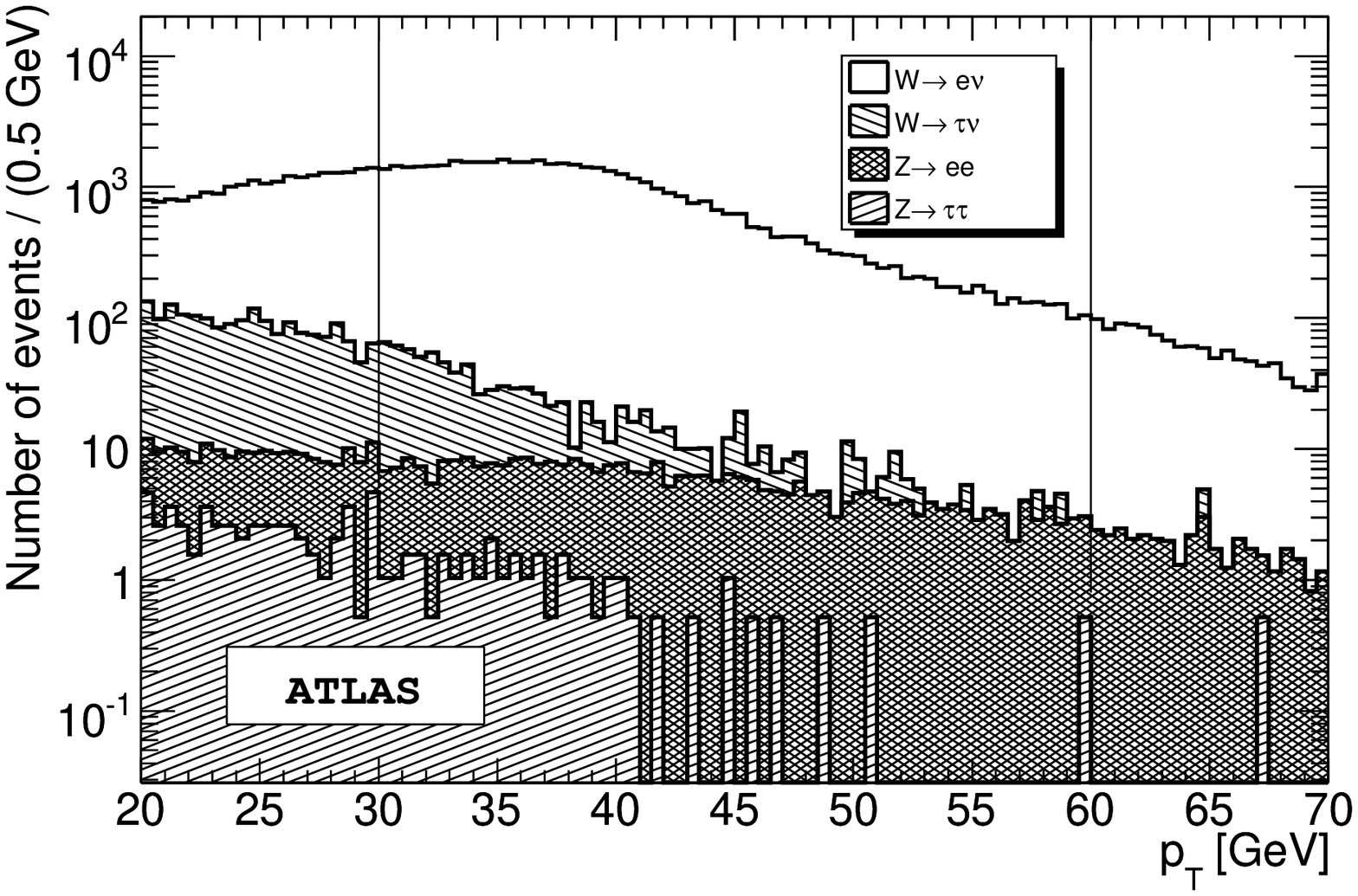}
\includegraphics[width=0.48\textwidth]{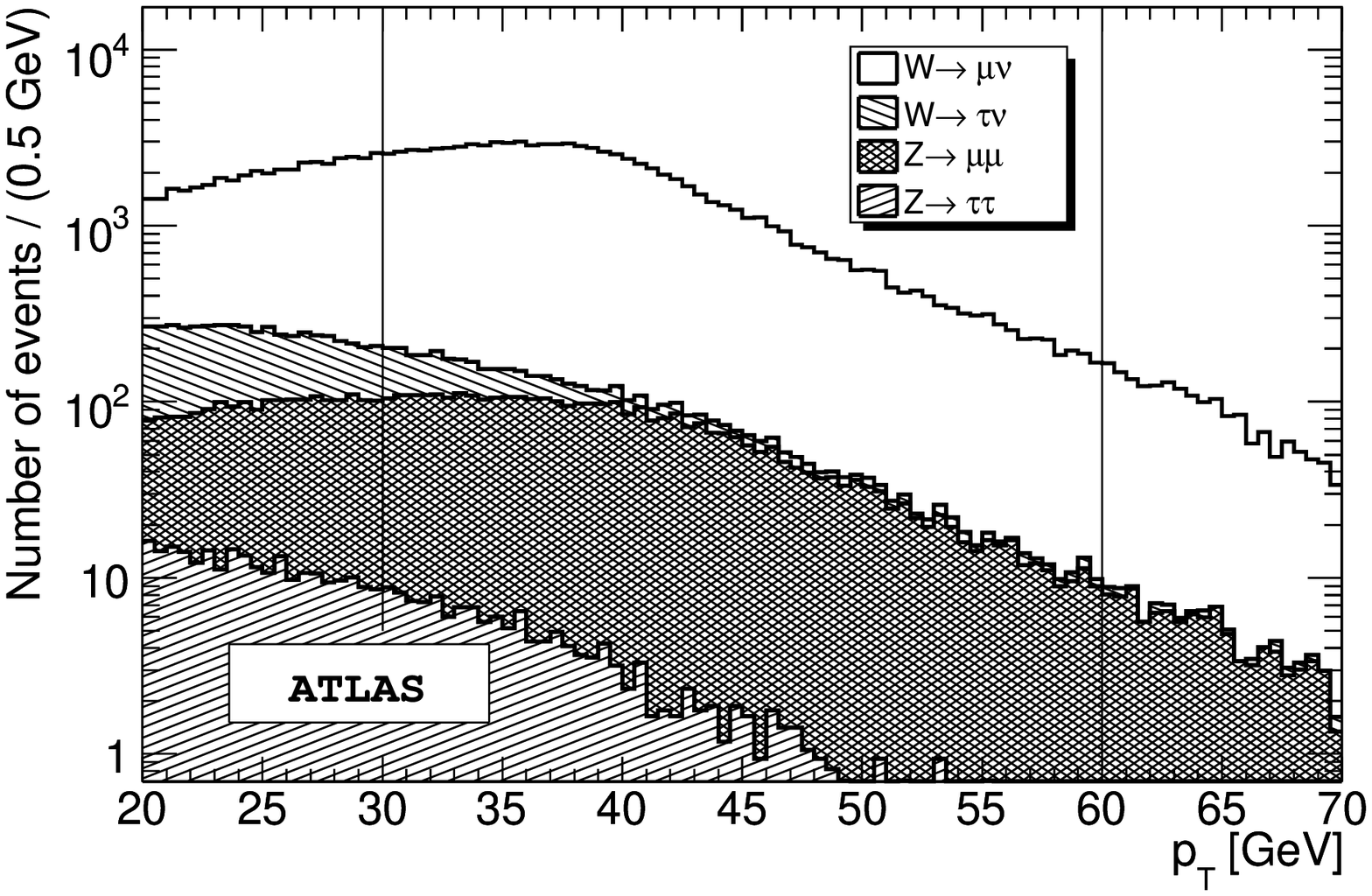}
\caption{\label{fig:background_shapes} Signal and backgrounds in the
\ptl\ distributions, for $\W\to e\nu$ (left) and $\W\to \mu\nu$
(right). The histograms correspond, from bottom to top, to $\Z \ra \tau\tau$, 
$\Z \ra \ell\ell$, $\W \ra \tau\nu$ and $\W\to \ell\nu$.} 
\end{center}
\end{figure}

\paragraph{\bf $\W \to \tau \nu$ events:}
The largest background is from $\W \to \tau \nu$ events, where the
$\tau$ decays into a lepton. This background is irreducible, as the
final state is identical to the signal; however, its \ptl\ and \mtw\
are on average lower, leaving a tail into the fitting range.
Though being the main background, its uncertainty is small, as only
$\tau$ decay parameters and the acceptance enter, with respective
uncertainties of 1\% and 2.5\%.\\

\paragraph{\bf $\Z \to \ell\ell$ events:}
The second largest background is from $Z \to \ell\ell$ events, where
one lepton is either undetected or not identified. This background can
be reduced using a \Z veto rejecting events, where the lepton and
a second isolated object (track and/or cluster) form an object with
an invariant mass between 80 and 100 \gev\ (see Figure~\ref{fig:zveto}).
Due to the high mass of the \Z boson, the \ptl\
distribution extends well into the fitting range. The \mtw\
distribution is again at low values, due to 
the smallness of missing momentum.
The size of this background has uncertainties from both the \W
to \Z cross section ratio $R_{\W\Z}$, and from the acceptance/veto
efficiency. It is expected to be larger for muons than for electrons,
as the former cannot be vetoed for $|\eta| > 2.7$.\\

\begin{figure}[tp]
\begin{center}
\includegraphics[width=0.6\textwidth]{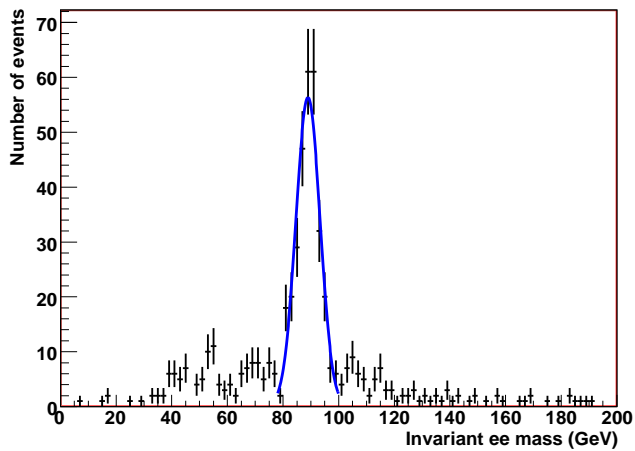}
\caption{\label{fig:zveto} Distribution of invariant mass between
lepton and a second isolated object (track and/or cluster) in $Z \to
\ell\ell$ events where only one lepton is identified. Events in the
range 80-100 \gev\ are rejected.} 
\end{center}
\end{figure}

\paragraph{\bf $Z \to \tau \tau$ events:} A small background originates from
the $\Z \to \tau \tau$ process, where one 
$\tau$ decays leptonically, while the other is not identified. While
the cross section for such a process is small, it contains
significant \met.\\

\paragraph{\bf Jet production:}
The QCD background cannot be obtained reliably from simulation. It
will thus have to be measured directly from data. For the Run I \W mass
measurement at CDF, this background could be estimated to a precision
of $\sim $50\%~\cite{phd:ASGordon}, limited by lepton identification
performances and statistics. At ATLAS, a precision of $\sim $10\% is
assumed in the electron channel, where this background is expected
to be significant. The assumed improvement is justified by the superior
granularity and resolution of the EM calorimeter~\cite{lhc:atlasTDR1}.
The muon final state is less contaminated by jet events, muons being
measured behind all calorimetry. A specific background is however
constituted by muons from hadron decays in flight. As we have no
measure of the uncertainty on this background, our results implicitly
assume it is small.
We stress that these estimates are essentially qualititative. A
realistic estimate of their impact on the measurement will only be
possible with data.\\ 

\paragraph{\bf Overall impact:}
We now estimate the overall impact of the backgrounds. The background
shapes can be empirically described by an exponential 
function in the fitting range, as illustrated in
Figure~\ref{fig:wtau_shapes} on the example of the $\W\ra\tau\nu$
background. The systematic uncertainty on \MW\ is then derived by
varying the function parameters within their uncertainties as
estimated above. The systematics uncertainty induced by the background
shapes amounts to 20\% of that induced by the normalizations.\\ 

\begin{figure}[tp]
\begin{center}
\includegraphics[width=0.75\textwidth]{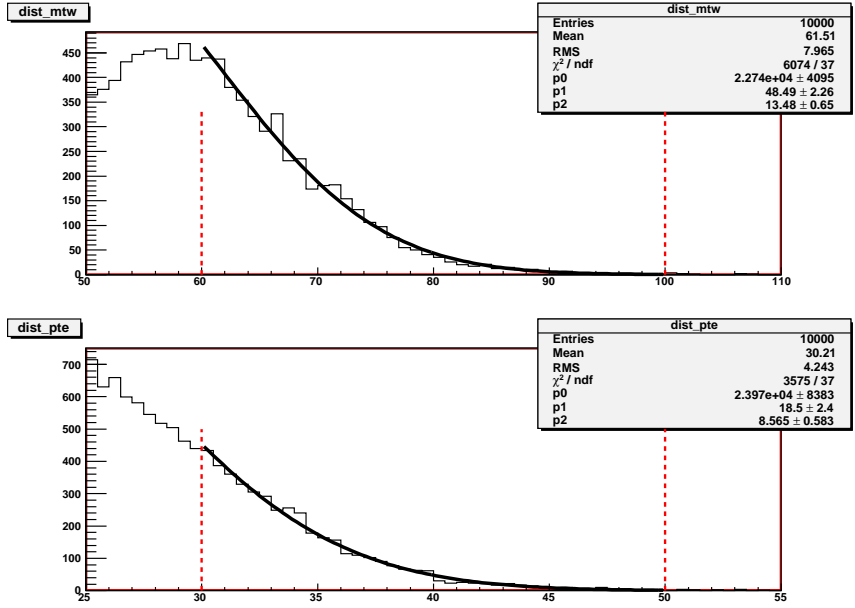}
\caption{\label{fig:wtau_shapes} $\W\ra\tau\nu$ background shape in the
        fitting range (indicated by dashed lines) for \mtw\ (top) and
	\ptl\ (bottom).}
\end{center}
\end{figure}

\noindent
The overall effect is obtained by repeating this procedure for all backgrounds.
Table \ref{tab:backgrounds} summarizes background uncertainty and
its impact on the \W mass determination.\\

\begin{table}[htbp]
\begin{center}
\begin{tabular}{lcccc}
\hline
\hline
  Background  &Variable &Error   &Derivative     &Impact (\mev)\\
\hline
  $W \to \tau \nu$
  &\mtw                 &2.5 \%  & -0.5 \mev/\%  &1.5\\
  &\ptl                 &2.5 \%  & -0.7 \mev/\%  &2.0 \\
\hline
  $Z \to \ell (\ell)$
  &\mtw                 &2.8 \% & 0.08 \mev/\%  &0.22\\
  &\ptl                 &2.8 \% & 0.09 \mev/\%  &0.26\\
\hline
  $Z \to \tau \tau$
  &\mtw                 &4.5 \%  &0.02 \mev/\%  &0.09\\
  &\ptl                 &4.5 \%  &0.03 \mev/\%  &0.14\\
\hline
  QCD events
  &\mtw                 &10 \%  & 0.04 \mev/\%  &0.40\\
  &\ptl                 &10 \%  & 0.05 \mev/\%  &0.50\\
\hline
  Total
  &\mtw                 &        &               &1.6\\
  &\ptl                 &        &               &2.1\\
\hline
\hline
\end{tabular}
\caption{\label{tab:backgrounds} Table of backgrounds along with
  its uncertainty, derivative, and impact on \MW. The overall
        systematic uncertainty from backgrounds is about 2~\MeV.}
\end{center}
\end{table}

\noindent
Combining the systematic errors from the backgrounds yields a total of
1.6 and 2.1 \MeV\ for the \mtw\ and \ptl\ distributions, respectively.

\subsection{Pileup and underlying event}
\label{subsec:pileup}
The soft hadronic activity accompanying the hard process (underlying
event), and the overlap with soft events produced in the same bunch
crossing (pile-up) generate additional particles that contribute to
the detector occupancy. In particular, the additional calorimetric
energy overlaps with the electron signal and distorts the electron scale
measurement.\\ 

\noindent
Typically, a soft event produces about 10 particles per unit rapidity
(integrated over $\phi$), with average transverse momentum $\pt \sim
500 \MeV$~\cite{lhc:atlasMinBias,lhc:atlasUE}. An electron cluster of
typical size $\delta\eta \times \delta\phi \sim 0.1 \times 0.1$ is
expected to contain about 40~\MeV\ of hadronic background, to be
subtracted from the electron signal.\\

\noindent
In particular, the hadronic background may have a non-negligible
$Q^2$-dependence, generating a non-universality between \W and
\Z events. These effects are small but need to be properly
accounted for when aiming at a precision on the absolute electron scale of 
$\delta\alpha/\alpha \sim 2 \times 10^{-5}$. \\

\noindent
This aspect was not studied here, but we follow the argument
of~\cite{lhc:atlasTDR1}. By measuring the energy flow away from any
high-\pt objects, as a function of $\eta$, independently in \W
and \Z events, a 2\% precision on the hadronic energy flow looks 
achievable. Such a result would bring down the size of the
effect from 40~\MeV\ to about 1~\MeV.\\

\noindent
We thus conclude that although soft hadronic interactions generate
shifts in the energy measurements that are large compared to the
statistical sensitivity to \MW, these shifts can be measured in the
data with sufficient accuracy. The final contribution to $\delta\MW$
is small.\\

\noindent 
This source of uncertainty affects the electron scale; the muon scale
is not affected. The impact on the recoil measurement is not discussed
here; this section is thus relevant for \ptl\ based measurements.

\subsection{Beam crossing angle}
\label{subsec:beamxing}
At the LHC, the proton beams are brought to collision at a crossing angle
of 142.5$\mu$rad~\cite{lhc:beam}. In terms of momentum, this translates into a
$7000~\gev \times 142.5 \times 10^{-6} \approx 1~\gev$ boost in the
horizontal plane ($x$-direction), per beam proton. However, in the
simulation protons collide head-on, giving rise to a systematic shift in
$p_x$ of all particles produced.\\

\noindent
Figure~\ref{fig:ptdiff} shows the difference in the transverse \W momentum
before and after taking this effect into account, $\Delta
p_{x}^\W = p_{x}^\W - p_{x}^{boost}$, which is expected to be up to
$\Delta p_{x}^\W = \MW \cdot 142.5 \times 10^{-6} \approx 11 \mev$.
However, since the \W boson line of flight has azimuthal symmetry, the
impact on the \W transverse momentum distribution is smaller, as most
of the effect is averaged out by the rotational symmetry.\\ 

\begin{figure}[tp]
\centering
  \setlength{\unitlength}{1mm}
  \begin{picture}(100,50)(0,0)
  \jput(0,0){\epsfig{file=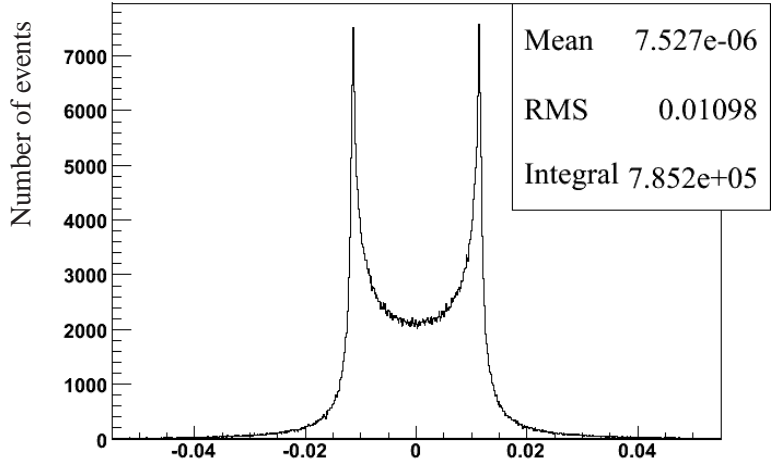,width=100mm}}
  \jput(0,33){\rotatebox{90}{Number of events}}
  \end{picture}
  \caption{\label{fig:ptdiff} Distribution of difference in the
    transverse \W momentum resulting from the boost, $\Delta p_{x}^\W
    = p_{x}^\W - p_{x}^{boost}$.}  
\end{figure}

\noindent
The size of the effect is estimated as usual, by including the
$p_{x}^{boost}$ in the pseudo-data and letting the templates
unchanged. We find that the effect is smaller than 0.1 \mev.

\section{Correlations}
\label{sec:correlations}
So far, all main sources of systematic uncertainties have been
investigated independently. Before we move to the combination of our
results, we need to address the question whether important
correlations are to be expected between the sources.
It is, however, beyond the scope of this work to discuss this issue
extensively, and we limit this section to the most important
examples. \\

\noindent
The uncertainty related to the absolute scale has the
strongest lever arm on the determination 
of~\MW\ ($\delta\MW /\delta\alpha = 1$). 
Therefore, we investigate below whether
uncertainties which affect the \W mass measurement can also bias
the absolute scale.

\subsection{Absolute scale vs. lepton reconstruction efficiency}

We repeat the procedure described in Section~\ref{subsec:lepscale}. As
before, \Z boson invariant mass templates are produced for
different scale and resolution hypotheses, and pseudo-data with scale
parameters to be determined. The impact of a \ptl-dependent lepton
reconstruction efficiency
is assessed by assuming perfect efficiency in the templates
($\epsilon=1$), and injecting the efficiency function discussed in
Section~\ref{subsec:lepeff} in the pseudo-data.\\

\noindent
The result is displayed in Figure~\ref{fig:correl1}. It appears that
the injected inefficiencies merely induce a reduction of statistics,
and hence some loss of precision in the scale determination, but no
appreciable bias: in spite of the reduction in
statistics, the reference invariant mass distribution is not
significantly distorted. Note that, since the efficiency is assumed perfect
in the templates, and realistic in the pseudo-data, any observed bias
would have been a large overestimation of the effect, representing
100\% uncertainty on the effect.\\ 

\begin{figure}[tp]
\begin{center}
\includegraphics[width=0.6\linewidth]{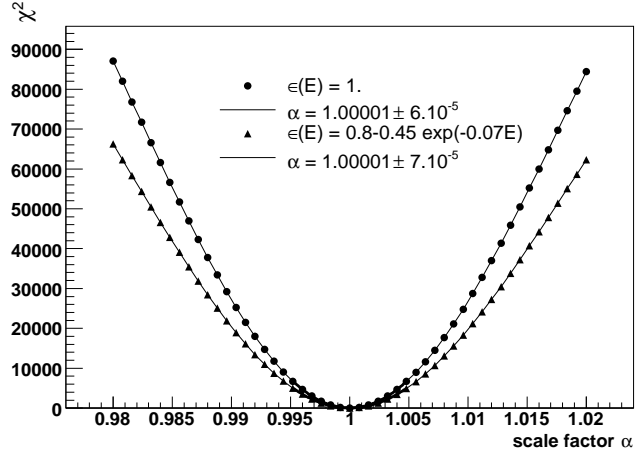}
\caption{\label{fig:correl1} Scale determination using \Z mass
templates assuming perfect identification efficiency. Inner parabola:
perfect efficiency is also assumed in the pseudo-data; outer parabola:
the pseudo-data incorporate a \pt-dependent efficiency.}  
\end{center} 
\end{figure}

\subsection{Absolute scale vs. PDFs}

Similarly as above, and also as in Section~\ref{subsec:systpdf}, we
compare \Z boson mass templates produced with the CTEQ6.1 central
set to pseudo-data produced with the 40 uncertainty sets.\\ 

\noindent
The results of the 40 fits are displayed in Figure~\ref{fig:correl2}, in
the form of biases with respect to position of the mass peak obtained
in the templates. The CTEQ6.1 uncertainty sets induce typical biases
of~$\sim$0.5 \MeV with respect to the central value. Summing over all
uncertainty sets gives a total scale uncertainty of about 2.5
\MeV. This translates into $\delta\MW \sim 2.2$ \MeV.\\

\begin{figure}
\begin{center}
\includegraphics[width=0.6\textwidth]{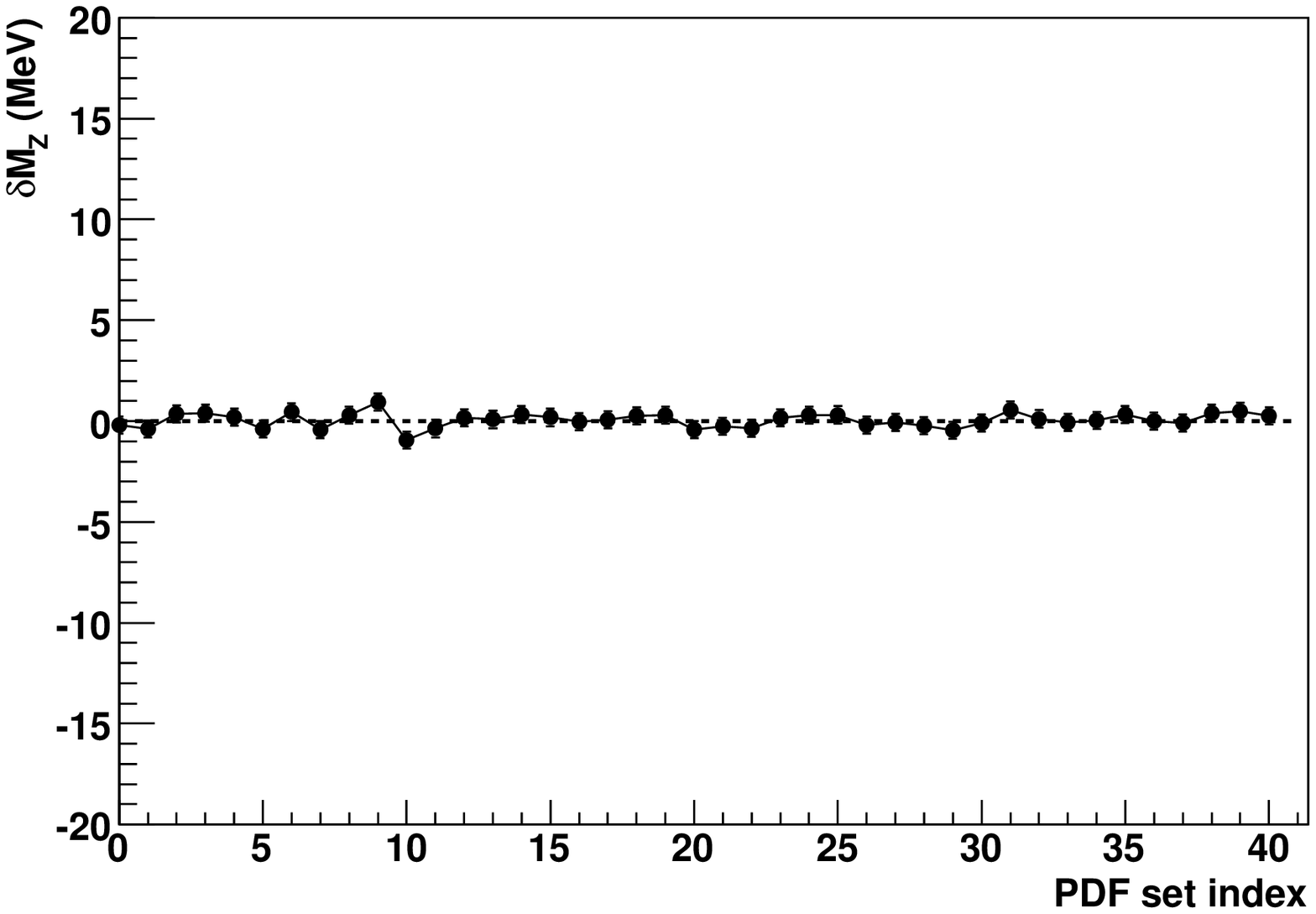}
\caption{\label{fig:correl2} Bias on \MZ\ obtained when varying the
proton PDFs within their uncertainties. Each point on the abscissa
correponds to a given PDF set: set 0 is the best fit, and gives 0
bias by definition; sets 1-40 are the uncertainty sets, each inducing
a given bias on \MZ. The total uncertainty is given by the
quadratic sum of the biases, giving $\delta\MZ \sim 2.5 \MeV$.} 
\end{center}
\end{figure}

\noindent
In other words, with current knowledge, the PDF
uncertainties induce a direct systematic uncertainty of about 25 \MeV
$via$ distortions of the \W distributions (\cf\
 Section~\ref{subsec:systpdf}), and an indirect uncertainty of 2.2 \MeV 
$via$ distortions of the \Z lineshape, propagating to the absolute
scale determination.\\

\noindent
Hence, the conclusions of Section~\ref{subsec:systpdf} are essentially
unchanged. Using measurements of the \Z boson distributions, the
PDF induced systematic uncertainty should drop to about 1 \MeV.

\subsection{Absolute scale vs. QED corrections}

QED corrections affect the determination of the absolute scale in two
ways. First, as was mentioned in Section~\ref{subsec:systqed}, the
observed \W and \Z decay lepton spectra are strongly
affected by photon emission. This effect needs to be taken into
account properly when producing the \Z mass templates.\\

\noindent
In muon final states, the theoretical distributions are based on the
final muons, after simulation of the QED photon emissions. Final state
electrons cannot be separated experimentally from the mostly
collinear photons. Hence, the simulation needs to reproduce this
recombination precisely. This demands precise theoretical control of
the photon distributions, an aspect which seems under sufficient
control (\cf\ Section~\ref{subsec:systqed}).
Likewise, a precise description of the detector geometry and EMC
shower development in the simulation are needed to properly simulate
the fraction of photon energy recombined in a given electron cluster.\\

\noindent
Secondly, as a consequence of the above, the absolute scale extracted
from \Z events actually corresponds to a mixture of photons and
electrons. In ATLAS, the EMC response to electrons and photons is
different by about 1\%, an effect coming from calorimeter geometry
(because their showers develop differently, electrons and photons of a
given energy do not ``feel'' the same sampling fraction) and from the
passive material in front of the EMC, which causes early showers or
conversions, with different probabilities for both particle
types~\cite{lhc:largTDR}. It is thus important to know whether \W and 
\Z behave similarly in this respect, and if any difference is
well understood theoretically.\\ 

\noindent
As is shown in Figure~\ref{fig:correl3}, the electron energy fraction
in EM clusters differs by about 0.6\% between \W and \Z
events, meaning that the energy scale measured in \Z events needs
to be corrected by a factor 1\% $\times$ 0.6\% = 6 $\times 10^{-5}$. 
Failing to take this factor into account would induce a bias of $\sim
5$~\MeV on the \MW\ fit. However, Figure~\ref{fig:correl3} also shows a
good stability of the theoretical prediction. Hence, although this
correction is not negligible, it does not carry a significant
uncertainty.\\

\begin{figure}[tp]
\begin{center}
\includegraphics[width=0.8\textwidth]{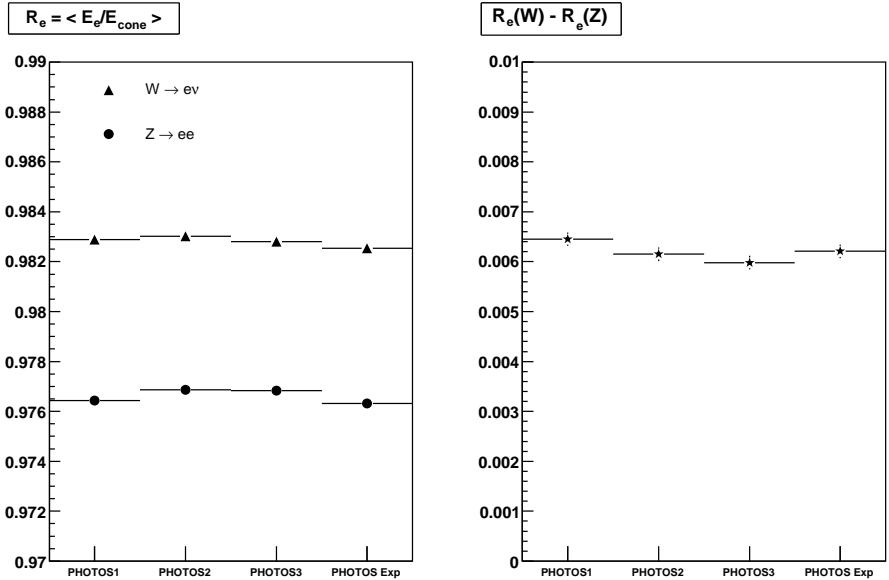}
\caption{\label{fig:correl3} For electron final states in \W and
\Z events, the energy fraction $R_e$ deposited by
electrons in reconstructed electromagnetic clusters (1-$R_e$ is photon
energy), for various {\tt PHOTOS} settings (see
Section~\ref{subsec:systqed}).}  
\end{center}
\end{figure}

\section{Impact on the \W mass measurement}
\label{sec:impact}
We summarize below our main results. Table~\ref{tab:results2} recalls
the main systematic contributions to the \ptl- and \mtw-based \MW\
measurement, with 10~\ifb\ of data. In both tables, numbers are given
for the electron and muon channels separately when applicable. \\

\noindent
The major difficulty is, as expected, the determination of the
absolute energy scale of the final state leptons
and the hadronic recoil. The analysis of the \Z
peak however allows to strongly constrain the lepton scale
uncertainty. The analysis is non trivial, because in addition to the
\Z mass parameters, many other effects enter the theoretical
description of the lineshape; most notably, QED radiation. Although
the effect is large, the theoretical understanding is adequate, as the
LEP1 \Z mass measurement indicates. The \Z mass measurement relies on
an analytical formulation of the inclusive radiation spectrum; the 
\W mass measurement at the LHC however requires a complete
Monte-Carlo implementation, providing an exclusive description of the
final state at the same level of precision. Such tools are critically
needed in the context of this measurement.\\

\noindent
The analysis of the transverse mass requires in addition a precise
calibration of the hadronic recoil using \Z events, and an unbiased
transport of the calibration to \W events. Such an algorithm is not
discussed here; the corresponding systematic uncertainty assumed here
is a compromise between the high statistical sensitivity of
the in situ calibration in ATLAS, and the actual result recently
obtained at the Tevatron~\cite{ex:cdfMwRun2}.\\

\noindent
The electron channel appears somewhat more difficult than
the muon channel. The first reason is the \pt-dependent electron
identification efficiency, which distorts the Jacobian distributions;
this effect is essentially absent in the muon channel. The second
reason is again related to QED radiation: since the muons do not
recombine with the emitted photons, the description of the effect is
purely theoretical. In the case of electrons, a large fraction of the
radiated energy is included in the electron cluster. Determining this
fraction requires a precise description of the detector geometry and 
reliable simulation of EM showers.\\

\noindent
We estimate that uncertainties related to the description of the \yw\
and \ptw\ distributions will be small once the \Z differential
cross-section will have been measured. As discussed in
Sections~\ref{subsec:systpdf} and~\ref{subsec:systqcd}, this result
relies on two assumptions. The first assumption concerns the light quark
flavour and charge symmetry in the low-$x$, high-$Q^2$ proton. We
estimated that relaxing these hypotheses within bounds allowed by the
existing data is unlikely to invalidate our result. Another assumption
is that the non-perturbative mechanisms controling the \ptw\ and \ptz\
distributions remain essentially universal, although heavy flavour
contributions to \W and \Z production  are different. The effect of
heavy flavours on the \ptw\ distribution has been
studied~\cite{th:Berge2005}, but a study comparing these effects on W
and \Z production is currently lacking. The largest remaining systematic
comes from the modeling of \ptw, in the \ptl-based measurement, contributing
a 3~\MeV\ uncertainty. The \mtw-based measurement is more stable in this
respect, but suffers additional experimental complications related to
the experimental control of the \met\ reconstruction.\\ 

\noindent
Backgrounds contribute an uncertainty $\delta\MW \sim 2$~\MeV. Of all
components, the background from jet production is the least well known,
but its contribution is expected to be small. We did not investigate
the possible impact of cosmic rays and hadron decays in flight, which
occur in the muon channels, but Tevatron experience indicates the
impact is small.\\

\noindent
All in all, a total uncertainty of about 7~\MeV\ can be achieved, in
each channel, using either the \ptl\ or the \mtw\ method, with
the equivalent of 10~\ifb\ of data. Most sources of systematic
uncertainty seem to scale with the accumulated \Z statistics;
notable exceptions are backgrounds, QED radiative corrections and
the underlying event. Their contribution to $\delta\MW$ is however subdominant.
Combining channels, and allowing for more data, we can therefore
expect further improvement.\\

\begin{table}[tp]
\begin{center}
\begin{tabular}{llcccll}
\hline
\hline
  Source         &  Effect       & $\partial\MW/\partial_{rel}\alpha$ (\MeV/\%) 
  & $\delta_{rel}\alpha$ (\%) & $\delta\MW$ (\MeV)\\ 

\hline
  Prod.\ Model   & \W width           & 1.2 & 0.4     & 0.5   \\
                 & \yw\ distribution  & $-$        & $-$       & 1   \\
                 & \ptw\ distribution & $-$        & $-$       & 3 \\
                 & QED radiation     & $-$        & $-$       & $<$1 (*)   \\
\hline
  Lepton measurement  & Scale \& lin.      & 800  & 0.005     & 4   \\
                      & Resolution         & 1    & 1.0       & 1   \\
                      & Efficiency         & $-$   & $-$       & 4.5 (e)
  ; $<$1 ($\mu$)  \\
\hline
  Recoil measurement  & Scale        & $-$    & $-$       & $-$   \\
                      & Resolution\  & $-$    & $-$       & $-$   \\
\hline
  Backgrounds    & $\W \ra \tau\nu$   & 0.15	& 2.5	    & 2.0   \\
                 & $\Z \ra \ell(\ell)$      & 0.08    & 2.8	  & 0.3   \\
                 & $\Z \ra \tau\tau$  & 0.03	& 4.5	    & 0.1   \\
                 & Jet events              & 0.05    & 10       & 0.5   \\
\hline
  Pile-up and U.E &  & & &                                              $<$1 (e); $\sim 0 (\mu)$\\
  Beam crossing angle & & & &                                            $<$0.1 \\
\hline
Total (\ptl) & & & & $\sim$7 (e); 6 ($\mu$) \\
\hline
\hline
\end{tabular}
\end{center}
\begin{center}
\end{center}

\begin{center}
\begin{tabular}{llcccll}
\hline
\hline
  Source         &  Effect       & $\partial\MW/\partial_{rel}\alpha$ (\MeV/\%) 
  & $\delta_{rel}\alpha$ (\%) & $\delta\MW$ (\MeV)\\ 
\hline
  Prod.\ Model   & \W width           & 3.2 & 0.4     & 1.3   \\
                 & \yw\ distribution  & $-$        & $-$       & 1   \\
                 & \ptw\ distribution & $-$        & $-$       & 1 \\
                 & QED radiation     & $-$        & $-$       & $<$1 (*)  \\
\hline
  Lepton measurement  & Scale \& lin.      & 800  & 0.005     & 4   \\
                      & Resolution         & 1    & 1.0       & 1   \\
                      & Efficiency         & $-$  & $-$       & 4.5 (e)
  ; $<$1 ($\mu$)  \\
\hline
  Recoil measurement  & Scale        & -200    & $-$      & $-$   \\
                      & Resolution   & -25     & $-$      & $-$   \\
                      & Combined     & $-$     & $-$      & 5 (**)   \\
\hline
  Backgrounds    & $\W \ra \tau\nu$   & 0.11	& 2.5	    & 1.5  \\
                 & $\Z \ra \ell(\ell)$      &-0.01    & 2.8	  & 0.2  \\
                 & $\Z \ra \tau\tau$  & 0.01	& 4.5	    & 0.1  \\
                 & Jet events              & 0.04    & 10        & 0.4  \\
\hline
  Pile-up and U.E &    & & &                                           $<$1 (e); $\sim 0 (\mu)$\\
  Beam crossing angle &     & & &                                       $<$0.1 \\
\hline
Total (\mtw) & & & & $\sim$8 (e); 7($\mu$) \\
\hline
\hline
\end{tabular}
\caption{\label{tab:results2} Breakdown of systematic uncertainties
  affecting the \MW\ measurement, when using the \ptl\ distribution
  (top) and the \mtw\ distribution (bottom). The
  projected values of $\delta_{rel}\alpha$ are given for a single
  channel and assume an integrated luminosity of 10~\ifb. The QED
  induced uncertainty (*) is realistic given the precision claimed for the
  \Z boson mass measurement at LEP1, but assumes that the needed
  theoretical tools will be implemented in time for the
  measurement. The recoil measurement uncertainty (**) has not
  explicitly been quantified here, but is conservatively extrapolated
  from recent Tevatron experience. See text for discussion.} 
\end{center}
\end{table}

\noindent
Let us briefly compare our results with the recent prospects presented
by the CMS Collaboration~\cite{lhc:cmsMw}. We base our comparison on
the \ptl-based \MW\ measurement and 10~\ifb\ of data. CMS claims
2~\MeV\ from the absolute scale, agreeing with our average scale result
of Section~\ref{sec:averagescale}. A simplified treatment of
non-linearities leaves a systematic uncertainty of 10~\MeV, and the
assumed 8\% relative knowledge on the resolution contributes 5~\MeV;
these numbers can be compared to the 4~\MeV\ we obtain in 
Section~\ref{sec:linearity}. We include a discussion of the
reconstruction efficiency uncertainty, which is omitted
in~\cite{lhc:cmsMw}. 
On the theoretical side, the present note and
Reference~\cite{lhc:cmsMw} agree on the initial uncertainties related
to PDFs and the description of the \W transverse momentum
distribution. Our improvements in this respect rely on an analysis of the
constraints provided by the analysis of the \Z boson differential
cross-section at the LHC. Finally, we claim a statistical sensitivity
of about 2~\MeV, compared to 15~\MeV\ in~\cite{lhc:cmsMw}. This is
explained by CMS choosing to base the \W templates on measured \Z events
($via$ the scaled observable method, or scaling the
kinematics event by event), thus paying for the smaller \Z boson
production rate. Such a procedure is in principle justified by the
reduction of other systematic uncertainties, but as we saw throughout
this paper this does not seem to be a worthy trade.

\section{Conclusions and perspectives}
\label{sec:conclusions}
We investigated the most important systematic uncertainties affecting
the \W mass determination at the LHC, and found that the analysis
of \Z production constrains the systematic uncertainties to a
total of about  7~\MeV\ per channel, exploiting 10~\ifb\ of 
data. Combining independent measurements may bring further improvement.\\  

\noindent
Among all investigated sources of systematic uncertainty, two items
in particular rely on assumptions. The first one concerns the treatment of
QED radiation. We argued that the theory is under very good control, 
having notably allowed a very precise \Z mass measurement at
LEP1, where QED effects are large, but the uncertainties finally have
an almost negligible contribution. To preserve  this situation at the
LHC, the \MW\ measurement requires QED simulation  tools providing the
same level of accuracy.\\

\noindent 
The second assumption concerns the effect of the light and heavy flavours in the
proton. Releasing the light flavour symmetry assumption in use in the
current global QCD fits will cause a decorrelation between \W
and \Z production at the LHC. This decorrelation can be expected to
be small, but will have to be measured at the LHC, notably 
using the rapidity-dependent \W charge asymmetry and the study of
associated \W/\Z + charm production. Similarly, heavy flavour PDFs
generate some decorrelation. This
decorrelation was verified to be small in the \yw\ and \yz\
distributions, and the same was assumed true for the \ptw\ and \ptz\
distributions. To verify this assumption requires a theoretical study
comparing the heavy flavours influence on soft gluon resummation in \W
and \Z events.\\ 
 
\noindent
A number of sources have not been studied explicitly, notably
the recoil measurement, affecting the \mtw distribution; the
underlying event, affecting the electron energy scale; and
\W polarization effects, affecting the leptonic angular
distributions. Other sources, like backgrounds from jets, cosmic
muons, or induced by the machine can only be studied reliably using
real data. We believe these mechanisms can be brought under sufficient
control, on the time scale of the LHC measurement of \MW.\\ 
 
\noindent 
The results presented here have only exploited \Z
boson measurements. Many other calibration processes exist, that
give additional constraints on the detector performance and on the
physics mechanisms influencing \W production. While first providing
a way to verify the robustness of the \Z-based calibrations, these
processes can help to reduce the uncertainties further in the case of
consistent results. We reserve these refinements to the analysis of
the forthcoming LHC data.

\section{Acknowledgements}

This work has continuously benefitted from feedback, suggestions and
discussions. Within the ATLAS Collaboration, we would like to thank
Lucia di Ciaccio, Amanda Cooper-Sarkar, Fares Djama, Daniel
Froidevaux, Joey Huston, Karl Jakobs, Max Klein, Ashutosh Kotwal,
Witek Krasny, Tom LeCompte, Dan Levin, Guillaume Unal,
and many others. We are also grateful to Philippe Charpentier and
Olivier Schneider from LHCb. From the theoretical community, we
acknowledge discussions with Stefan Berge, Walter Giele, Staszek
Jadach, Pavel Nadolsky, Fred Olness, Fulvio Piccinini, and Zbyszek Was. 

\bibliographystyle{h-physrev}
\bibliography{%
atlaspaper}

\end{document}